\documentclass{aastex631}

\usepackage{xcolor}   
\newcommand{\rev}[1]{{\color{black} #1}} 
\usepackage{float}
\usepackage{wasysym}
\usepackage{xcolor}
\usepackage{rotating}
\usepackage{longtable}
\usepackage{multirow}

\begin{document}

\title{Apache Point Observatory follow-up of ACcelerating Candidate ExopLanet host Stars \\(APO ACCELS): Ages for 166 Accelerating Stars in the Northern Hemisphere}

\author[0000-0003-2461-6881]{Anne E. Peck}
\affiliation{Department of Astronomy, New Mexico State University, P.O. Box 30001, MSC 4500, Las Cruces, NM 88003, USA}

\author[0000-0001-6975-9056]{Eric L. Nielsen}
\affiliation{Department of Astronomy, New Mexico State University, P.O. Box 30001, MSC 4500, Las Cruces, NM 88003, USA}
 
\author[0000-0002-4918-0247]{Robert J. De Rosa}
\affiliation{European Southern Observatory, Alonso de C\'{o}rdova 3107, Vitacura, Santiago, Chile}

\author[0000-0001-5684-4593]{William Thompson}
\affiliation{National Research Council, Herzberg Astronomy and Astrophysics, Victoria, British Columbia}

\author[0000-0003-1212-7538]{Bruce Macintosh}
\affiliation{Department of Astronomy and Astrophysics, UC Santa Cruz, Santa Cruz CA 95064}

\author[0009-0008-9687-1877]{William Roberson}
\affiliation{Department of Astronomy, New Mexico State University, P.O. Box 30001, MSC 4500, Las Cruces, NM 88003, USA}

\author[0000-0002-9156-9651]{Adam J. R. W. Smith}
\affiliation{Department of Astronomy, New Mexico State University, P.O. Box 30001, MSC 4500, Las Cruces, NM 88003, USA}

\author[0000-0003-3906-9518]{Jessica Klusmeyer}
\affiliation{Department of Astronomy, New Mexico State University, P.O. Box 30001, MSC 4500, Las Cruces, NM 88003, USA}

\author[0009-0006-4626-832X]{Asif Abbas}
\affiliation{Department of Astronomy, New Mexico State University, P.O. Box 30001, MSC 4500, Las Cruces, NM 88003, USA}

\author[0000-0001-9659-7486]{Jason Jackiewicz}
\affiliation{Department of Astronomy, New Mexico State University, P.O. Box 30001, MSC 4500, Las Cruces, NM 88003, USA}

\author[0000-0002-9771-9622]{Jon Holtzman}
\affiliation{Department of Astronomy, New Mexico State University, P.O. Box 30001, MSC 4500, Las Cruces, NM 88003, USA}

\author[0009-0000-8603-169X]{Hannah Gallamore}
\affiliation{Department of Astronomy, New Mexico State University, P.O. Box 30001, MSC 4500, Las Cruces, NM 88003, USA}


\author[0000-0002-0457-2941]{Marah Brinjikji}
\affiliation{School of Earth and Space Exploration, Arizona State University, Tempe, AZ 85287, USA}

\author{Jennifer Patience}
\affiliation{School of Earth and Space Exploration, Arizona State University, Tempe, AZ 85287, USA}

\author[0000-0002-9242-9052]{Jayke S. Nguyen}
\affiliation{Department of Astronomy \& Astrophysics, UC San Diego, La Jolla, CA 92093, USA}

\author[0000-0001-7443-6550]{Alex Madurowicz}
\affiliation{Department of Physics, Stanford University, Stanford, CA 94305, USA}


\author[0000-0002-8711-7206]{Dmitry Savransky}
\affiliation{Sibley School of Mechanical and Aerospace Engineering, Cornell University, Ithaca, NY 14853, USA}



\begin{abstract}

Directly imaged substellar companions with well-constrained ages and masses serve as vital empirical benchmarks for planet formation and evolution models. Potential benchmark companions can be identified from astrometric accelerations of their host stars. We use \textit{Gaia} DR3 and \textit{Hipparcos} astrometry to identify 166 northern hemisphere stars with astrometric accelerations consistent with a substellar companion between $0.5"$ and $1"$. For this accelerating sample we identify young stars using APO/ARCES spectra and $TESS$ light curves. From spectroscopic screening of the sample, we measure ages for 24 stars with detectable amounts of lithium, place lower age limits on 135 stars with lithium non-detections, and measure ages from $R^{'}_{HK}$ for \rev{34} stars. 129 stars have  $TESS$ light curves from which we measure ages for 20 stars with rotation rates $<15$ days, and we identify 3 eclipsing binaries. We present \rev{median ages and confidence intervals of age posteriors} for the entire sample and discuss how the overall age distribution of our sample compares to a uniform star formation rate in the solar neighborhood. We identify \rev{47 stars with median ages $<2$ Gyr, 31 stars with median ages $<1$ Gyr, and 14 stars with median ages $<0.5$ Gyr}, making them high-priority targets for direct imaging follow-up.

\end{abstract}



\section{Introduction} \label{sec:intro}

Direct imaging of exoplanets allows for demographic studies of wide-separation giant planets and for analysis of these planets' atmospheres through spectra and photometry \rev{(e.g. \citealt{2019AJ....158...13N, 2021A&A...651A..72V})}. Each direct imaging detection allows for detailed characterization of the companion \rev{(e.g. \citealt{2008Sci...322.1348M, 2010Natur.468.1080M, 2009A&A...506..927L, 2015Sci...350...64M, 2024Natur.633..789M})}. While wide-separation giant planets have relatively low occurrence rates \rev{(e.g. \citealt{2019AJ....158...13N, 2021A&A...651A..72V})}, additional companions can be identified from astrometric accelerations of their host stars \rev{(e.g. \citealt{2021ApJS..254...42B})}. A targeted survey informed by astrometric accelerations, radial velocities, and stellar ages can complement previous blind surveys by increasing the number of directly imaged substellar companions. Planets detected via astrometry have the added benefit of measurable dynamical masses making them key benchmarks for planet formation and evolution models \rev{(e.g. \citealt{2003IAUS..211..325A, 2021ApJ...920...85M, 2024ApJ...975...59M})}, especially if their ages are known \rev{(e.g. \citealt{2020AJ....159...71N, 2023A&A...672A..94D})}. The mass of a companion can be determined by combining imaging and astrometry, and the companion's age can be measured from observations of the host star. A broader sample of benchmark exoplanets will help constrain planet formation and evolution models.
	
 \textit{Gaia} astrometry is expected to be sensitive to planets between $\sim1-15 \ M_\textrm{Jup}$ with periods between $\sim0.5-10 $ yr around a wide range of stellar host masses \citep{2014ApJ...797...14P}. \textit{Gaia} Data Release 4 (DR4) is scheduled for mid-2026 and is set to include a list of astrometrically detected exoplanets between $\sim1-5$ AU  detected using only \textit{Gaia} \citep{2016A&A...595A...1G}.
 Prior to the release of DR4, substellar companions can be detected by combining positions, parallaxes, and proper motions from the \textit{Hipparcos} Catalog \citep{1997A&A...323L..49P, 1997A&A...323L..61V} and \textit{Gaia} Data Release 3 (DR3) \citep{2023A&A...674A...1G}. Using astrometry the influence of a massive planet on the path of a star can be observed, and longer-period companions can be detected by linking these two missions \rev{(e.g. \citealt{2014ApJ...797...14P, 2021ApJS..254...42B, 2024A&A...684C...2K})}. Velocity is measured as proper motion for stars in both \textit{Hipparcos} (1991.25) and \textit{Gaia} DR3 (2016.0) as well as from the difference between the two catalog positions. An acceleration of the photocenter can be identified from discrepancies between the three velocity vectors. Combined \textit{Hipparcos} and \textit{Gaia} astrometry yield typical acceleration precision as low as $\sim$1\,m\,s$^{-1}$\,yr$^{-1}$ for nearby stars, and is sensitive to longer periods than \textit{Gaia} alone \citep{2021ApJS..254...42B}. 

Several substellar companions have been detected using this method. The $3-7 \ M_\textrm{Jup}$ planet AF Lep b was identified by \rev{the} acceleration of its young host star, a member of the $\sim24$ Myr $\beta$ Pictoris moving group \citep{2015MNRAS.454..593B}, before being detected by direct imaging \citep{2023A&A...672A..94D, 2023A&A...672A..93M, 2023ApJ...950L..19F}. Higher mass substellar companions have been imaged around older accelerating stars including the $13.9-16.1 \ M_\textrm{Jup}$ companion to the $\le 500$ Myr HIP 99770 \citep{2023Sci...380..198C}, the \rev{$\sim28 \ M_ \textrm{Jup}$} companion to the $\sim750$ Myr HIP 21152 \citep{2022ApJ...934L..18K}, and the $\sim31 \ M_\textrm{Jup}$ companion to the \rev{$\lesssim1.5$} Gyr HIP 5319 \citep{2022AJ....164..152S}. Since these accelerations are from astrometric measurements of position and velocity with a limited time baseline, there is a degeneracy between the companion mass and orbital parameters, especially the period. For example, the acceleration of the $\sim500$ Myr HR 1645 appeared to be consistent with a brown dwarf companion on a $\sim100$ yr orbit but was instead shown to be caused by a stellar binary with a $<1$ yr orbit \citep{2019AJ....158..226D}.

Planets cool monotonically over time, with younger, more massive planets being brighter. Given current ground-based direct imaging sensitivity, substellar companions with a contrast ratio brighter than $10^{-6}$ can generally be detected to within $\sim0.5''$ of their host stars \rev{(e.g. \citealt{2014SPIE.9148E..0JM, 2023ASPC..534..799C})}. Depending on the luminosity and distance of the host star, a wide separation ($\sim10-30$ AU) giant planet can be detected around stars younger than a few hundred Myr, and brown dwarfs around stars younger than $\sim1$ Gyr. Thus, measuring accurate ages of host stars and identifying young systems is crucial when selecting high-priority direct imaging targets. We measure the ages of 166 northern hemisphere accelerating stars so that we can identify the youngest systems and prioritize them for direct imaging follow-up.

We use the following methods to derive the ages of our sample stars.
 \begin{enumerate}
     \item  {\bf Ca H and K emission ($R'_{HK}$):} The calcium H and K emission lines are observed as cores within deep absorption lines. The absorption lines form in the optically thick photosphere and the emission cores are generated in the optically thin chromosphere via magnetic heating. Thus, these emission cores trace magnetic activity in stars. As a star spins down with age, its magnetic and chromospheric activity decreases \rev{(e.g. \citealt{1971BAAS....3Q.455S})}. The youngest stars have the fastest rotation and, thus, the strongest magnetic activity, which causes the strongest calcium emission. $R'_{HK}$ is calibrated as an age tracer for F5-K2 stars \rev{(e.g. \citealt{1991ApJ...375..722S, 2008ApJ...687.1264M, 2020ApJ...898...27S})}. 
     
     \item {\bf Lithium:} Throughout the main sequence lifetime of solar-type and lower mass stars, convection transports surface lithium to deeper, hotter layers of the star where it is destroyed via fusion. Lithium depletion is most efficient in low-mass stars that are either fully convective or have thick convective envelopes \citep{1993AJ....105.2299S, 2021A&A...654A.137L, 2010ARA&A..48..581S} making it a useful age tracer for G, K, and M stars. Lithium equivalent width is a powerful age tracer for very young late-type (later than $\sim$K5) stars due to their rapid lithium depletion, but after about $150 $ Myr only equivalent width upper limits can be established for these stars.
     
     \item {\bf Gyrochronology:} Stars later than $\sim$F7 spin down as they age due to angular momentum loss via magnetic braking \citep{1967ApJ...150..551K}. Rotation periods can be measured from flux variations due to star spots rotating on and off the disk of the star. Longer rotation periods correspond to older ages.
     
     \item {\bf Color-Magnitude Diagram Positions:} After reaching the zero-age main sequence, the position of a star on the CMD does not change significantly during the first third of its main sequence lifetime. During the final two-thirds of its main sequence lifetime, the core becomes more helium-rich, increasing the star’s luminosity. This results in detectable movement across the CMD. Age posteriors for these stars can be derived by comparing their photometry to theoretical isochrones \rev{(e.g. \citealt{2013ApJ...776....4N, 2019AJ....158...13N})}. This method is especially powerful for A and F-type stars which have shorter main-sequence lifetimes between $\sim1-5$ Gyr.  

 \end{enumerate}
 
\begin{figure*}[h!]
\centering
 \includegraphics[scale = 0.5]{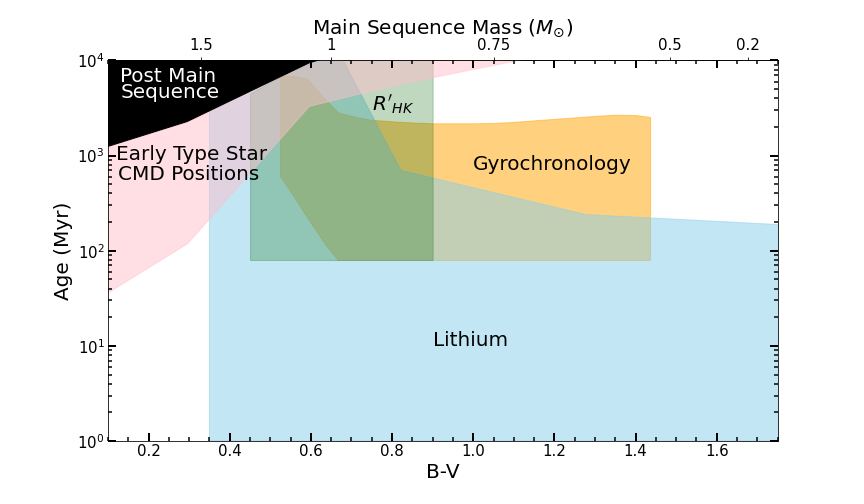}
 
 \caption{Age tracers used in this work for young, low-mass stars. This approximate schematic shows which combinations of age and mass are robustly covered by different age tracers. Each shaded region indicates where each tracer is most robust but does not include the regions where limits can be derived from each tracer (e.g. above the blue shaded region lithium is not expected to be detectable, but upper limits on lithium abundance can be used to derive lower limits on the age).}
 \label{tracers}
\end{figure*}

 An approximate schematic of which tracer can produce the most precise age for different stars is shown in Figure \ref{tracers}. Overlap allows for robust age measurements from multiple tracers for some stars. Significant scatter is observed in the age indicators for a given age and mass, which can be quantized to produce a posterior probability distribution on age. The precision of a given age tracer depends on the mass and age of the star. Several tools have been developed to derive age posteriors from these tracers such as \texttt{BAFFLES} \citep{2020ApJ...898...27S}, \texttt{EAGLES} \citep{2023MNRAS.523..802J}, \texttt{stardate} \citep{2019AJ....158..173A}, and \texttt{gyro-interp} \citep{2023ApJ...947L...3B}. We present \rev{median ages and $68\%$ and $95\%$ confidence intervals} for 166 northern hemisphere accelerating stars based on these four age tracers below.
 
\section{Sample, observations, and data reduction} \label{sec:data}

\begin{figure*}[hb!]
\centering
 \includegraphics[scale = 0.25]{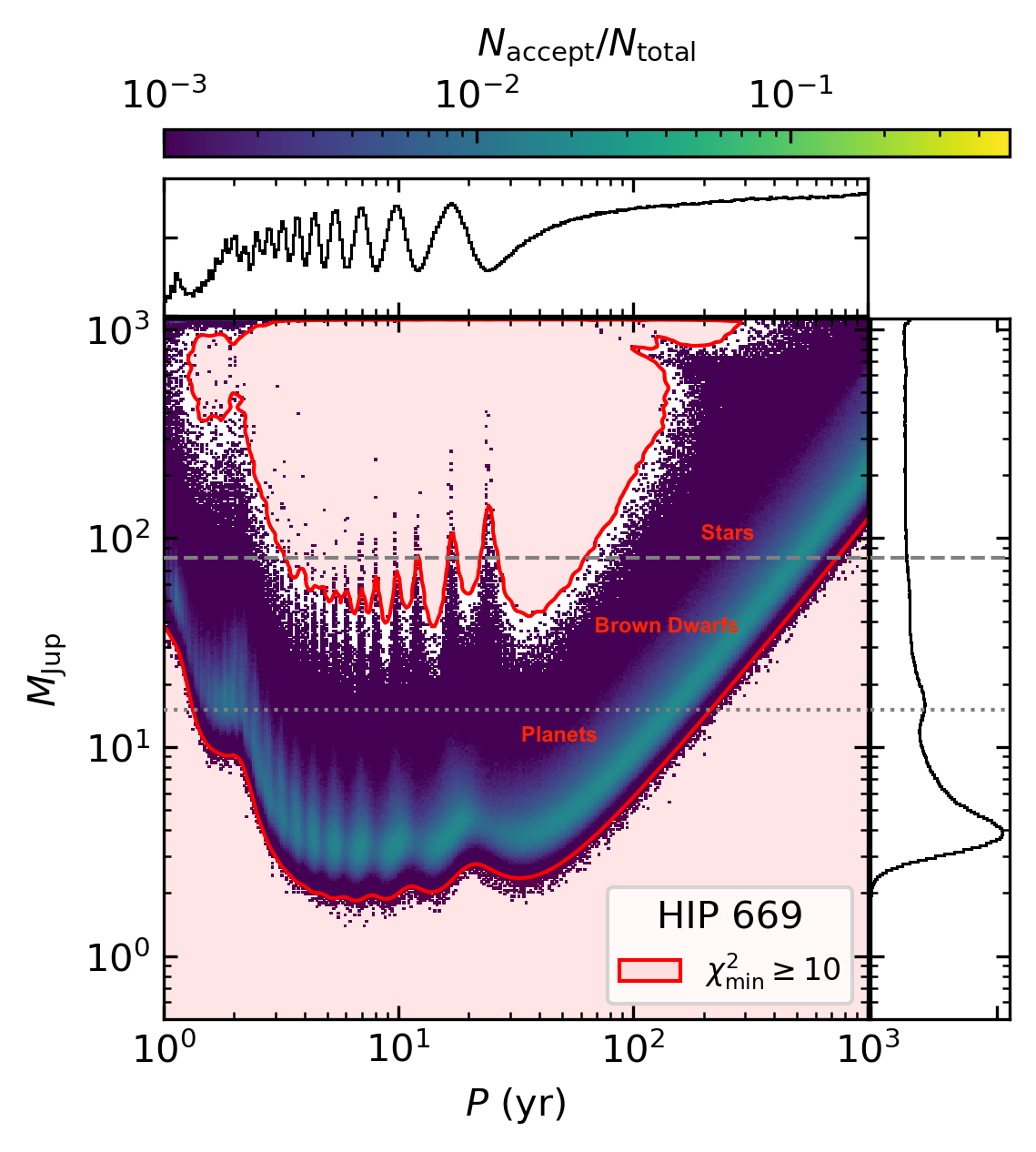}
  
 \caption{Results from rejection sampling of combined \textit{Hipparcos} and \textit{Gaia} DR3 astrometry show the possible companion mass-period combinations responsible for the observed acceleration of HIP 669. Candidate companions include stellar binaries, brown dwarfs, and wide-separation giant planets. The near-infrared contrast of these substellar companions depends on the age of the star. }
 \label{dot_ex}
\end{figure*}

We identify candidate exoplanet hosts by examining the \textit{Hipparcos}-\textit{Gaia} Catalog of Accelerations presented in \cite{2021ApJS..254...42B} created using the \textit{Gaia} DR3 catalog. We select stars with a $>4\sigma$ proper motion anomaly, combining the significance of the three pairs of proper motions. We exclude known close binaries reported in the Washington Double Star Catalog (WDS) \citep{2001AJ....122.3466M} or the Ninth Catalog of Spectroscopic Binary Orbits (SB9) \citep{2004A&A...424..727P} and stars with obvious companions in archival imaging data. Of the remaining stars, we reject those whose astrometric signal is consistent with a stellar companion between $0.1-1.0''$. We determine the allowed combinations of companion mass and orbital period using the rejection sampling algorithm described in detail in \cite{2023A&A...672A..94D}. We compare simulated astrometric measurements from the simulated photocenter motion caused by an orbiting companion to the observed accelerations. \rev{We use the scan timings and scan angles of the \textit{Hipparcos} and \textit{Gaia} satellites to forward model the position and proper motion of the photocenter measured by the two missions. The photocenter motion is modeled as a combination of the linear motion of the system through space and the orbit of the primary star about the system barycenter.} Figure \ref{dot_ex} shows an example of the allowed mass-period parameter space for a star in our sample (HIP 669) based on the observed astrometric acceleration.

\begin{figure*}[ht!]
\centering
 \includegraphics[scale = 0.35]{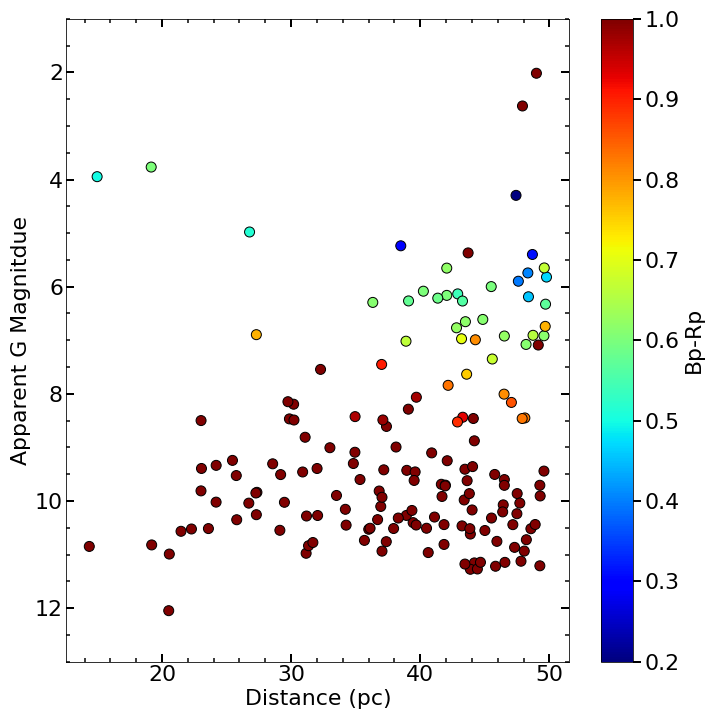}
  \includegraphics[scale = 0.35]{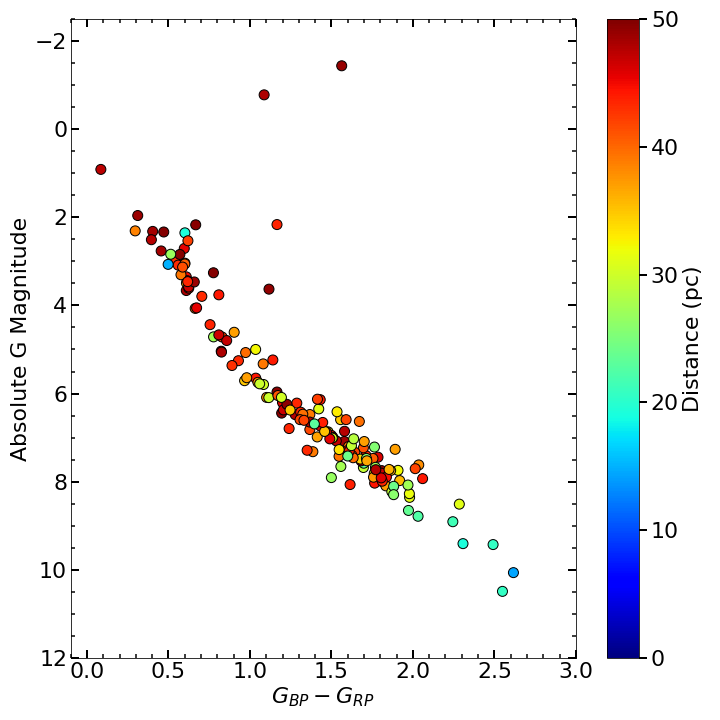}
 \caption{{\bf Left:} Apparent \textit{Gaia} G magnitude of the accelerating stars as a function of distance. {\bf Right:} Color-magnitude diagram of the sample. The median of our sample is $G_{BP}-G_{RP} = 1.4$ ($\sim$K5), 41.5 pc, and $G=9.6$. Our sample covers a wide range of spectral types, so we require a variety of age tracers to derive ages for each star. }
 \label{sample}
\end{figure*}

We limit our sample to stars observable from Apache Point Observatory (APO), i.e. stars with declinations $>-30^{\circ}$. We also limit the sample to stars within 50 pc to maximize the angular separation between the star and the potential companion to maximize the chance of the companion being directly imaged. We do not include stars that have archival radial velocities (e.g. from ESO/HARPS or Keck/HIRES). Figure \ref{sample} shows the distance, color, and magnitudes of our sample. The majority of our sample is between 30-50 pc with apparent G magnitudes between 6 and 11.

 Age posteriors for lithium and $R^{'}_{HK}$ from \texttt{BAFFLES} \citep{2020ApJ...898...27S} are calculated as a function of $B-V$. We adopt the $B-V$ values and associated errors that are reported in the \textit{Hipparcos} catalog \citep{1997A&A...323L..61V}. Five stars in our sample (HIP 33368, HIP 46385, HIP 63661, HIP 69860, and HIP 81655) do not have a $B-V$ reported in the \textit{Hipparcos} Catalog. An additional five stars (HIP 9067, HIP 111976, HIP 59953, HIP 1068, and HIP 1771) have very high uncertainties ($\sim$0.5 magnitudes) on their reported B-V colors in the \textit{Hipparcos} catalog. We derive $B-V$ for these targets by fitting a relation to the $B-V$ / $G_{BP}-G_{RP}$ color-color plot (Figure \ref{bmv} in Appendix). We use the $B-V$ values from the \textit{Hipparcos} Catalog \citep{1997A&A...323L..61V} and the $G_{BP}-G_{RP}$ values from \textit{Gaia} DR3 \citep{2023A&A...674A...1G}. We filter the \textit{Gaia} catalog for all stellar entries that were cross-matched to \textit{Hipparcos} by the \textit{Gaia} team, merged the \textit{Hipparcos}/\textit{Tycho} data, we exclude variable stars, and then apply the following filters: $RUWE \leq 1.6$, $Gmag<13$, $Gmag_{err}<0.01$, $Hpmag_{err}<0.1$, $(B-V)_{err}<0.1$, $B-V\neq0.0$, and $B-V<2.0$.

 We find the best-fitting relationship is
 \begin{equation}
 B-V = -0.102(G_{BP}-G_{RP})^{4} + 0.226(G_{BP}-G_{RP})^{3}+0.017(G_{BP}-G_{RP})^{2}+0.688(G_{BP}-G_{RP})+0.002  
\end{equation} Based on the scatter in the relationship, we adopt a 0.1 magnitude uncertainty on the derived $B-V$ colors.
 
 Gyrochronological age posteriors from \texttt{gyro-interp} depend on effective temperature \citep{2023ApJ...947L...3B}. We determine effective temperatures for our sample by interpolating the grid from \cite{2013ApJS..208....9P} to the measured \textit{Gaia} $G_{BP}-G_{RP}$ colors. To estimate the uncertainty on the effective temperature we find the range of effective temperatures corresponding to the range of $G_{BP}-G_{RP}\pm0.1$.


To measure lithium equivalent widths and $R^{'}_{HK}$, we obtain spectra using the Astronomy Research Consortium Echelle Spectrograph (ARCES) on the APO 3.5 meter telescope. ARCES is a multi-use spectrograph with an average spectral resolution $R = \frac{\Delta\lambda}{\lambda} = 35,000$ covering 3,200 \r{A} to 10,000 \r{A} \citep{2003SPIE.4841.1145W}. We achieve SNRs between $\sim20$ for our faintest targets and $\sim200$ for our brightest stars.
The observing campaign began in October 2020 and is ongoing. We aim to observe each star at least three times over several months to years so that we can vet for binaries, which we will discuss in detail in future work. As of writing, we have observed each star at least once and 98 out of the 166 stars at least three times. We reduce the data using the python package, \texttt{pyvista}\footnote{https://pyvista.readthedocs.io/}. Each spectrum is bias-corrected, extracted, flat-fielded, and continuum-corrected. Wavelength solutions are extracted from the ThAr taken nearest in time to the science frame and are then applied to the science spectrum.  

We measure rotation periods from $TESS$ light curves reduced with the SPOC pipeline  \citep{2016SPIE.9913E..3EJ}. All light curves with cadences between 20-1800s are downloaded from the Mikulski Archive for Space Telescopes (MAST). We found light curves for 129 of our sample stars and identify periodic signals in the light curves of \rev{34} of these 129 stars.

\section{Lithium} \label{sec:lithium}

We measure lithium 6707.79 \r{A} equivalent widths from ARCES spectra so that we can use them and the python package \texttt{BAFFLES} \citep{2020ApJ...898...27S} to derive age posteriors. \rev{First, we bring the spectrum to the approximate rest frame using the \textit{Gaia} DR3 radial velocity measurement \citep{2023A&A...674A...1G} for the star and correct for the Earth's motion by applying a barycentric correction.} To find the lithium 6707.79 \r{A} feature and measure its equivalent width, we fit a double Gaussian with a slope to the spectrum with the functional form
\begin{equation}
  \rev{f_{mod} = (m(\lambda-6707.0)+b)(C_{1}e^{-0.5(6707.79 -RV_{s}-\lambda)^{2}/\sigma^{2}}+C_{2}e^{-0.5(\lambda_{ref}-RV_{s}-\lambda)^{2}/\sigma^{2}})}
\end{equation}

\noindent where $\lambda$ is the wavelength array, $m$ and $b$ are the slope and intercept of the continuum; $\lambda_{ref}$ is the central wavelength of some reference feature; $C_{1}$ is the depth of the lithium feature; $C_{2}$ is the depth of the reference line; $\sigma$ represents the standard deviation of both Gaussians; $RV_{s}$ represents an additional radial velocity shift to the \textit{Gaia} DR 3 measurement. \rev{We include this additional RV offset term in our fit to account for potential orbiting companions or any radial velocity drift in ARCES.} For most stars, we adopt $\lambda_{ref} = 6705.1$  \r{A}, a prominent Fe line. For lower mass stars where the 6705.1 \r{A} Fe line is not present, we adopt $\lambda_{ref} =  6717.8$  \r{A}, a prominent Ca line. \rev{The reference line prevents the fit from selecting a value for the radial velocity, $RV_{s}$ that shifts another line (or noise trough) to the location of the lithium line. This also prevents the fit from adjusting the width of the lithium feature to match a noise trough.} This is especially useful in cases where the lithium line is weak or not detectable. When $\lambda_{ref} = 6705.1$ \r{A} we use the region of spectra from 6704 \r{A} to 6710  \r{A}. When $\lambda_{ref} = 6717.8$ \r{A} we use the region of the spectrum from 6706 \r{A} to 6720  \r{A}. Both of these regions fall on a single ARCES order.

Since the 6707.79 \r{A} feature is a blend of the $^6Li$ and $^7Li$ doublets, we find the effective center of the entire feature by fitting the double Gaussian to a narrow region of spectrum from 6704 \r{A} to 6710 \r{A} for several stars where $\lambda_{ref} = 6705.1$ \r{A} such that

\begin{equation}
 \rev{ f_{mod} = (m(\lambda-6707.0)+b)(C_{1}e^{-0.5(6707.79 + s-RV_{s}-\lambda)^{2}/\sigma^{2}}+C_{2}e^{-0.5(6705.105-RV_{s}-\lambda)^{2}/\sigma^{2}})}
\end{equation}
 
\noindent Here, $s$ is the deviation of the center of the lithium feature from 6707.79 \r{A}.
 
 We use the python package \textit{emcee} \citep{2013PASP..125..306F} to run Markov Chain Monte Carlo (MCMC) fits on several spectra and found that $s$ for the best fit was consistently 0.0119 \r{A} regardless of spectral type, so we adopt 6707.819 \r{A} as the effective center of the lithium feature. 
 
With Equation 3 and \textit{emcee}, we run MCMC to find the best-fitting parameters and estimates of their uncertainties for each spectrum. The uncertainties in each pixel of the spectrum are assumed to follow a Poisson distribution and include estimates of the gain and read noise. 

\rev{First, we run a preliminary MCMC with 100 walkers and 10,000 steps to find good starting positions. Then, a second round of MCMC is run from these starting positions, with 100 walkers, for 50,000 steps with a 2,000 step burn-in, which was sufficient to achieve convergence for most stars. 16 stars with non-detections required 100,000 steps with a 10,000 step burn-in, and 2 additional stars required 400,000 steps with a 200,000 step burn-in. We assessed convergence both by visual inspection and by comparing the posteriors at the start and end of each chain. We also compute the integrated autocorrelation time (IAT) of our chains, which has a maximum value of 1.2; we therefore estimate that there are at least 35,000 independent samples of the posterior, consistent with the start and end of each chain being independent. Comparing the median of each parameter at the first 10\% of each chain to the final 10\% we find the largest fractional difference to be 9\%. A similar test comparing the standard deviation from the first and last 10\% yields a maximum fractional difference of 6\%. These results are consistent with convergence.} To calculate the lithium equivalent width, we integrate a Gaussian with parameters from the MCMC chains 
\begin{equation}
 \Sigma_{Li} = \sqrt{2\pi}C_{1}\sigma
\end{equation} which we converted to an equivalent width by scaling the area by the height of the nearby continuum, $H$
\begin{equation}
 EW_{Li} = \frac{\sqrt{2\pi}C_{1}\sigma}{H}
\end{equation}

To find the height of the nearby continuum, we evaluate a line defined by the slope ($m$) and intercept ($b$) of the fit at 6707.819 \r{A} (the center of the lithium feature). We use our chains to compute the posterior on equivalent width and report the median. We take the standard deviation of the equivalent width posterior from the converged chain as our uncertainty.

For stars that do not have a clear lithium feature, we use the same MCMC fitting method to calculate upper limits on the lithium equivalent width. For upper limits, $C_{1}$ in Equation 3 approaches zero, and $\sigma$ and $RV_{s}$ are set almost entirely by the reference line. We identify upper limits by measuring the skew of the equivalent widths' distribution. If the skew exceeds 0.225 or the mean of the distribution is below 5 m\r{A}, we flag the spectrum as a candidate upper limit. We confirm our upper limit selection by visually inspecting the spectrum. We calculate the 1, 2, and 3$\sigma$ upper limits by calculating the 68th, 95th, and 99.7th percentiles of the cumulative distribution function. We over-plot Gaussians with $\sigma$ determined by the reference line and depths set by the 1, 2, and 3$\sigma$ upper limits to visually assess the corresponding upper limit on the equivalent width. We report the $2\sigma$ upper limits.

\begin{figure*}[ht !]
\centering
 \includegraphics[scale = 0.3]{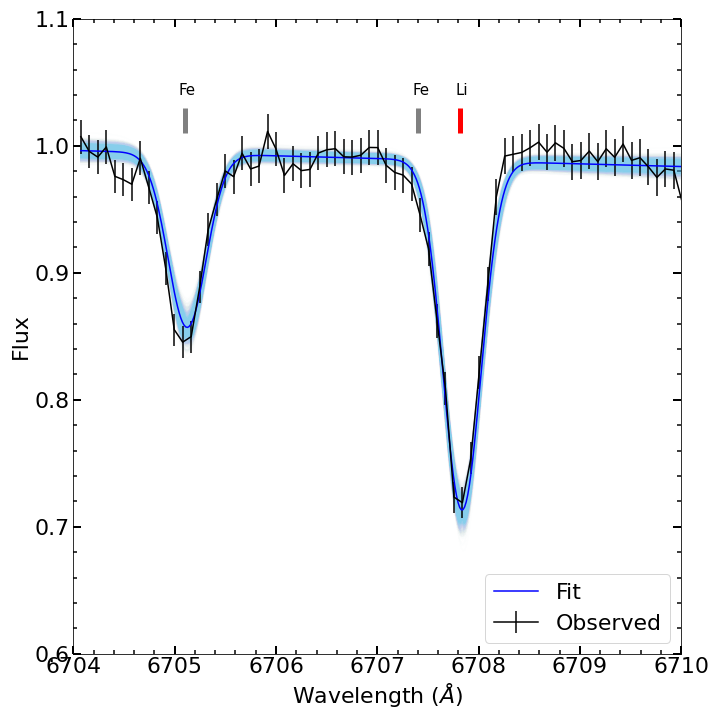}
 \includegraphics[scale = 0.3]{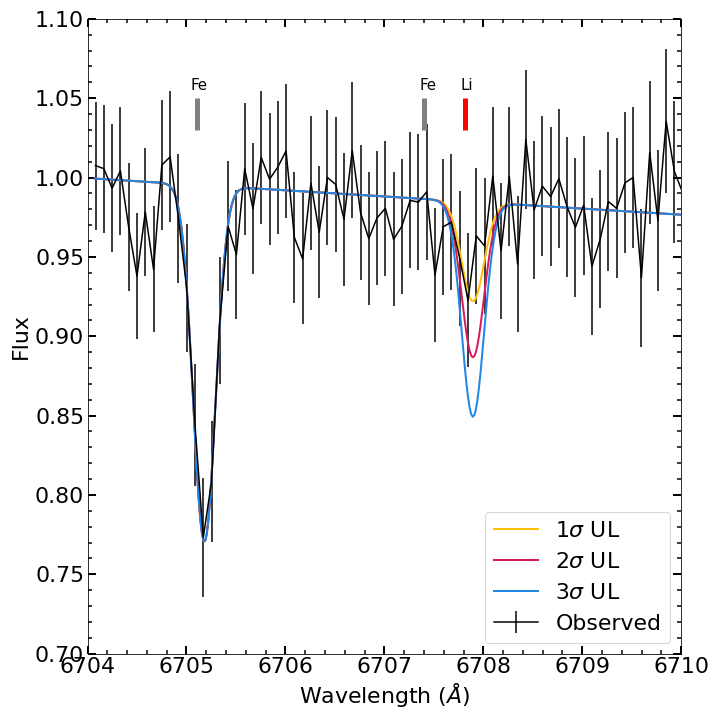}
 \caption{Our double Gaussian fitting technique allows us to robustly measure lithium equivalent widths and, in cases where lithium is not detected, place upper limits on the lithium equivalent widths. {\bf Left:} A spectrum of HIP 41277 ($SNR\sim75$) covers both the Fe 6705.1 \r{A} line (used as a reference line) and the Li 6707.79 \r{A} feature. Black points with error bars are the APO/ARCES data, the dark blue line represents the highest likelihood fit from the MCMC, and the light blue lines are random draws from the posterior. We find a lithium equivalent width for HIP 41277 of $125\pm{5}$ m\r{A}. {\bf Right:} An example spectrum ($SNR\sim30$) of an upper limit for lithium, from HIP 110663. Data are again black points with error bars, while yellow, red, and blue represent 1, 2, and 3$\sigma$ upper limits. We find a 2$\sigma$ upper limit of $30$ m\r{A} for HIP 110663, which we use to set a lower limit on the age of the star.}
 \label{li_spec_examples}
 
\end{figure*}

We calculate the equivalent width for each star using both spectral orders that include the lithium 6707.8 \r{A} feature. \rev{We adopt the equivalent width measurement or upper limit from the order/observation combination that yields the best fit to the data. This is generally the segment with the highest signal-to-noise ratio and no cosmic rays. We confirm our selection by visually assessing the fits. Each fit typically returns similar measurements of the lithium equivalent width, with error bars scaling with SNR as we would expect.} Figure \ref{li_spec_examples} shows examples of a lithium detection and a lithium upper limit determined from this process.

\begin{figure*}[ht!]
\centering
 
 \includegraphics[scale = 0.375]{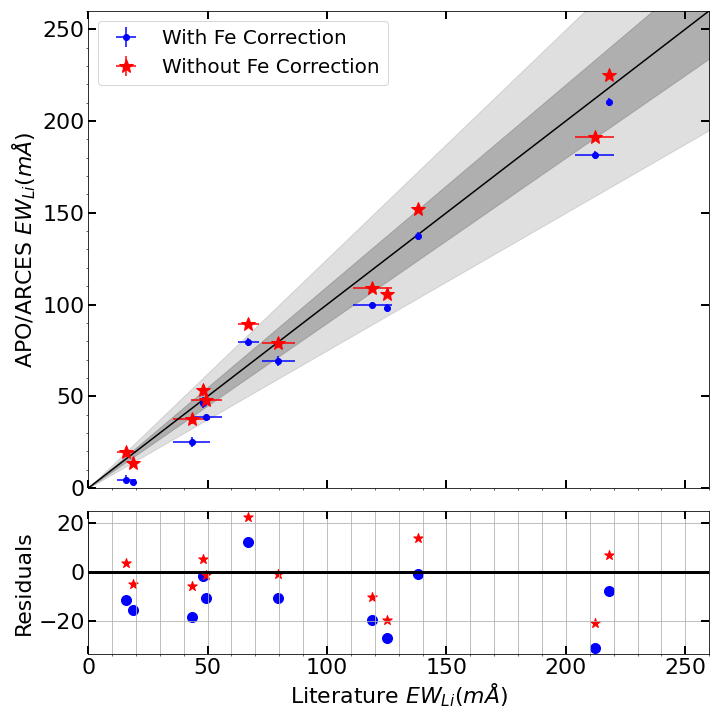}
 \caption{Comparing our lithium equivalent width measurements to results from the literature for 12 stars (red stars) shows good agreement. The black line shows the 1-to-1 relation, with $\pm10\%$ in dark gray and $\pm25\%$ in light gray. We find that applying a correction for the 6707.4 \r{A} iron line \citep{1993AJ....105.2299S} results in a systematic underestimate of the literature abundances (blue points). We conclude that our Gaussian fitting technique is robust to contamination from the 6707.4 \r{A} iron line, so we do not include an additional correction to our reported equivalent width values. We take literature lithium equivalent widths for HIP 104075, HIP 105232, HIP 113905, HIP 35628, HIP 97779 from \cite{2009A&A...504..829G}, HIP 115147, HIP 43410, HIP 52462, HIP 71631 from \cite{2003A&A...399..983W}, HIP 42438 from \cite{2007AJ....133.2524W}, and HIP 43726 from \cite{2005PASJ...57...45T}.}
 \label{li_check}
 
\end{figure*}
For stars with large amounts of lithium or high rotational velocities, the lithium 6707.8 \r{A} feature can be blended with the 6707.4 \r{A} Fe line. Past work has introduced a correction factor to equivalent widths to subtract the contribution of the iron line from the blended equivalent width. For example \cite{1993AJ....105.2299S} calculate iron 6707.4 \r{A} equivalent widths as a function of $B-V$, and subtract this from their blended equivalent widths. Similar methods are used by \cite{1993ASSL..183..303F, 2003A&A...399..983W, 2005PASJ...57...45T, 2007AJ....133.2524W, 2009A&A...504..829G}. To assess the degree to which including an iron blend correction improves our lithium equivalent width measurements, we compiled literature values where available for our target stars and S-value calibration targets (discussed below) from \cite{2013A&A...551L...8P}. In  Figure \ref{li_check}, we compare our measured lithium equivalent widths with and without the \cite{1993AJ....105.2299S} correction to these literature values for 12 stars. We find that our method with no correction applied yields more consistent equivalent widths compared to these literature values. Applying the \cite{1993AJ....105.2299S} correction systematically under-predicts the literature values. We attribute this to our method of fitting a Gaussian to the full absorption profile and taking the integral of the Gaussian, rather than summing \rev{fluxes in a specific wavelength range} directly from the spectrum. As a result, since the asymmetric wing from the iron line introduces a negligible effect on the final equivalent width, we do not apply a correction to our measured equivalent widths. 

\begin{figure*}[ht!]
\centering
 
 \includegraphics[scale = 0.375]{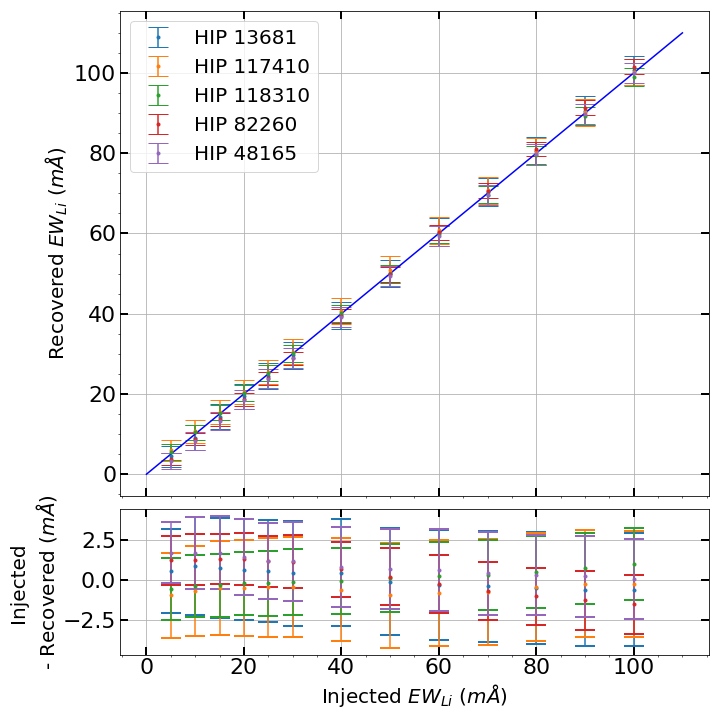}
 \caption{\rev{We confirm that our lithium fitting routine reliably recovers accurate equivalent widths with injection/recovery tests on 5 stars. We chose these stars since they had no detectable lithium, and because they cover a range of SNR and spectral types. The recovered equivalent widths are all within $\lesssim$1 m\r{A} of the injected value, and consistent with reported error bars.}}
 \label{recovery}
 
\end{figure*}

\rev{To ensure there are no systematic offsets in our lithium fitting technique, we inject a Gaussian of a given equivalent width into an ARCES spectrum with no detectable lithium. Then we feed the injected spectrum through our fitting routine and recover the equivalent width. We conduct this test on five spectra with moderate to high signal-to-noise corresponding to stars at a range of spectral types. For each star, we repeat the injection/recovery test for 13 different values of injected lithium equivalent widths from 5-100 m\r{A}. Figure \ref{recovery} compares the recovered equivalent widths to the injected lines, which agree to $\lesssim$1 m\r{A}, well within the reported error bars for each measurement. The main source of this disagreement is noise in each spectrum with the noise being on average slightly above or slightly below the continuum at the location of the lithium line, resulting in a slight underestimate or overestimate of the equivalent width, respectively.}

\begin{figure*}[h!]
\centering
 
 \includegraphics[scale = 0.375]{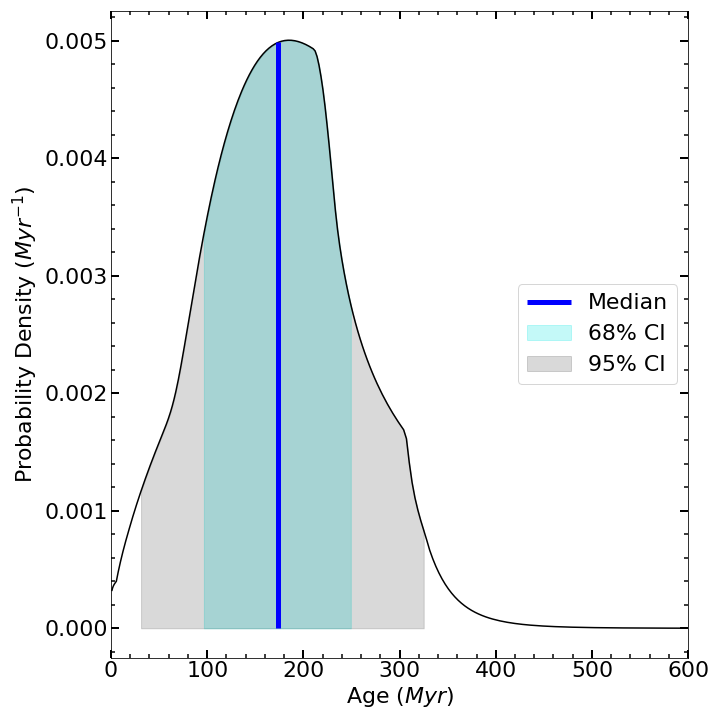}
 \caption{An example of an age posterior for one star in our sample, HIP 41277, using \texttt{BAFFLES} \citep{2020ApJ...898...27S}. From an adopted B-V of $1.03\pm{0.015}$ and a lithium equivalent width of $125\pm{5}$ m\r{A}, we derive an age posterior with a median age of 174 Myr and $68\%$ confidence interval between 97 and 253 Myr. Lithium measurements \rev{and median ages, and $68\%$ and $95\%$ confidence intervals} for the full sample are reported in Table \ref{tab:lithium}.}
 \label{li_post_example}
 
\end{figure*}

\rev{We calculate age posteriors from lithium equivalent widths and uncertainties for our target stars using Bayesian Ages For Field LowEr-mass Stars (\texttt{BAFFLES}) \citep{2020ApJ...898...27S}. \texttt{BAFFLES} utilizes a Bayesian framework that is calibrated on stars in clusters and associations of known ages. Functional forms were fit both to lithium equivalent width as a function of $B-V$ for each calibration cluster, and the intrinsic scatter in equivalent width. Final age posteriors represent a combination of the astrophysical scatter in the age/equivalent width relation and the measurement error \citep{2020ApJ...898...27S}. \texttt{BAFFLES} uses a uniform star formation history as the age prior. The uncertainty on age is largely driven by the observed astrophysical scatter in lithium equivalent width at a constant age, with a much smaller contribution from the measurement uncertainty for a single star.} \texttt{BAFFLES} is calibrated to calculate age posteriors from lithium for stars with $B-V$ between 0.35 and 1.9 from both measurements and upper limits. Figure \ref{li_post_example} shows an example \texttt{BAFFLES} posterior for HIP 41277, where an equivalent width of $125\pm{5}$ m\r{A} translates to an age posterior of $174^{+79}_{-77}$ Myr. We detect lithium in 26 stars, and determine upper limits for 140 stars; of the full sample, \rev{159} stars are in the \texttt{BAFFLES} $B-V$ range.  We report $B-V$, lithium equivalent width measurements (or upper limits), and \rev{median ages and confidence intervals} from \texttt{BAFFLES} (where applicable) for our sample in Table \ref{tab:lithium}.

\clearpage

\section{Calcium Emission} \label{sec:rhk}

\begin{figure*}[ht !]
\centering
 \includegraphics[scale = 0.4]{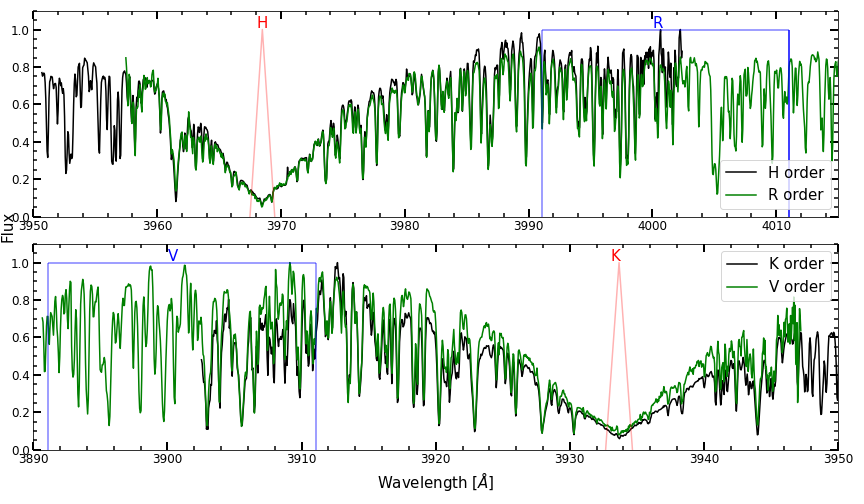}
 \includegraphics[scale = 0.4]{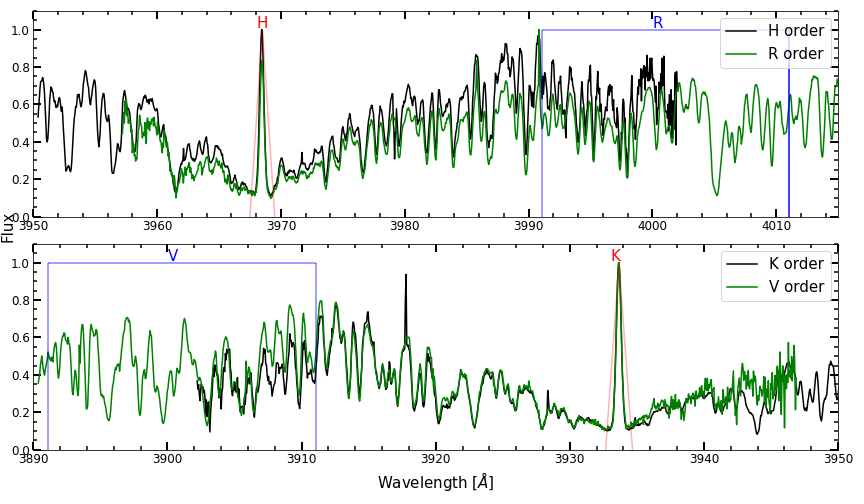}
 \caption{\rev{The S-value is measured from spectra using triangular transmission profiles (red lines) centered on the Ca II H and K emission lines, and rectangular profiles (blue lines) to measure the nearby continuum level. The top two panels show an example of a low-activity star with no significant H and K emission (HIP 102040), and the bottom two panels show HIP 115147, a star with strong H and K emission.}}
 \label{RHK_examples}
\end{figure*}

We also use \texttt{BAFFLES} to derive age posteriors from calcium H (3968.47 \r{A}) and K (3933.67 \r{A}) emission ($R'_{HK}$) for stars with spectral types between F5 and K2. $R'_{HK}$ is derived from the S-value index, which is the ratio of calcium emission flux to nearby continuum. S-values were initially measured photometrically following the methods of the Mt. Wilson Ca II H \& K survey that ran from 1966 to 1983 \citep{1991ApJS...76..383D}. The emission flux is measured through two triangular bandpasses, and the continuum flux is measured through two rectangular bandpasses on either side of the emission cores, as shown in Figure \ref{RHK_examples}. 

Subsequently, the photometric S-value index is most commonly measured from spectra with the S-value index calibrated for each instrument. To calibrate APO/ARCES, we observe 99 stars with published S-values from \cite{2013A&A...551L...8P}, which compiled published S-values from the literature including \cite{1991ApJS...76..383D}, \cite{2004ApJS..152..261W}, and \cite{2003AJ....126.2048G}. We follow a prescription similar to that in \cite{2004ApJS..152..261W} with some modifications to calibrate the APO S-values. \rev{We use a 1-parameter fit (compared to the 4-parameter fit of \rev{\citealt{2004ApJS..152..261W})} with the form \begin{equation}\label{eq:s}
 S = a\left(\frac{H+K}{R+V}\right)
\end{equation}}
$H$ and $K$ are the summed counts from convolving the spectrum with the triangular transmission profiles, while $R$ and $V$ are the summed counts from convolving the continuum with the rectangular transmission profiles (taken from \rev{\citealt{1991ApJS...76..383D})}.

 \rev{We shift the spectra to the rest frame so the calcium emission cores or the centers of the absorption lines are centered in the narrow triangular band passes. Using the $G_{BP}-G_{RP}$ to $B-V$ conversion described in Section \ref{sec:data}, and the grid from \cite{2013ApJS..208....9P} we find effective temperatures, masses, and radii for our targets, assuming they are on the main sequence. We find the synthetic spectrum with solar metallicity from \cite{2013A&A...553A...6H} with the closest effective temperature and surface gravity. We convolve the synthetic spectrum with a Gaussian to match the resolution of ARCES. Then, we shift the ARCES spectra to the rest frame by cross correlating the observed spectrum with the theoretical spectral template.}

\begin{figure*}[ht!]
\centering
  \includegraphics[scale = 0.5]{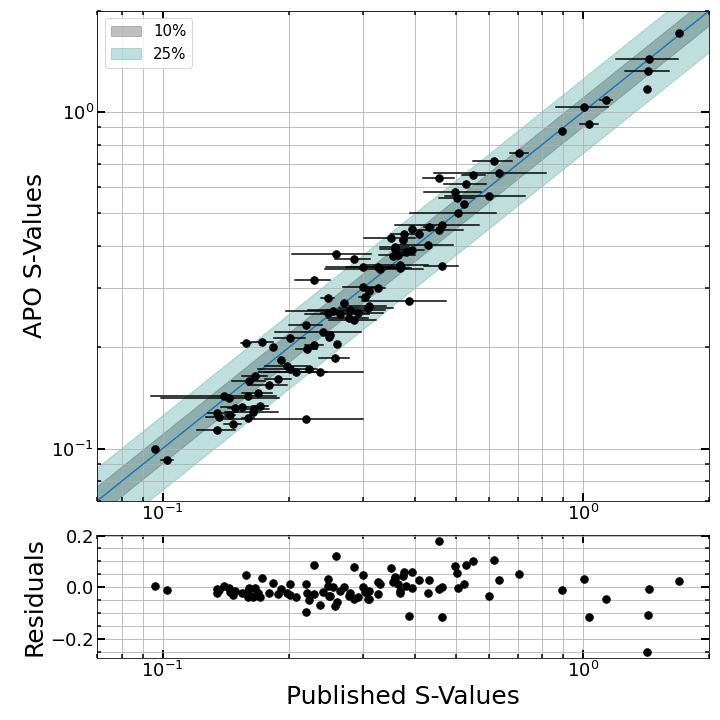}
 
 \caption{\rev{We use a 1 parameter fit to calibrate chromospheric S-values from APO/ARCES to values reported in the literature. We observe a  $\sim15\%$ scatter in the relationship, which we largely attribute to the intrinsic variability in chromospheric activity for each star. Error bars represent the maximum and minimum S-value reported in the literature for each star. We report S-values and $log(R'_{HK})$ for all the accelerating stars in Table \ref{tab:rhk}.}}
 \label{sval_calib}

\end{figure*}

\rev{Literature S-values are often not given with error bars, and generally variation over the star's activity cycle is expected to be larger than measurement errors. Given the lack of reliable error bars we do not attempt to derive a posterior probability distribution on the fit parameter. Instead, we assume error bars of unity for each measurement, and find the value of $a$ in Equation \ref{eq:s} that minimizes the $\chi^2$. We compute this minimum value over a uniform grid for $a$ with $10^6$ elements between 0 and 40, finding a best-fit value of $a = 18.3$, which results in a $\sim$15\% scatter on the relationship between APO S-values and the literature S-values. The fit is shown in Figure \ref{sval_calib}, and we attribute this $\sim$15\% scatter mainly to variable emission over a stellar cycle.}

\rev{We investigated more complicated fits, such as including additional parameters to weight the H and K bands or the R and V bands differently, or attempting to further correct the continuum by comparing to theoretical spectra, or performing different fits for different spectral type bins. None of these more complicated methods significantly improved the scatter, and some resulted in systematic offsets from the fit at larger S-values. As a result, we adopt the more straightforward approach outlined above.}

The transformation from S-value to $R'_{HK}$  is \begin{equation}\label{eq:r}
R'_{HK} = 1.34\times10^{-4}C_{Cf}S - R_{phot}
\end{equation} \noindent where $C_{Cf}$ is the color correction factor, a third-order polynomial in $B-V$  \citep{1982PhDT.......153M}, and $R_{phot}$ is another third-order polynomial in $B-V$ that corrects for the photospheric contribution to the emission in the Ca H and K lines \citep{1984ApJ...276..254H}. 
\begin{figure*}[ht!]
\centering
  \includegraphics[scale = 0.375]{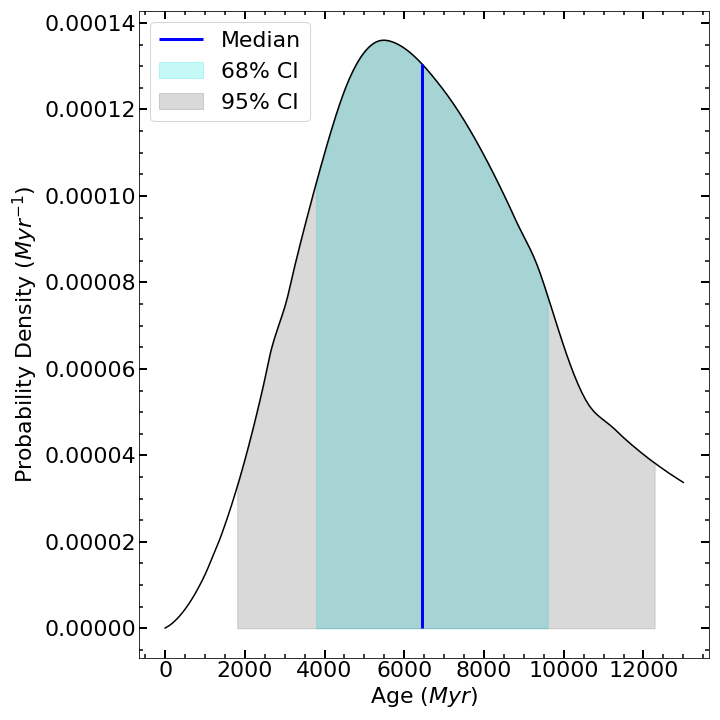}
 
 \caption{An example age posterior for HIP 669, using our APO/ARCES measurement and \texttt{BAFFLES} \citep{2020ApJ...898...27S}. For this relatively inactive star, with $log(R'_{HK}) = -4.88$ and $B-V=0.62\pm0.015$, we derive an age posterior with a median age of 6.5 Gyr and $68\%$ confidence interval between 3.8 and 9.7 Gyr.}
 \label{RHK_post_example}

\end{figure*}
S and $R'_{HK}$ are given for each of our target stars in Table \ref{tab:rhk}, as well as \rev{median ages and $68\%$ and $95\%$ confidence intervals} for the 34 stars between F5 and K2 derived using \texttt{BAFFLES} \citep{2020ApJ...898...27S}. \rev{Similar to lithium, \texttt{BAFFLES} calculates age posteriors from $R'_{HK}$ values in this spectral type range based on empirical fits to $R'_{HK}$ as a function of age and the intrinsic scatter as observed in clusters and associations of known ages.} Figure \ref{RHK_post_example} shows an example age posterior for HIP 669, with a $log(R'_{HK})$ of $-4.88$ that gives an age posterior of $6.5^{+3.2}_{-2.7}$ Gyr.

\section{Gyrochronology} \label{sec:gyro}

We calculate \rev{gyrochronological ages} using the python package \texttt{gyro-interp} which computes age posteriors by interpolating between rotation sequence models for open clusters with well-defined ages \citep{2023ApJ...947L...3B}. \texttt{gyro-interp} is calibrated for stars with $3800 K < T_{eff} < 6200 K$. We measure rotation periods from $TESS$ light curves processed with the SPOC pipeline \citep{2016SPIE.9913E..3EJ}, which were available for 129 of our target stars. We visually inspect each light curve and its corresponding Lomb-Scargle periodogram \citep{1976Ap&SS..39..447L, 1982ApJ...263..835S} to make an initial estimate for the rotation period. \rev{Then, to further validate and refine the selected period}, we use the non-linear least squares fitting python package \textit{LMFIT} \citep{2016ascl.soft06014N} to fit a sinusoid to the light curve. We run \textit{LMFIT} to find the sinusoid's amplitude, phase, vertical offset, and period. We limit the period to a narrow range around the initial estimate to prevent the fit from getting stuck at a local minimum at a harmonic of the true period. \rev{We do not use MCMC or another sampling algorithm to find the rotation period because a sinusoid is only an approximation of spot modulation and because photometric measurements are correlated over small time spans.} 

To ensure we select an actual rotation signal we compare our adopted period to peaks in the corrected (PDCSAP) and the uncorrected (SAP) periodograms. The period must appear as a local maximum in both periodograms to be accepted. We do not require the period to be the global maximum of either periodogram.

We differentiate between ``flat" light curves and robust variability signals by comparing $\chi^{2}$ for a flat line and the best-fitting sinusoid. Based on visual inspection of light curves from our target stars, we identify light curves with $\Delta \chi^{2} \gtrsim500$ as having a variability signal, and classify all other light curves as flat. Regardless of their $\Delta  \chi^{2}$, \rev{25} stars have variability amplitudes $\lesssim0.1\%$ and are taken to be flat. We also visually confirm variability in each light curve. We do not measure rotation periods for 8 stars that share a $TESS$ pixel with another star that is $\gtrsim2.5\%$ as bright as the target star in the \textit{Gaia} G band. We also visually inspect the target-pixel files (TPFs) for bright stars on the edges of nearby pixels.

For stars observed in multiple $TESS$ sectors, we adopt the period from the \rev{light curve with the fewest and/or smallest down-link gaps, which we determine by visual inspection. If more than one curve has similar amounts of data gaps, we select the light curve with the highest signal-to-noise.} We use the other sectors to further validate our adopted rotation periods. Some stars are not variable in some sectors but are highly variable in others. We attribute this to spot evolution, such that spots are present in one sector but not present in another. We adopt the rotation period from the sector displaying variability.

\rev{Measuring rotation period uncertainties can be challenging, and there are several different methods commonly used. For example, \cite{2021AJ....162..147H, 2019ApJS..244...21S, 2021ApJS..255...17S} estimate rotation uncertainties by the width of the periodogram peak, while \cite{2020ApJ...903...99H} use the width of the autocorrelation peak. We estimate uncertainties on the measured rotation periods with a similar method, by fitting a Gaussian to the exponentiated Lomb-Scargle periodogram peak, which is proportional to the likelihood of a single-sinusoid model at each frequency \citep{2018ApJS..236...16V, 2020sdmm.book.....I}. \cite{2018ApJS..236...16V} cautions that this method does not take into account the possibility of having selected the wrong periodogram peak. Therefore, we validate our period selection in two ways: visual inspection and false alarm probability (FAP). For all our measured periods, the largest FAP we calculate is $0.5\%$, and all the light curves cover more than one period making visual identification of the period relatively robust. We propagate the $\sigma$ of the best fitting Gaussian in frequency space to period, such that
\begin{equation}
    \sigma_{P} = \sigma_{f}P^{2}
\end{equation}}

\begin{figure*}[ht!]
\centering
 \includegraphics[scale = 0.375]{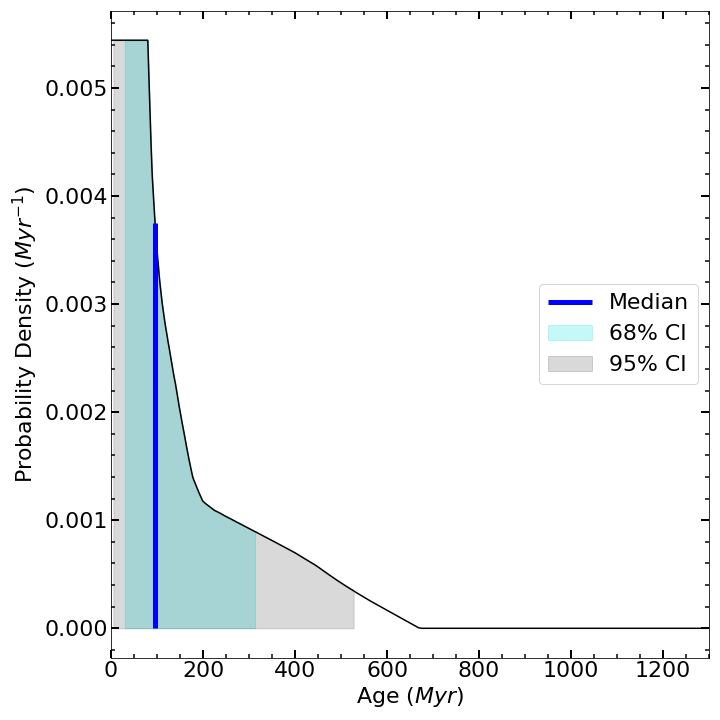}

 \caption{We calculate age posteriors from rotation periods using \texttt{gyro-interp} \citep{2023ApJ...947L...3B}. The age posterior for HIP 41277 with an adopted $T_{eff}$ of $4551\pm195$K and a measured rotation period of $3.70\pm0.24$ days has a median age of 96 Myr and a $68\%$ confidence interval between 30 and 316 Myr.}
 \label{gyro-post}
\end{figure*}

We use \texttt{gyro-interp} \citep{2023ApJ...947L...3B} to turn these rotation periods and uncertainties into age posteriors, with an example \rev{gyrochronological age} posterior shown in Figure \ref{gyro-post}. \rev{\texttt{gyro-interp} generates age posteriors from rotational periods in a similar manner to how \texttt{BAFFLES} generates posteriors from lithium and calcium measurements. A reference sample of stars in clusters and associations of known ages was used to find the distribution in rotation period as a function of age and temperature, as well as the astrophysical scatter. The generated age posterior is then a combination of the rotation period and errors and the astrophysical scatter. As with lithium, the dominant source of uncertainty in the final posterior is astrophysical scatter rather than measurement uncertainty on the period.}

\begin{figure*}[ht!]
\centering
 \includegraphics[scale = 0.35]{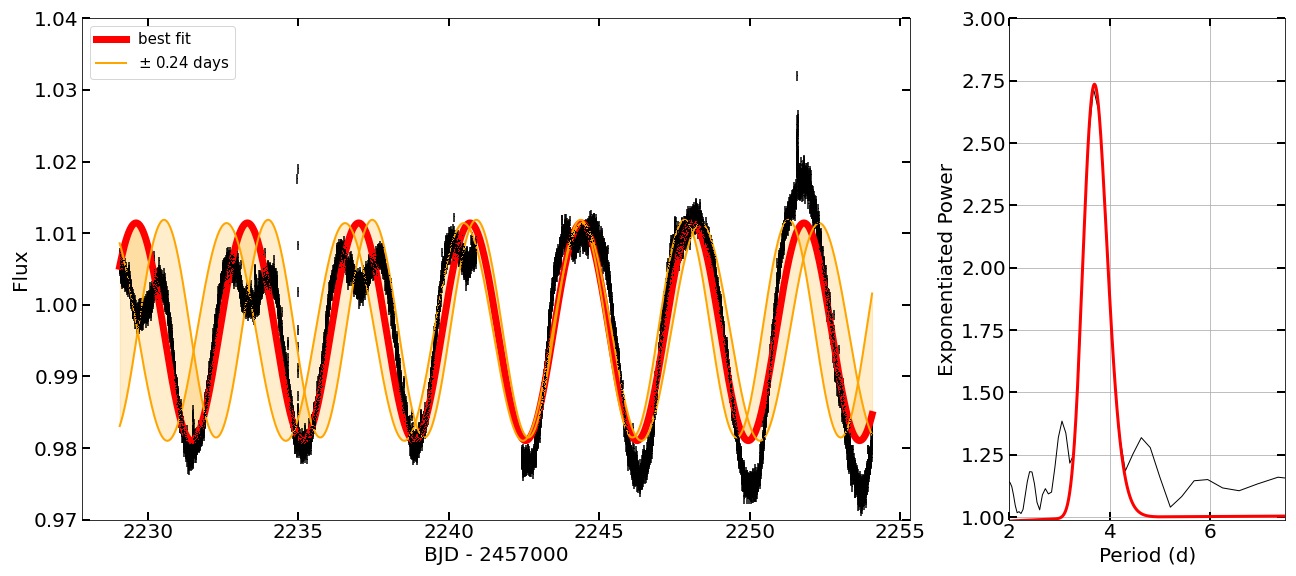}

 \caption{\rev{We measure rotation periods for the accelerating stars by analyzing their periodograms and fitting sinusoids to their lightcurves. The $TESS$ light curve for HIP 41277 is shown in black with the best fitting sinusoid with a period of 3.7 days overplotted in red. The orange curves represent the best-fitting period adjusted by the uncertainty measured from the width of the exponentiated periodogram peak. The right-hand panel shows the exponentiated Lomb-Scargle periodogram with a significant peak at $\sim3.7$ days and the best fitting Gaussian.}}
 \label{gyro-TESS}
\end{figure*}

\subsection{Complex light curves}

Several $TESS$ light curves of our targets do not exhibit straightforward rotation signals. In this section, we briefly describe the different types of variability we observe \rev{(e.g. \citealt{2013MNRAS.432.1203M}).}

\begin{figure*}[ht!]
\centering
 \includegraphics[scale = 0.3]{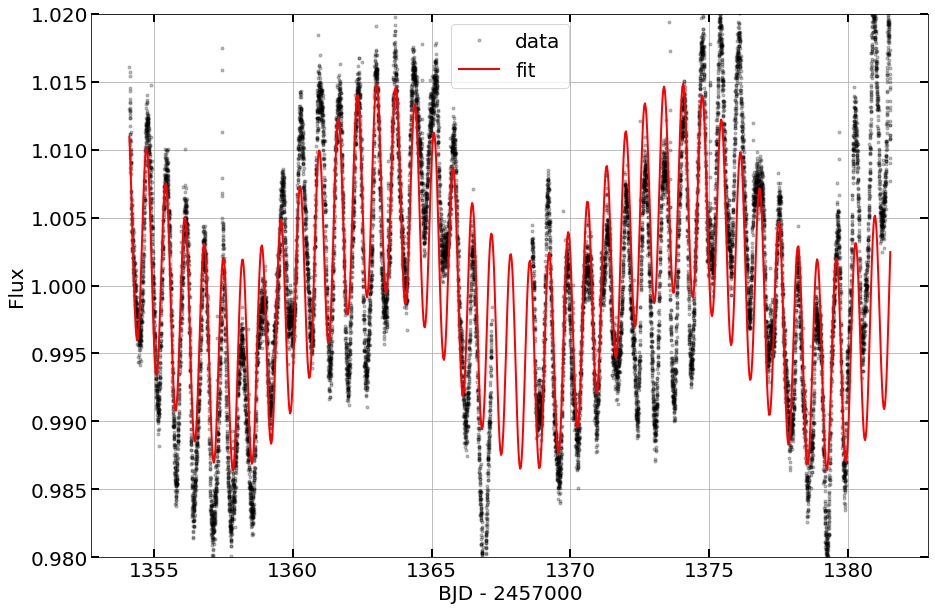}

 \caption{We fit a two-component sinusoid model (red) to a light curve (black) for HIP 117410. We find the periods of the two components to be $10.65\pm1.07$ days and $0.69\pm0.07$ days}
 \label{2_periods}
\end{figure*}

{\bf Double periods - } HIP 117410 shows two distinct periods in its light curves, but is not on obviously contaminated pixels. We interpret this as a rotation signal of two stars on the same pixels. To determine each period, we visually assess each light curve and make an initial estimate for each period. Then, we fit a double sinusoid to the flux of the form, 
\begin{equation}
    F = A_{1}\sin(2\pi f_{1} t + \phi_{1}) +  A_{2}\sin(2\pi f_{2} t + \phi_{2})
\end{equation}
 to each light curve using \textit{LMFIT}. $f_{1}$ and $f_{2}$ are the frequencies, $A_{1}$ and $A_{2}$ are the amplitudes, and $\phi_{1}$ and $\phi_{2}$ are the phase shifts of each of the component sinusoids. We find the best fitting periods for each component to be $10.65\pm1.07$ days and $0.69\pm0.07$ days. The best-fitting double sinusoid for HIP 117410 is shown in Figure \ref{2_periods}. 
 
 HIP 117410 (WDS 23484-1259) is a $\sim1"$ double star with a flux ratio of $\sim4$ in $V$ band \citep{2001AJ....122.3466M}, so this pair is most likely the source of the two signals. This pair was not flagged as a contaminated pixel due to the absence of a \textit{Gaia} G band measurement for the secondary. Early analysis of the APO/ARCES radial velocities of HIP 117410 (to be discussed in detail in a future paper) suggests that the system is not a short-period stellar binary.
 
 We cannot determine which period corresponds to which star, or directly measure $B-V$ for the secondary with the current data. Instead, we derive an age posterior for each period with the B-V of the primary, then marginalize over each star by summing the two posteriors and normalizing the resulting posterior so that it integrates to 1 (Table \ref{table:117410}).

\begin{table}[ht!]
\centering

 \begin{tabular}{|c|c|c|c|}

     \hline
        $ P_{rot}$ & Age & $68\%$ CI & $95\%$ CI \\
          (d) & (Myr) & (Myr) & (Myr)  \\
       \hline

      $10.65\pm1.07$ & 415.99  & 214.20--967.37  & 146.59--11492.34\\
      
       $0.69\pm0.07$ & 105.19 & 32.77--337.67 & 5.56--542.49 \\
      
      Combined & 254.07 & 64.24--625.41 & 10.11--1400.23 \\
      
     \hline

\end{tabular}

\caption{The age posteriors for HIP 117410 for both periods and the combined posteriors are listed. We adopt the combined age posterior for this system.}

\label{table:117410}
\end{table}

{\bf Eclipsing Binaries - }\rev{We identify HIP 16247, HIP 45731, and HIP 49577 as eclipsing binaries. HIP 45731 and HIP 16247 were previously analyzed in \cite{2022ApJS..258...16P} and \cite{2023A&A...672A.119G}, respectively.}

\begin{figure*}[ht!]
\centering
 \includegraphics[scale = 0.35]{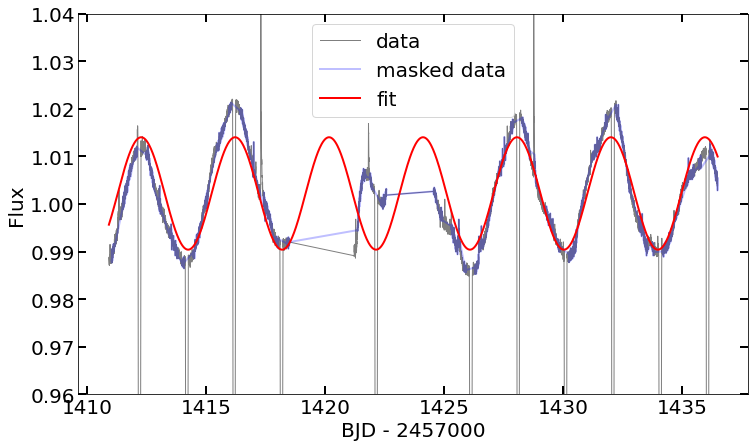}

 \caption{We fit a sinusoid to the underlying rotation signal of an eclipsing binary system. The light curve for HIP 16247 is shown in gray. The masked data with the eclipses removed is shown in purple, and the best-fitting sinusoid is shown in red. This star is tidally locked with $P_{orb} \approx P_{rot} \approx 3.98$ days.}
 \label{EB_ex}
\end{figure*}

We calculate the orbital period and rotational periods separately for each of these three systems. To find the orbital period we phase-fold the light curve based on the highest periodogram peak, verifying that the eclipses overlap, and the primary and secondary eclipses are not repeated in the folded light curve. If the phase folded curve has too many eclipses or the eclipses do not overlap, we try the next harmonic. We estimate the uncertainty of the orbital period by altering the period until the eclipses no longer overlap in the phase-folded light curve. To measure the underlying rotational period, we mask out the eclipses and then use the method described above for single stars. Table \ref{table:EBS} gives the orbital periods ($P_{o}$) and rotational periods ($P_{r}$) for each system as well as approximate depths of the primary ($D_{p}$) and secondary ($D_{s}$) eclipses. Figure \ref{EB_ex} shows the light curve for the eclipsing binary HIP 16247 with the best fitting sinusoid to the underlying rotation plotted in red. \rev{The data and the sinusoid fit are out of phase with each other between $1420 < BJD - 2457000 < 1425$. We attribute this to the fact that a simple sinusoid does not account for spot evolution and changing amplitudes, which can shift the phase of the periodicity, but a sinusoid is sufficient for capturing the periodicity of the underlying rotation rate.}

\begin{table}[ht!]
\centering

 \begin{tabular}{|c|c|c|c|c|c|}

     \hline
      Target & WDS & $ P_{o}$ & $D_{p}$ & $D_{s}$ & $P_{r}$\\
            & &(d) & & & (d)  \\
       \hline
    
      HIP 16247 & 03294-2406 &  $3.98\pm0.004$ &  $\sim28\%$ &  $\sim22\%$ & $3.95\pm0.4$ \\
 
      HIP 45731 & 09194+6203&  $2.92\pm0.003$&  $\sim15\%$ &  $\sim14\%$ & $2.92\pm0.3$ \\
  
      HIP 49577 & -- &  $16.85\pm0.17$ &  $\sim24\%$ &  $\sim10\%$ & $7.2\pm0.72$ \\
     \hline

\end{tabular}

\caption{The estimated parameters of the three eclipsing binaries in our sample.}
\label{table:EBS}
\end{table}
\begin{figure*}[hb!]
\centering
 \includegraphics[scale = 0.3]{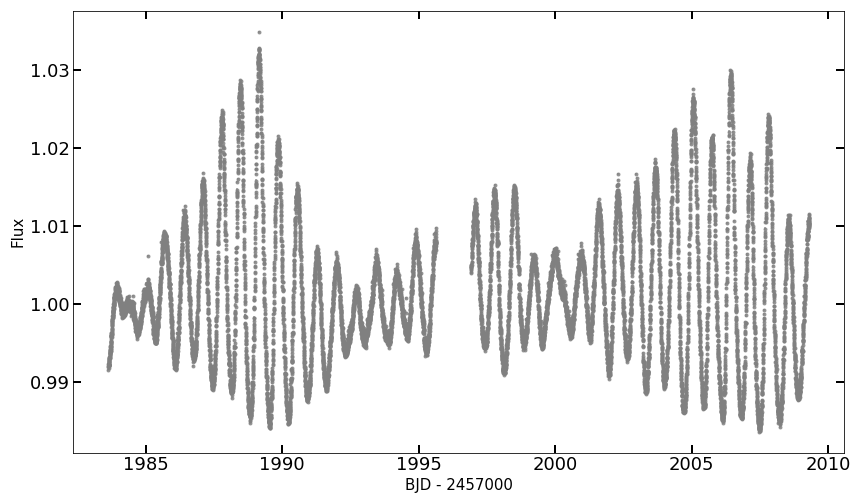}

\includegraphics[scale = 0.3]{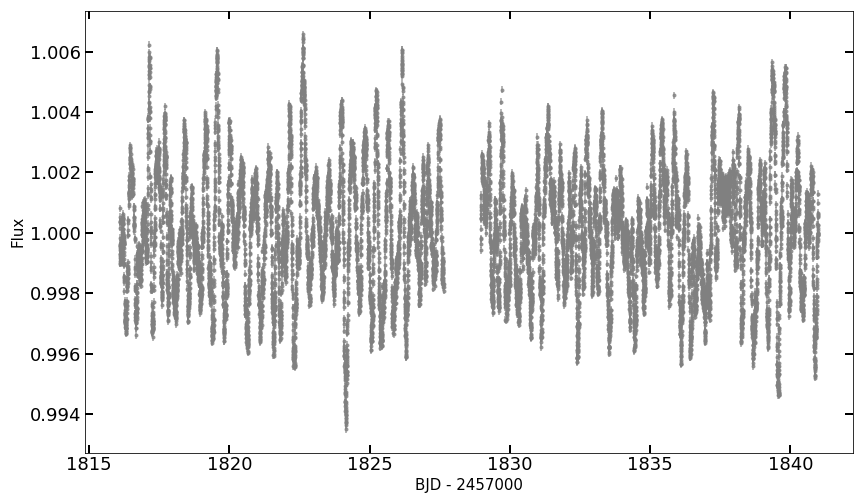}
 \caption{Several of our target stars exhibit photometric variability that is inconsistent with rotation. {\bf Top:} A light curve of HIP 59767, a $\gamma$ Doradus variable \citep{2016MNRAS.458.2307K}, shows a beat pattern making it difficult to measure an underlying rotation period. {\bf Bottom:} A light curve of HIP 22361 shows stochastic variability that is not attributed to rotation.}
 
 \label{notrot_ex}
\end{figure*}

In close binaries, the stars can be spun up by tidal forces, so that the rotation period reflects tidal evolution rather than spin-down following the empirical gyrochronology relations. For HIP 1627 and HIP 45731, the rotation periods and orbital periods we measured are approximately the same suggesting that both systems are tidally locked. Therefore, we do not derive \rev{gyrochronological ages} for these systems. For HIP 49577, the rotation rate we derived is significantly different than the orbital period we derived. From the rotation period and $B-V$ of the primary star, we measure a median age of 211 Myr and $68\%$ confidence interval of $105-422$ Myr. However, while this system is not tidally locked, we cannot rule out tidal forces affecting the rotational evolution of the system. 

{\bf Other Variable Stars - }We identify several stars that exhibit variability that does not resemble the rotational modulation of star spots. We use the examples in \cite{2013MNRAS.432.1203M} to visually classify these stars as stochastic or beat pattern variables. HIP 11090 and HIP 59767 both display a beat pattern in their light curves. HIP 21066, HIP 41274, HIP 22361, and HIP 59199 exhibit stochastic variability. An example of each is shown in Figure \ref{notrot_ex}. Along with not displaying clear rotation signals, most of these stars are outside the \texttt{gyro-interp} temperature range and are therefore not candidates for deriving \rev{gyrochronological ages}. HIP 41274 is within the \texttt{gyro-interp} temperature range, but it is a RS-CVn variable \citep{2022MNRAS.512.4835M}, so it is also not a candidate for deriving a \rev{gyrochronological age}. The close binary to HIP 41274 could be the source of its observed acceleration.

\section{Combined Age Posteriors} \label{sec:total_ages}

For stars with $B-V\le0.58$ (corresponding to an F9.5 spectral type and approximate main-sequence mass of $1.1M_{\odot}$), we apply additional constraints on their ages based on stellar lifetimes and their CMD positions \rev{(e.g. \citealt{2013ApJ...776....4N, 2019AJ....158...13N})}. Using this technique, we also produce age posteriors for stars too massive for Li, $R^{'}_{HK}$, or gyrochronology. CMD age posteriors are especially useful in cases where we can only derive a lower limit on the age from a lithium equivalent width upper limit because the stellar lifetime can be used to constrain the upper bound on the age (e.g. late F stars). To calculate CMD position-based age posteriors, we follow the method from \cite{2019AJ....158...13N}, which uses a Bayesian approach to \rev{compute ages from stellar models and optical photometry. For each star, MCMC is used to explore the 5-dimensional parameter space in stellar mass, age, metallicity, initial rotation, inclination of the rotation axis, and parallax. We use the same priors on these parameters as in \cite{2019AJ....158...13N}, and a Gaussian likelihood for the $Gaia$ $G_B$, $G$, and $G_R$ catalog photometry and errors. The model predictions at each MCMC step are generated, as in \cite{2019AJ....158...13N}, from a combination of MESA \citep{2011ApJS..192....3P} evolutionary models with ATLAS9 model atmospheres \citep{2004A&A...419..725C}.} As in \cite{2019AJ....158...13N}, for stars with $B-V\leq0.27$, we include a prior based on the metallicity distribution of stars $<1$ Gyr from \cite{2011A&A...530A.138C}.

\begin{figure*}[hb!]
\centering
 \includegraphics[scale = 0.35]{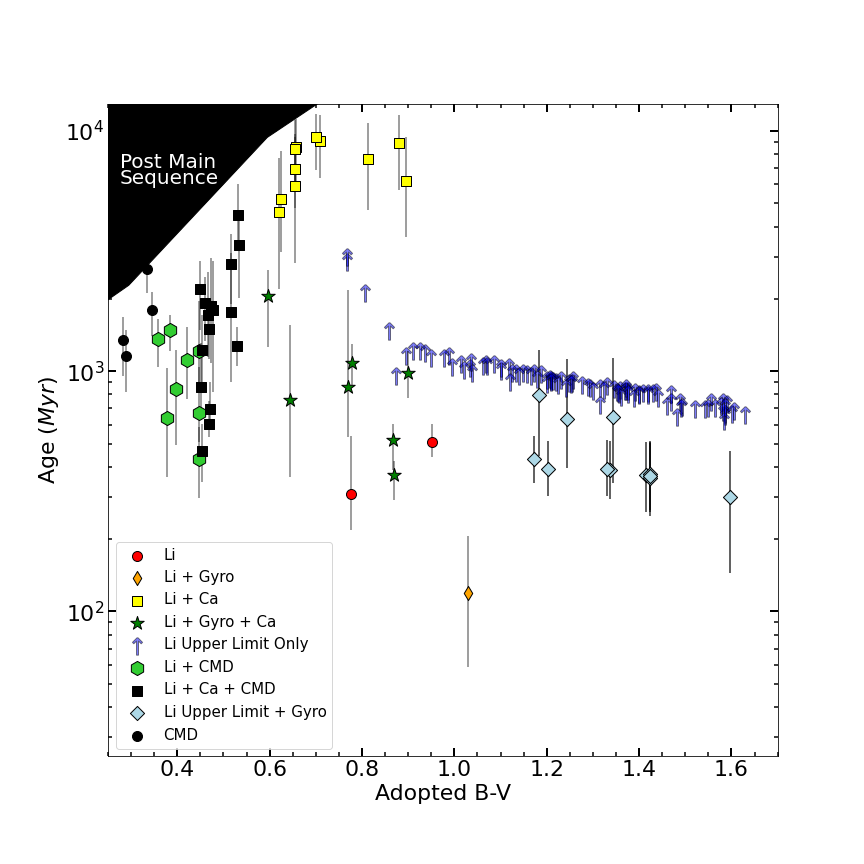}

 \caption{The medians and $68\%$ confidence intervals for our derived age posteriors for the full sample of 166 stars, as a function of $B-V$. The symbol and color of each star represent which age tracer (or tracers) was used to derive the age posterior. The error bars represent the $68\%$ confidence intervals.}
 \label{bmv_age}
\end{figure*}

For \rev{55} stars more than one age tracer is available, and we improve our constraints by taking the product of each age posterior for a given star. In Table \ref{tab:general} we present our full 166-star sample, including basic properties and \rev{final median ages and $68\%$ and $95\%$ confidence intervals for each star}. Figure \ref{bmv_age} shows the age posteriors for the sample as a function of $B-V$.

{\setlength{\tabcolsep}{4pt}\begin{longrotatetable}
\begin{deluxetable}{|c|c|c|c|c|c|c|c|c|c|c|c|}
\centering
\tabletypesize{\scriptsize}
\tablecaption{Our complete sample of accelerating stars with age posteriors and age tracers used. Ages come from a combination of lithium (Li), lithium upper limits (Li UL),\\ calcium emission (R'HK), rotation periods (Gyro), and CMD positions (CMD). Age posteriors are either presented as a median with $68\%$ and $95\%$ confidence intervals or as $68\%$,\\ $95\%$ and $99.7\%$ lower limits.}
\tablehead{
\colhead{}  & \colhead{}  &\colhead{}   & \colhead{}  & \colhead{}  & \colhead{}  & \colhead{}  & \colhead{}    & \colhead{Median Age}  & \colhead{$68\%$ CI}   & \colhead{$95\%$ CI} \\
\colhead{Target}  & \colhead{R.A.}  &\colhead{Dec.}   & \colhead{G}  & \colhead{Adopted}   & \colhead{Simbad} & \colhead{Dist}  & \colhead{Tracers}  & \colhead{ or $68\%$ LL}  & \colhead{or $95\%$ LL}   & \colhead{or $99.7\%$ LL} \\
\colhead{} & \colhead{}  & \colhead{} & \colhead{mag}  & \colhead{$B-V$}  &\colhead{Spec. Type}   & \colhead{(pc)}  &\colhead{} & \colhead{(Myr)} & \colhead{(Myr)} & \colhead{(Myr)} }

\startdata
HIP 143 &  00 01 49.35 &  +21 27 45.01 &  $ 9.9811 \pm 0.0002 $ &  $ 1.033 \pm 0.019 $ &  K3 & $ 43.4 \pm  0.59 $&  Li UL & $\ge 4442.08 $ & $\ge 1034.76 $ & $\ge 484.61 $ \\
HIP 669 &  00 08 16.36 &  -14 49 28.16 &  $ 6.8956 \pm 0.0002 $ &  $ 0.624 \pm 0.015 $ &   G1V & $ 27.31 \pm  0.02 $&  Li, R'HK &  5234.81 &  3132.71-8290.74 &  1783.08-11668.88 \\
HIP 1068 &  00 13 15.83 &  +69 19 36.76 &  $ 12.0476 \pm 0.0015 $ &  $ 1.315 \pm 0.1 $ &   M4 & $ 20.52 \pm  0.05 $&  Li UL & $\ge 4217.04 $ & $\ge 720.14 $ & $\ge 138.08 $ \\
HIP 1412 &  00 17 40.90 &  -08 40 56.14 &  $ 10.2711 \pm 0.0006 $ &  $ 1.414 \pm 0.005 $ &  K7V & $ 32.05 \pm  0.09 $&  Li UL, Gyro &  367.99 &  258.82-508.19 &  148.22-612.04 \\
HIP 1539 &  00 19 12.39 &  -03 03 13.01 &  $ 10.2808 \pm 0.0004 $ &  $ 1.375 \pm 0.005 $ &   K5V & $ 31.19 \pm  0.02 $&  Li UL & $\ge 4294.84 $ & $\ge 828.91 $ & $\ge 280.46 $ \\
HIP 1771 &  00 22 25.19 &  +06 40 03.16 &  $ 10.6176 \pm 0.0005 $ &  $ 1.489 \pm 0.1 $ &   K7V & $ 43.88 \pm  0.09 $&  Li UL & $\ge 4234.75 $ & $\ge 744.83 $ & $\ge 164 $ \\
HIP 2426 &  00 30 55.58 &  +73 22 14.89 &  $ 10.0247 \pm 0.0002 $ &  $ 1.359 \pm 0.02 $ &   K & $ 29.48 \pm  0.01 $&  Li UL & $\ge 4298.94 $ & $\ge 834.64 $ & $\ge 285.77 $ \\
HIP 2843 &  00 36 01.85 &  -05 34 14.58 &  $ 6.6114 \pm 0.0002 $ &  $ 0.466 \pm 0.007 $ &   F6V & $ 44.84 \pm  0.06 $&  Li UL, R'HK, CMD &  1715.26 &  1150.23-2585.26 &  808.98-3587.63 \\
HIP 3008 &  00 38 15.29 &  +52 19 55.52 &  $ 9.8115 \pm 0.0003 $ &  $ 1.424 \pm 0.02 $ &   M0.0V & $ 23.02 \pm  0.01 $&  Li UL, Gyro &  360.24 &  250.15-502.94 &  128.65-609.82 \\
HIP 3293 &  00 42 01.42 &  +40 41 17.71 &  $ 6.9164 \pm 0.0003 $ &  $ 0.452 \pm 0.009 $ &   F5 & $ 49.58 \pm  0.07 $&  Li, R'HK, CMD &  862.22 &  651.1-1311.35 &  521.75-2077.47 \\
HIP 3633 &  00 46 34.04 &  +14 19 31.61 &  $ 9.6005 \pm 0.0001 $ &  $ 0.814 \pm 0.061 $ &   K3V & $ 46.51 \pm  0.21 $&  Li UL, R'HK &  7664.88 &  4696.93-10795.26 &  2383.87-12584.18 \\
HIP 3724 &  00 47 48.47 &  +06 01 54.92 &  $ 10.963 \pm 0.0004 $ &  $ 1.4 \pm 0.02 $ &   M0.0V & $ 40.62 \pm  0.05 $&  Li UL & $\ge 4277.49 $ & $\ge 804.65 $ & $\ge 255.04 $ \\
HIP 4024 &  00 51 34.56 &  -14 53 26.46 &  $ 9.4072 \pm 0.0002 $ &  $ 1.069 \pm 0.009 $ &   K3.5IV-V & $ 43.46 \pm  0.15 $&  Li UL & $\ge 4471.25 $ & $\ge 1075.56 $ & $\ge 530.93 $ \\
HIP 5319 &  01 08 01.34 &  +32 00 43.66 &  $ 6.1346 \pm 0.0003 $ &  $ 0.398 \pm 0.01 $ &  F5IV & $ 42.9 \pm  0.06 $&  Li UL, CMD &  843.22 &  495.26-1230.45 &  219.62-1644.67 \\
HIP 5763 &  01 13 58.86 &  +16 29 40.26 &  $ 9.3939 \pm 0.0002 $ &  $ 1.22 \pm 0.015 $ & K6V & $ 32.01 \pm  0.03 $&  Li UL & $\ge 4348.13 $ & $\ge 903.42 $ & $\ge 352.3 $ \\
HIP 6776 &  01 27 06.32 &  +34 22 39.18 &  $ 6.1623 \pm 0.0003 $ &  $ 0.453 \pm 0.007 $ &   F7IVsv & $ 42.05 \pm  0.05 $&  Li UL, R'HK, CMD &  1220.64 &  830.65-1719.17 &  617.88-2140.42 \\
HIP 6796 &  01 27 27.60 &  +23 14 40.42 &  $ 9.4186 \pm 0.0004 $ &  $ 1.157 \pm 0.074 $ &  K7V & $ 37.17 \pm  0.04 $&  Li UL & $\ge 4400.69 $ & $\ge 976.89 $ & $\ge 409.71 $ \\
HIP 6833 &  01 27 55.56 &  -01 59 29.35 &  $ 8.4502 \pm 0.0002 $ &  $ 0.656 \pm 0.002 $ &  G5V & $ 48.1 \pm  0.06 $&  Li, R'HK &  6929.95 &  4767.61-9685.66 &  2971.34-12173.29 \\
HIP 6943 &  01 29 26.45 &  +42 16 54.85 &  $ 7.078 \pm 0.0003 $ &  $ 0.453 \pm 0.011 $ &   F5 & $ 48.19 \pm  0.06 $&  Li, R'HK, CMD &  466.38 &  344.95-600.94 &  208.59-1059.12 \\
HIP 7287 &  01 33 53.62 &  +27 14 08.74 &  $ 10.5178 \pm 0.0002 $ &  $ 1.2 \pm 0.02 $ &   K5 & $ 48.57 \pm  0.22 $&  Li UL & $\ge 4334.42 $ & $\ge 884.25 $ & $\ge 332.72 $ \\
HIP 8653 &  01 51 31.19 &  -07 44 23.57 &  $ 8.4607 \pm 0.0003 $ &  $ 0.644 \pm 0.019 $ &   G5V & $ 47.89 \pm  1.09 $&  Li, Gyro, R'HK &  759.94 &  364.45-1561.09 &  185.93-2455.98 \\
HIP 9067 &  01 56 43.73 &  +29 48 39.04 &  $ 10.8658 \pm 0.0003 $ &  $ 1.47 \pm 0.1 $ &  K5V & $ 47.3 \pm  0.03 $&  Li UL & $\ge 4275.7 $ & $\ge 801.85 $ & $\ge 231.94 $ \\
HIP 9989 &  02 08 40.08 &  -00 21 35.61 &  $ 10.4039 \pm 0.0003 $ &  $ 1.248 \pm 0.002 $ &   K5V & $ 39.45 \pm  0.03 $&  Li UL & $\ge 4340.02 $ & $\ge 892.08 $ & $\ge 341.68 $ \\
HIP 10050 &  02 09 23.11 &  +17 13 27.11 &  $ 6.2937 \pm 0.0003 $ &  $ 0.471 \pm 0.001 $ &   F5V & $ 36.32 \pm  0.04 $&  Li, R'HK, CMD &  694.63 &  612.9-897.62 &  509.72-1344.86 \\
HIP 10532 &  02 15 41.72 &  -09 00 47.23 &  $ 8.877 \pm 0.0002 $ &  $ 0.88 \pm 0.033 $ &   K0V & $ 44.19 \pm  0.03 $&  Li UL, R'HK &  8922.3 &  5685.95-11625.28 &  2863.63-12794.61 \\
HIP 11090 &  02 22 50.30 &  +41 23 46.65 &  $ 5.7433 \pm 0.0009 $ &  $ 0.289 \pm 0.006 $ &  F0III-IV & $ 48.33 \pm  0.14 $&  CMD &  1155.1 &  820.23-1479.52 &  520.53-1744.73 \\
HIP 11815 &  02 32 23.18 &  +03 22 57.48 &  $ 11.2101 \pm 0.0014 $ &  $ 1.357 \pm 0.02 $ &  K5 & $ 49.26 \pm  0.05 $&  Li UL & $\ge 4276.65 $ & $\ge 803.47 $ & $\ge 253.41 $ \\
HIP 12837 &  02 45 01.09 &  -22 09 59.51 &  $ 7.0178 \pm 0.0002 $ &  $ 0.517 \pm 0.008 $ &   F7V & $ 38.9 \pm  0.03 $&  Li, R'HK, CMD &  1771.7 &  899.96-3111.02 &  386.64-4669.22 \\
HIP 13681 &  02 56 14.10 &  -11 50 49.84 &  $ 9.8618 \pm 0.0003 $ &  $ 1.062 \pm 0.015 $ &   -- & $ 47.51 \pm  0.04 $&  Li UL & $\ge 4456.4 $ & $\ge 1054.8 $ & $\ge 509.53 $ \\
HIP 13782 &  02 57 23.76 &  -23 51 43.79 &  $ 5.399 \pm 0.0005 $ &  $ 0.238 \pm 0.003 $ &   A5IV/V & $ 48.69 \pm  0.15 $&  CMD &  757.31 &  595.43-917.22 &  443.02-1075.26 \\
HIP 15211 &  03 16 05.89 &  -05 51 15.85 &  $ 9.9066 \pm 0.0002 $ &  $ 0.936 \pm 0.007 $ &  K3 & $ 49.29 \pm  0.04 $&  Li UL & $\ge 4551.82 $ & $\ge 1187.49 $ & $\ge 609.82 $ \\
HIP 16247 &  03 29 22.87 &  -24 06 03.09 &  $ 8.8107 \pm 0.0005 $ &  $ 1.02 \pm 0.037 $ &  K3VkFe+0.4 & $ 31.08 \pm  0.02 $&  Li UL & $\ge 4428.57 $ & $\ge 1015.84 $ & $\ge 461.81 $ \\
HIP 16709 &  03 34 59.64 &  +34 36 39.51 &  $ 10.3002 \pm 0.0004 $ &  $ 1.225 \pm 0.02 $ &   M0 & $ 41.1 \pm  0.65 $&  Li UL & $\ge 4331.28 $ & $\ge 879.86 $ & $\ge 328.75 $ \\
HIP 17118 &  03 39 58.81 &  +63 52 13.47 &  $ 6.7675 \pm 0.0002 $ &  $ 0.478 \pm 0.023 $ &   F5 & $ 42.81 \pm  0.03 $&  Li, R'HK, CMD &  1790.5 &  818.76-2885.78 &  405.89-3903.25 \\
HIP 17213 &  03 41 13.71 &  +75 23 56.51 &  $ 9.6211 \pm 0.0006 $ &  $ 1.1 \pm 0.015 $ &   K0 & $ 43.63 \pm  0.08 $&  Li UL & $\ge 4420.43 $ & $\ge 1004.51 $ & $\ge 453.14 $ \\
HIP 18527 &  03 57 43.92 &  -20 16 04.05 &  $ 8.6088 \pm 0.0004 $ &  $ 0.899 \pm 0.021 $ &   K2VkFe-0.5 & $ 37.38 \pm  0.03 $&  Li UL, Gyro, R'HK &  981.67 &  770.11-1216.88 &  619.34-1469.49 \\
HIP 19739 &  04 13 55.87 &  +82 55 06.98 &  $ 10.1025 \pm 0.0005 $ &  $ 1.424 \pm 0.02 $ &   M0.0V & $ 36.94 \pm  0.17 $&  Li UL, Gyro &  372.38 &  265.37-510.32 &  199.63-612.75 \\
HIP 19912 &  04 16 18.56 &  +08 43 09.96 &  $ 9.4283 \pm 0.0002 $ &  $ 1.148 \pm 0.002 $ &   K7V & $ 38.96 \pm  0.26 $&  Li UL & $\ge 4408.38 $ & $\ge 987.64 $ & $\ge 433.87 $ \\
HIP 20419 &  04 22 25.69 &  +11 18 20.56 &  $ 9.3585 \pm 0.0004 $ &  $ 1.183 \pm 0.005 $ &   K2 & $ 44.05 \pm  0.48 $&  Li UL, Gyro &  797.7 &  444.12-1230.22 &  313.95-1599.99 \\
HIP 20648 &  04 25 29.38 &  +17 55 40.46 &  $ 4.2957 \pm 0.001 $ &  $ 0.049 \pm 0.007 $ &   A2IV-Vs & $ 47.42 \pm  0.31 $&  CMD &  602.61 &  530.07-670.92 &  456.89-742.38 \\
HIP 21066 &  04 30 57.17 &  +10 45 06.36 &  $ 6.9211 \pm 0.0002 $ &  $ 0.472 \pm 0.013 $ &   F7V & $ 46.51 \pm  0.06 $&  Li UL, R'HK, CMD &  1864.19 &  1085.7-2959.28 &  650.55-3268.37 \\
HIP 21152 &  04 32 04.80 &  +05 24 36.14 &  $ 6.2678 \pm 0.0002 $ &  $ 0.42 \pm 0.014 $ &   F5V & $ 43.27 \pm  0.05 $&  Li UL, CMD &  1116.14 &  765.3-1521.58 &  423.86-1861.12 \\
HIP 22361 &  04 48 50.35 &  +75 56 28.39 &  $ 5.8984 \pm 0.0004 $ &  $ 0.283 \pm 0.004 $ &   A9IV & $ 47.6 \pm  0.06 $&  CMD &  1350.34 &  959.72-1684.76 &  602.75-1979.1 \\
HIP 23208 &  04 59 38.25 &  -10 59 02.26 &  $ 9.3009 \pm 0.0004 $ &  $ 1.18 \pm 0.006 $ &  K5Vk & $ 34.82 \pm  0.05 $&  Li UL & $\ge 4365.63 $ & $\ge 927.89 $ & $\ge 376.02 $ \\
HIP 24017 &  05 09 44.50 &  +64 55 10.16 &  $ 6.2661 \pm 0.0002 $ &  $ 0.448 \pm 0.004 $ &   F3V & $ 39.09 \pm  0.03 $&  Li UL, CMD &  1216.34 &  828.62-1964.68 &  633.89-2692.15 \\
HIP 24177 &  05 11 29.73 &  +10 07 15.66 &  $ 11.156 \pm 0.0005 $ &  $ 1.461 \pm 0.005 $ &  M0V & $ 44.19 \pm  0.18 $&  Li UL & $\ge 4216.67 $ & $\ge 719.62 $ & $\ge 159.14 $ \\
HIP 24292 &  05 12 47.95 &  +16 18 25.43 &  $ 10.1675 \pm 0.0003 $ &  $ 1.23 \pm 0.015 $ &  K7V & $ 44.01 \pm  0.03 $&  Li UL & $\ge 4357.55 $ & $\ge 916.6 $ & $\ge 365.54 $ \\
HIP 24457 &  05 14 49.83 &  +00 42 31.67 &  $ 9.4575 \pm 0.0002 $ &  $ 1.118 \pm 0.005 $ &  K5V & $ 39.6 \pm  0.02 $&  Li UL & $\ge 4445.1 $ & $\ge 1038.96 $ & $\ge 482.73 $ \\
HIP 25606 &  05 28 14.72 &  -20 45 33.99 &  $ 2.6272 \pm 0.0015 $ &  $ 0.807 \pm 0.027 $ &  G5II-IIIa & $ 47.92 \pm  0.47 $&  Li UL & $\ge 5405.2 $ & $\ge 2111.49 $ & $\ge 1021.51 $ \\
HIP 27021 &  05 43 52.79 &  -21 16 47.04 &  $ 6.9723 \pm 0.0001 $ &  $ 0.531 \pm 0.006 $ &   F8/G0V & $ 43.2 \pm  0.03 $&  Li, R'HK, CMD &  4464.59 &  3159.78-5991.22 &  2077.47-7143.76 \\
HIP 28823 &  06 05 03.38 &  +42 58 53.88 &  $ 5.8206 \pm 0.0006 $ &  $ 0.358 \pm 0.005 $ & F3V & $ 49.78 \pm  0.22 $&  Li UL, CMD &  1360.86 &  1038.87-1642.63 &  703.98-1867.14 \\
HIP 29208 &  06 09 35.91 &  +05 40 08.05 &  $ 8.194 \pm 0.0002 $ &  $ 0.895 \pm 0.022 $ &   G0 & $ 30.19 \pm  0.02 $&  Li UL & $\ge 4526.8 $ & $\ge 1152.02 $ & $\ge 568.66 $ \\
HIP 33282 &  06 55 23.46 &  +44 25 19.30 &  $ 10.7232 \pm 0.0004 $ &  $ 1.344 \pm 0.035 $ &  K6V & $ 48.22 \pm  0.03 $&  Li UL, Gyro &  647.7 &  341.31-1139.51 &  248.03-1546.97 \\
HIP 33368 &  06 56 22.10 &  +46 59 26.75 &  $ 10.4419 \pm 0.0005 $ &  $ 1.346 \pm 0.1 $ &  K7V & $ 47.17 \pm  0.06 $&  Li UL & $\ge 4308.04 $ & $\ge 847.34 $ & $\ge 288 $ \\
HIP 34804 &  07 12 17.20 &  +59 15 51.95 &  $ 10.2697 \pm 0.0003 $ &  $ 1.591 \pm 0.082 $ & K4 & $ 38.93 \pm  0.7 $&  Li UL & $\ge 4224.76 $ & $\ge 730.29 $ & $\ge 157.52 $ \\
HIP 36081 &  07 26 03.61 &  -15 39 57.28 &  $ 6.7409 \pm 0.0002 $ &  $ 0.62 \pm 0.005 $ &  G1/2V & $ 49.68 \pm  0.46 $&  Li, R'HK &  4585.75 &  2189.37-7747.86 &  1010.09-11446.52 \\
HIP 37267 &  07 39 10.33 &  +29 54 32.16 &  $ 10.4391 \pm 0.0006 $ &  $ 1.277 \pm 0.022 $ &   K5V & $ 48.9 \pm  0.03 $&  Li UL & $\ge 4328.24 $ & $\ge 875.61 $ & $\ge 325.41 $ \\
HIP 40518 &  08 16 23.06 &  -17 22 03.83 &  $ 10.0726 \pm 0.0003 $ &  $ 1.134 \pm 0.004 $ &  -- & $ 46.43 \pm  0.6 $&  Li UL & $\ge 4406.56 $ & $\ge 985.1 $ & $\ge 430.98 $ \\
HIP 41274 &  08 25 14.08 &  -07 10 12.85 &  $ 7.0892 \pm 0.0006 $ &  $ 0.952 \pm 0.011 $ &  K0V & $ 49.14 \pm  0.07 $&  Li &  505.8 &  439.44-602.43 &  335.92-820.22 \\
HIP 41277 &  08 25 16.90 &  +04 15 12.72 &  $ 9.9158 \pm 0.002 $ &  $ 1.03 \pm 0.015 $ &  K8V & $ 41.68 \pm  0.08 $&  Li, Gyro &  119.62 &  58.63-206.48 &  17.47-300.01 \\
HIP 41319 &  08 25 49.87 &  +17 02 46.57 &  $ 5.9986 \pm 0.0003 $ &  $ 0.448 \pm 0.005 $ &  F5IIIm? & $ 45.5 \pm  0.07 $&  Li, CMD &  668.73 &  509.36-1040.76 &  400.75-1505.21 \\
HIP 42231 &  08 36 37.96 &  +39 02 13.55 &  $ 11.2749 \pm 0.0003 $ &  $ 1.241 \pm 0.015 $ &   -- & $ 43.88 \pm  0.15 $&  Li UL & $\ge 4308.4 $ & $\ge 847.87 $ & $\ge 296.87 $ \\
HIP 42343 &  08 37 57.55 &  -08 52 53.72 &  $ 6.9078 \pm 0.0002 $ &  $ 0.516 \pm 0.001 $ &   F6V & $ 48.73 \pm  0.05 $&  Li, R'HK, CMD &  2791.29 &  1896.43-3724.4 &  1184.85-4152.26 \\
HIP 42507 &  08 40 00.26 &  -06 28 33.06 &  $ 9.2424 \pm 0.0003 $ &  $ 1.361 \pm 0.001 $ &  K6V & $ 25.46 \pm  0.09 $&  Li UL & $\ge 4259.1 $ & $\ge 778.95 $ & $\ge 226.84 $ \\
HIP 43745 &  08 54 35.72 &  +63 45 29.74 &  $ 10.4509 \pm 0.0002 $ &  $ 1.423 \pm 0.02 $ &   K5V & $ 34.26 \pm  0.02 $&  Li UL, Gyro &  366.55 &  260.53-505.31 &  168.77-610.43 \\
HIP 44162 &  08 59 39.91 &  -19 12 28.95 &  $ 6.0845 \pm 0.0003 $ &  $ 0.461 \pm 0.005 $ &   F5V & $ 40.24 \pm  0.13 $&  Li UL, R'HK, CMD &  1925.01 &  1340.03-2478.25 &  901.96-3477.51 \\
HIP 44953 &  09 09 30.16 &  -00 23 54.35 &  $ 8.4254 \pm 0.0002 $ &  $ 0.767 \pm 0.002 $ &   G8V & $ 34.95 \pm  0.03 $&  Li UL & $\ge 5993.75 $ & $\ge 2846.19 $ & $\ge 1323.58 $ \\
HIP 45621 &  09 17 55.38 &  -03 23 14.09 &  $ 7.5442 \pm 0.0003 $ &  $ 0.869 \pm 0.013 $ &   K1/2V & $ 32.28 \pm  0.04 $&  Li, Gyro, R'HK &  371.47 &  291.48-423.05 &  144.29-489.92 \\
HIP 45731 &  09 19 22.86 &  +62 03 16.89 &  $ 10.5142 \pm 0.0006 $ &  $ 1.586 \pm 0.02 $ &   M1.0Ve & $ 37.94 \pm  0.03 $&  Li UL & $\ge 4167.6 $ & $\ge 650.87 $ & $\ge 91.63 $ \\
HIP 46385 &  09 27 30.56 &  +50 39 12.45 &  $ 10.9763 \pm 0.0005 $ &  $ 1.586 \pm 0.1 $ &   M2.5V & $ 31.16 \pm  0.05 $&  Li UL & $\ge 4175.04 $ & $\ge 661.4 $ & $\ge 92.44 $ \\
HIP 47013 &  09 34 53.52 &  +72 12 20.44 &  $ 5.6501 \pm 0.0005 $ &  $ 0.53 \pm 0.003 $ &   F7V & $ 49.6 \pm  0.08 $&  Li UL, R'HK, CMD &  1275.94 &  1054.99-1523.22 &  861.32-1758.82 \\
HIP 47261 &  09 37 58.33 &  +22 31 23.15 &  $ 9.4588 \pm 0.0003 $ &  $ 1.21 \pm 0.015 $ &   K4V & $ 30.89 \pm  0.02 $&  Li UL & $\ge 4351.49 $ & $\ge 908.11 $ & $\ge 356.79 $ \\
HIP 47300 &  09 38 21.77 &  +40 14 23.25 &  $ 5.237 \pm 0.0007 $ &  $ 0.223 \pm 0.004 $ &   F0V & $ 38.49 \pm  0.12 $&  CMD &  718.62 &  500.33-934.73 &  298.93-1134.22 \\
HIP 47387 &  09 39 14.59 &  +66 50 58.14 &  $ 9.6883 \pm 0.0002 $ &  $ 1.07 \pm 0.047 $ &   K2 & $ 41.63 \pm  0.54 $&  Li UL & $\ge 4450.55 $ & $\ge 1046.57 $ & $\ge 489.6 $ \\
HIP 47539 &  09 41 31.44 &  -06 23 35.83 &  $ 11.2695 \pm 0.0004 $ &  $ 1.353 \pm 0.011 $ & -- & $ 44.42 \pm  0.37 $&  Li UL & $\ge 4288.12 $ & $\ge 819.51 $ & $\ge 270.07 $ \\
HIP 47741 &  09 43 55.61 &  +26 58 08.40 &  $ 10.8486 \pm 0.0004 $ &  $ 1.542 \pm 0.039 $  & M3.5V & $ 14.37 \pm  0.01 $&  Li UL & $\ge 4202.01 $ & $\ge 698.97 $ & $\ge 130.85 $ \\
HIP 48165 &  09 49 03.16 &  +07 49 08.68 &  $ 9.5048 \pm 0.0002 $ &  $ 0.925 \pm 0.015 $  & G5 & $ 45.77 \pm  0.48 $&  Li UL & $\ge 4574.64 $ & $\ge 1219.3 $ & $\ge 638.59 $ \\
HIP 48629 &  09 54 59.84 &  +40 23 07.28 &  $ 10.7594 \pm 0.0002 $ &  $ 1.581 \pm 0.197 $  & K5V & $ 37.37 \pm  0.04 $&  Li UL & $\ge 4234.47 $ & $\ge 744.26 $ & $\ge 152.31 $ \\
HIP 49046 &  10 00 26.71 &  +27 16 01.61 &  $ 10.5279 \pm 0.0002 $ &  $ 1.372 \pm 0.015 $  & M0.5V & $ 36.01 \pm  0.02 $&  Li UL & $\ge 4292.26 $ & $\ge 825.31 $ & $\ge 276.52 $ \\
HIP 49526 &  10 06 43.80 &  +41 42 52.65 &  $ 10.5229 \pm 0.0003 $ &  $ 1.492 \pm 0.02 $  & M1.0V & $ 22.28 \pm  0.02 $&  Li UL & $\ge 4210.41 $ & $\ge 710.86 $ & $\ge 147.69 $ \\
HIP 49577 &  10 07 13.74 &  -14 18 16.54 &  $ 9.6168 \pm 0.0004 $ &  $ 1.331 \pm 0.014 $ & K5V & $ 39.53 \pm  0.64 $&  Li UL & $\ge 4322.44 $ & $\ge 867.5 $ & $\ge 318.63 $ \\
HIP 49910 &  10 11 23.70 &  -01 39 44.69 &  $ 11.1457 \pm 0.0004 $ &  $ 1.391 \pm 0.023 $ &   -- & $ 46.55 \pm  0.04 $&  Li UL & $\ge 4286.95 $ & $\ge 817.89 $ & $\ge 268.72 $ \\
HIP 51208 &  10 27 34.85 &  +12 58 48.14 &  $ 10.2039 \pm 0.0003 $ &  $ 1.2 \pm 0.015 $ &  K4V & $ 46.38 \pm  0.46 $&  Li UL & $\ge 4361.89 $ & $\ge 922.66 $ & $\ge 371.01 $ \\
HIP 52339 &  10 41 48.44 &  +12 06 31.90 &  $ 10.9365 \pm 0.0003 $ &  $ 1.42 \pm 0.015 $  & K7V & $ 37.04 \pm  0.03 $&  Li UL & $\ge 4281.04 $ & $\ge 809.62 $ & $\ge 260.19 $ \\
HIP 53175 &  10 52 39.61 &  +00 29 01.95 &  $ 10.1771 \pm 0.0004 $ &  $ 1.293 \pm 0.005 $  & K4V & $ 39.35 \pm  0.03 $&  Li UL & $\ge 4312.99 $ & $\ge 854.28 $ & $\ge 304.58 $ \\
HIP 53869 &  11 01 16.35 &  +29 35 27.47 &  $ 9.4421 \pm 0.0003 $ &  $ 0.868 \pm 0.045 $  & K3V & $ 49.58 \pm  0.06 $&  Li UL, Gyro, R'HK &  515.87 &  388.61-604.35 &  245.7-648.74 \\
HIP 54094 &  11 04 07.15 &  +53 22 55.47 &  $ 9.7081 \pm 0.0001 $ &  $ 1.1 \pm 0.02 $ &   -- & $ 46.51 \pm  0.03 $&  Li UL & $\ge 4441.41 $ & $\ge 1033.83 $ & $\ge 481.2 $ \\
HIP 54199 &  11 05 16.57 &  +65 48 49.51 &  $ 8.2868 \pm 0.0001 $ &  $ 0.874 \pm 0.016 $  & G5 & $ 39.08 \pm  0.45 $&  Li UL & $\ge 4383.16 $ & $\ge 952.11 $ & $\ge 383.88 $ \\
HIP 55192 &  11 18 00.48 &  +35 26 41.81 &  $ 8.0635 \pm 0.0003 $ &  $ 0.776 \pm 0.015 $ &   G5 & $ 39.7 \pm  0.17 $&  Li &  308.96 &  218.51-539.13 &  66.28-1459.69 \\
HIP 55409 &  11 20 51.76 &  -23 13 02.42 &  $ 7.8415 \pm 0.0002 $ &  $ 0.658 \pm 0.003 $ &   G3/5V & $ 42.16 \pm  0.04 $&  Li UL, R'HK &  8592.32 &  5708.12-11271.52 &  3349.2-12735.77 \\
HIP 57361 &  11 45 34.44 &  -20 21 12.38 &  $ 10.8203 \pm 0.0004 $ &  $ 1.482 \pm 0.015 $ &   M2.5V & $ 19.2 \pm  0.06 $&  Li UL & $\ge 4164.34 $ & $\ge 646.46 $ & $\ge 85.43 $ \\
HIP 57571 &  11 48 01.50 &  +71 21 28.14 &  $ 10.754 \pm 0.0002 $ &  $ 1.369 \pm 0.02 $ &  -- & $ 45.93 \pm  0.35 $&  Li UL & $\ge 4295.72 $ & $\ge 830.14 $ & $\ge 281.31 $ \\
HIP 57645 &  11 49 13.15 &  -20 20 34.67 &  $ 8.9933 \pm 0.0002 $ &  $ 0.895 \pm 0.015 $ &  G3V & $ 38.12 \pm  0.03 $&  Li UL, R'HK &  6204.18 &  3637.04-9422.01 &  1781.04-12308.42 \\
HIP 57984 &  11 53 37.20 &  +49 36 01.07 &  $ 10.5512 \pm 0.0002 $ &  $ 1.316 \pm 0.015 $ &  K5 & $ 45.0 \pm  0.15 $&  Li UL & $\ge 4310.95 $ & $\ge 851.43 $ & $\ge 302.07 $ \\
HIP 59000 &  12 05 50.66 &  -18 52 30.92 &  $ 9.3933 \pm 0.0007 $ &  $ 1.336 \pm 0.014 $ &  K5V & $ 23.06 \pm  0.05 $&  Li UL, Gyro &  388.45 &  293.22-514.42 &  219.44-613.01 \\
HIP 59198 &  12 08 22.20 &  -00 28 57.36 &  $ 10.5473 \pm 0.0004 $ &  $ 1.409 \pm 0.015 $ &  K4V & $ 29.13 \pm  0.08 $&  Li UL & $\ge 4283.83 $ & $\ge 813.52 $ & $\ge 264.26 $ \\
HIP 59199 &  12 08 24.81 &  -24 43 43.95 &  $ 3.9466 \pm 0.0013 $ &  $ 0.334 \pm 0.015 $ &   F1V & $ 14.98 \pm  0.04 $&  CMD &  2665.78 &  2119.54-3334.23 &  1360.6-3849.28 \\
HIP 59247 &  12 09 13.08 &  +19 28 03.61 &  $ 10.4409 \pm 0.0003 $ &  $ 1.29 \pm 0.015 $ &   K7V & $ 41.84 \pm  0.03 $&  Li UL & $\ge 4317.08 $ & $\ge 860.01 $ & $\ge 310.2 $ \\
HIP 59767 &  12 15 20.27 &  +87 42 00.42 &  $ 6.1886 \pm 0.0008 $ &  $ 0.346 \pm 0.011 $ &   F1IV & $ 48.38 \pm  0.05 $&  CMD &  1793.47 &  1401.11-2138.93 &  945.46-2464.7 \\
HIP 59905 &  12 17 15.20 &  -07 48 43.93 &  $ 10.0396 \pm 0.0003 $ &  $ 1.125 \pm 0.015 $ &   K4V & $ 47.71 \pm  0.05 $&  Li UL & $\ge 4406.67 $ & $\ge 985.26 $ & $\ge 430.87 $ \\
HIP 59953 &  12 17 48.39 &  +46 37 20.73 &  $ 10.7356 \pm 0.0002 $ &  $ 1.602 \pm 0.1 $ &   M0.5V & $ 35.68 \pm  0.02 $&  Li UL & $\ge 4177.73 $ & $\ge 665.13 $ & $\ge 96.84 $ \\
HIP 60829 &  12 28 04.06 &  -15 39 07.60 &  $ 8.1604 \pm 0.0002 $ &  $ 0.71 \pm 0.012 $ &   G6V & $ 47.06 \pm  0.06 $&  Li UL, R'HK &  9106.19 &  6391.7-11635.89 &  4156.82-12789.88 \\
HIP 61947 &  12 41 46.81 &  +43 02 26.34 &  $ 11.1443 \pm 0.0003 $ &  $ 1.582 \pm 0.082 $ &   K5V & $ 44.66 \pm  0.04 $&  Li UL & $\ge 4183.93 $ & $\ge 673.77 $ & $\ge 104.87 $ \\
HIP 62325 &  12 46 22.54 &  +09 32 22.86 &  $ 5.3697 \pm 0.0004 $ &  $ 0.989 \pm 0.007 $ &   G9 & $ 43.71 \pm  0.12 $&  Li UL & $\ge 4529.3 $ & $\ge 1156.33 $ & $\ge 585.8 $ \\
HIP 62627 &  12 49 56.23 &  +71 11 39.30 &  $ 9.707 \pm 0.0001 $ &  $ 0.995 \pm 0.003 $ &   K8 & $ 49.26 \pm  0.07 $&  Li UL & $\ge 4447.71 $ & $\ge 1042.53 $ & $\ge 480.59 $ \\
HIP 63419 &  12 59 45.50 &  -04 25 49.04 &  $ 8.4878 \pm 0.0004 $ &  $ 0.779 \pm 0.021 $ &   K1V & $ 37.11 \pm  0.04 $&  Li, Gyro, R'HK &  1081.58 &  874.63-1294.12 &  662.26-1603.98 \\
HIP 63421 &  12 59 47.34 &  -06 51 37.39 &  $ 11.1775 \pm 0.0005 $ &  $ 1.354 \pm 0.01 $ &   K5V & $ 43.46 \pm  0.04 $&  Li UL & $\ge 4285.0 $ & $\ge 815.16 $ & $\ge 265.57 $ \\
HIP 63661 &  13 02 51.36 &  -02 40 26.25 &  $ 10.9351 \pm 0.0003 $ &  $ 1.468 \pm 0.1 $ &   K5V & $ 48.05 \pm  0.06 $&  Li UL & $\ge 4236.36 $ & $\ge 747.11 $ & $\ge 166.58 $ \\
HIP 65120 &  13 20 43.83 &  +10 52 33.31 &  $ 10.8352 \pm 0.0003 $ &  $ 1.49 \pm 0.02 $ &  M0V & $ 31.35 \pm  0.03 $&  Li UL & $\ge 4206.97 $ & $\ge 706.06 $ & $\ge 142.13 $ \\
HIP 65602 &  13 27 02.90 &  -24 17 25.56 &  $ 8.4679 \pm 0.0002 $ &  $ 0.911 \pm 0.002 $ &  K2+V & $ 29.87 \pm  0.03 $&  Li UL & $\ge 4574.65 $ & $\ge 1219.41 $ & $\ge 641.92 $ \\
HIP 65651 &  13 27 35.55 &  -01 50 16.98 &  $ 10.3186 \pm 0.0002 $ &  $ 1.203 \pm 0.02 $ &   K4 & $ 45.52 \pm  0.05 $&  Li UL, Gyro &  392.63 &  301.17-513.07 &  126.99-611.37 \\
HIP 65706 &  13 28 17.76 &  +30 02 45.98 &  $ 10.5091 \pm 0.0003 $ &  $ 1.577 \pm 0.082 $ &   K5V & $ 40.49 \pm  0.09 $&  Li UL & $\ge 4215.51 $ & $\ge 717.64 $ & $\ge 144.27 $ \\
HIP 66262 &  13 34 49.79 &  +47 22 34.35 &  $ 9.8141 \pm 0.0001 $ &  $ 1.165 \pm 0.067 $ &  K4V & $ 36.82 \pm  0.36 $&  Li UL & $\ge 4386.44 $ & $\ge 956.97 $ & $\ge 393.3 $ \\
HIP 66315 &  13 35 29.15 &  +46 33 30.88 &  $ 10.4651 \pm 0.0002 $ &  $ 1.039 \pm 0.121 $ &  K3 & $ 43.23 \pm  0.03 $&  Li UL & $\ge 4412.74 $ & $\ge 985.51 $ & $\ge 395.09 $ \\
HIP 66587 &  13 38 58.68 &  -06 14 12.46 &  $ 10.0215 \pm 0.0003 $ &  $ 1.419 \pm 0.009 $ &   M0.5V & $ 24.19 \pm  0.01 $&  Li UL & $\ge 4273.57 $ & $\ge 799.18 $ & $\ge 249.51 $ \\
HIP 66828 &  13 41 47.23 &  +21 29 08.18 &  $ 10.3186 \pm 0.0002 $ &  $ 1.33 \pm 0.02 $ &  K5V & $ 38.3 \pm  0.03 $&  Li UL, Gyro &  393.32 &  301.47-515.92 &  223.58-613.17 \\
HIP 69311 &  14 11 12.64 &  +43 06 23.96 &  $ 9.5996 \pm 0.0002 $ &  $ 1.172 \pm 0.07 $ &   K5V & $ 35.34 \pm  0.01 $&  Li UL, Gyro &  431.78 &  342.15-536.81 &  251.18-620.33 \\
HIP 69333 &  14 11 24.19 &  +30 05 02.44 &  $ 9.8627 \pm 0.0002 $ &  $ 1.173 \pm 0.004 $ &  K5V & $ 43.8 \pm  0.03 $&  Li UL & $\ge 4402.4 $ & $\ge 979.3 $ & $\ge 426.36 $ \\
HIP 69860 &  14 17 47.84 &  +21 26 01.58 &  $ 10.8113 \pm 0.0005 $ &  $ 1.629 \pm 0.1 $ &  M1.0V & $ 41.84 \pm  0.21 $&  Li UL & $\ge 4171.9 $ & $\ge 656.97 $ & $\ge 92.18 $ \\
HIP 69963 &  14 18 59.04 &  +38 38 26.33 &  $ 10.7754 \pm 0.0004 $ &  $ 1.39 \pm 0.02 $ &   M1.0V & $ 31.69 \pm  0.05 $&  Li UL & $\ge 4252.64 $ & $\ge 769.92 $ & $\ge 217.68 $ \\
HIP 70472 &  14 24 49.86 &  -17 27 08.09 &  $ 10.1555 \pm 0.0004 $ &  $ 1.257 \pm 0.01 $ &   K7V & $ 34.2 \pm  0.05 $&  Li UL & $\ge 4358.91 $ & $\ge 918.5 $ & $\ge 368.06 $ \\
HIP 71243 &  14 34 11.70 &  +32 32 04.12 &  $ 6.2142 \pm 0.0001 $ &  $ 0.45 \pm 0.003 $ &  F5V & $ 41.36 \pm  0.03 $&  Li UL, R'HK, CMD &  2202.78 &  1488.73-2873.15 &  911.32-3350.92 \\
HIP 71515 &  14 37 33.07 &  -16 32 27.62 &  $ 8.4606 \pm 0.0005 $ &  $ 0.95 \pm 0.002 $ &   K2/3V & $ 44.11 \pm  0.3 $&  Li UL & $\ge 4516.62 $ & $\ge 1138.55 $ & $\ge 566.76 $ \\
HIP 71899 &  14 42 23.10 &  +21 17 35.10 &  $ 7.3501 \pm 0.0002 $ &  $ 0.533 \pm 0.01 $ &   F8 & $ 45.58 \pm  0.04 $&  Li, R'HK, CMD &  3344.07 &  2011.77-4550.69 &  935.24-5494.66 \\
HIP 71957 &  14 43 03.62 &  -05 39 29.53 &  $ 3.7671 \pm 0.002 $ &  $ 0.385 \pm 0.006 $ &   F2V & $ 19.16 \pm  0.14 $&  Li, CMD &  1483.42 &  1211.67-1716.6 &  893.61-1957.38 \\
HIP 73121 &  14 56 40.64 &  +21 04 16.72 &  $ 9.1024 \pm 0.0002 $ &  $ 1.035 \pm 0.009 $ &   K & $ 40.88 \pm  0.16 $&  Li UL & $\ge 4483.97 $ & $\ge 1093.34 $ & $\ge 544.57 $ \\
HIP 73183 &  14 57 25.90 &  +55 54 39.01 &  $ 9.9311 \pm 0.0006 $ &  $ 1.555 \pm 0.197 $ &  K5V & $ 37.04 \pm  0.4 $&  Li UL & $\ge 4236.44 $ & $\ge 747.08 $ & $\ge 152.29 $ \\
HIP 73252 &  14 58 15.67 &  +59 35 00.43 &  $ 9.5255 \pm 0.0002 $ &  $ 1.565 \pm 0.082 $ &  K5V & $ 25.75 \pm  0.01 $&  Li UL & $\ge 4205.13 $ & $\ge 703.31 $ & $\ge 129.34 $ \\
HIP 73787 &  15 04 53.98 &  -18 35 27.36 &  $ 9.006 \pm 0.0004 $ &  $ 1.21 \pm 0.008 $ &  K5+Vk & $ 33.0 \pm  0.64 $&  Li UL & $\ge 4358.02 $ & $\ge 917.25 $ & $\ge 366.04 $ \\
HIP 74926 &  15 18 39.55 &  -18 37 35.71 &  $ 9.8387 \pm 0.0003 $ &  $ 1.214 \pm 0.009 $ &   K4V & $ 27.35 \pm  0.1 $&  Li UL & $\ge 4350.77 $ & $\ge 907.11 $ & $\ge 356.02 $ \\
HIP 79203 &  16 09 55.27 &  -18 20 26.25 &  $ 6.3266 \pm 0.0003 $ &  $ 0.447 \pm 0.007 $ &   F3V & $ 49.7 \pm  0.17 $&  Li, CMD &  430.94 &  297.69-584.21 &  170.71-1095.7 \\
HIP 79593 &  16 14 20.73 &  -03 41 39.56 &  $ 2.0164 \pm 0.0055 $ &  $ 1.584 \pm 0.01 $ &   M0.5III & $ 49.0 \pm  1.33 $&  Li UL & $\ge 4144.8 $ & $\ge 619.13 $ & $\ge 62.8 $ \\
HIP 81655 &  16 40 48.88 &  +36 19 00.08 &  $ 10.5661 \pm 0.0003 $ &  $ 1.605 \pm 0.1 $ &   M2V & $ 21.47 \pm  0.02 $&  Li UL & $\ge 4191.94 $ & $\ge 684.91 $ & $\ge 114.09 $ \\
HIP 82260 &  16 48 28.31 &  -16 20 04.33 &  $ 7.4515 \pm 0.0002 $ &  $ 0.767 \pm 0.002 $ &   G8IV-V & $ 37.01 \pm  0.18 $&  Li UL & $\ge 6115.02 $ & $\ge 2980.3 $ & $\ge 1383.07 $ \\
HIP 83676 &  17 06 08.20 &  -06 10 02.07 &  $ 8.489 \pm 0.0003 $ &  $ 1.015 \pm 0.015 $ & K3V & $ 30.22 \pm  0.02 $&  Li UL & $\ge 4468.19 $ & $\ge 1071.18 $ & $\ge 513.21 $ \\
HIP 83929 &  17 09 19.61 &  +41 26 09.85 &  $ 10.4487 \pm 0.0004 $ &  $ 1.597 \pm 0.197 $ & K4V & $ 39.68 \pm  0.06 $&  Li UL, Gyro &  298.37 &  145.14-467.64 &  84.98-595.32 \\
HIP 87370 &  17 51 07.73 &  -22 55 15.61 &  $ 6.991 \pm 0.0001 $ &  $ 0.656 \pm 0.015 $ &   G3V & $ 44.27 \pm  0.19 $&  Li, R'HK &  8392.46 &  5631.85-11082.4 &  3331.8-12696.46 \\
HIP 88962 &  18 09 33.26 &  -12 02 19.98 &  $ 9.8511 \pm 0.0005 $ &  $ 1.372 \pm 0.004 $ &   K7V & $ 27.27 \pm  0.01 $&  Li UL & $\ge 4310.61 $ & $\ge 850.97 $ & $\ge 302.95 $ \\
HIP 89320 &  18 13 30.98 &  +81 04 32.78 &  $ 10.5168 \pm 0.0003 $ &  $ 1.586 \pm 0.197 $ &   K8 & $ 43.84 \pm  0.02 $&  Li UL & $\ge 4209.43 $ & $\ge 709.43 $ & $\ge 119.63 $ \\
HIP 89449 &  18 15 18.29 &  +18 29 59.98 &  $ 9.5074 \pm 0.0003 $ &  $ 1.323 \pm 0.019 $ &  K7V & $ 29.19 \pm  0.01 $&  Li UL & $\ge 4298.3 $ & $\ge 833.75 $ & $\ge 284.08 $ \\
HIP 90306 &  18 25 31.92 &  +38 21 12.67 &  $ 10.5152 \pm 0.0003 $ &  $ 1.442 \pm 0.02 $ &   K6V & $ 23.6 \pm  0.01 $&  Li UL & $\ge 4254.11 $ & $\ge 771.96 $ & $\ge 220.21 $ \\
HIP 93072 &  18 57 32.65 &  -19 02 46.86 &  $ 10.3501 \pm 0.0005 $ &  $ 1.437 \pm 0.004 $ &   K9Vk & $ 25.79 \pm  0.01 $&  Li UL & $\ge 4285.25 $ & $\ge 815.5 $ & $\ge 266.15 $ \\
HIP 98007 &  19 55 01.54 &  +04 03 43.66 &  $ 9.0905 \pm 0.0001 $ &  $ 0.978 \pm 0.003 $ & K3 & $ 34.94 \pm  0.21 $&  Li UL & $\ge 4510.78 $ & $\ge 1130.48 $ & $\ge 561.44 $ \\
HIP 99969 &  20 16 55.42 &  +06 55 18.29 &  $ 9.2486 \pm 0.0001 $ &  $ 1.14 \pm 0.015 $ & K4V & $ 42.09 \pm  0.11 $&  Li UL & $\ge 4388.19 $ & $\ge 959.42 $ & $\ge 405.74 $ \\
HIP 100133 &  20 18 45.79 &  -00 39 26.33 &  $ 10.2561 \pm 0.0003 $ &  $ 1.43 \pm 0.012 $ &   M0V & $ 27.31 \pm  0.27 $&  Li UL & $\ge 4281.86 $ & $\ge 810.76 $ & $\ge 261.29 $ \\
HIP 101852 &  20 38 21.10 &  -04 30 42.73 &  $ 7.6339 \pm 0.0003 $ &  $ 0.597 \pm 0.012 $ &  G0V & $ 43.59 \pm  0.09 $&  Li UL, Gyro, R'HK &  2050.45 &  1262.36-2646.59 &  580.98-3021.11 \\
HIP 102119 &  20 41 42.24 &  -22 19 20.51 &  $ 9.3352 \pm 0.0013 $ &  $ 1.121 \pm 0.009 $ &  K6Ve & $ 24.2 \pm  0.02 $&  Li UL & $\ge 4347.61 $ & $\ge 902.7 $ & $\ge 349.81 $ \\
HIP 104687 &  21 12 22.60 &  -15 00 00.36 &  $ 8.0059 \pm 0.0007 $ &  $ 0.654 \pm 0.041 $ &  G3V & $ 46.49 \pm  0.05 $&  Li, R'HK &  5895.89 &  2829.05-9452.36 &  855.62-12158.07 \\
HIP 105504 &  21 22 07.77 &  -10 30 47.89 &  $ 9.7045 \pm 0.0002 $ &  $ 1.255 \pm 0.015 $ &  K7 & $ 41.95 \pm  0.06 $&  Li UL & $\ge 4341.38 $ & $\ge 893.98 $ & $\ge 343.5 $ \\
HIP 107317 &  21 44 12.99 &  +06 38 29.28 &  $ 10.9909 \pm 0.0006 $ &  $ 1.522 \pm 0.021 $ & M3.0Ve & $ 20.57 \pm  0.02 $&  Li UL & $\ge 4199.9 $ & $\ge 696.16 $ & $\ge 131.93 $ \\
HIP 108036 &  21 53 17.77 &  -13 33 06.36 &  $ 4.9792 \pm 0.0009 $ &  $ 0.378 \pm 0.011 $ &   F2V & $ 26.78 \pm  0.08 $&  Li, CMD &  641.55 &  361.52-1031.51 &  166.74-1407.73 \\
HIP 109807 &  22 14 26.74 &  +02 42 24.12 &  $ 9.8969 \pm 0.0003 $ &  $ 1.252 \pm 0.005 $ &  K5V & $ 33.51 \pm  0.02 $&  Li UL & $\ge 4335.15 $ & $\ge 885.27 $ & $\ge 334.91 $ \\
HIP 110106 &  22 18 13.19 &  -03 10 19.84 &  $ 10.5093 \pm 0.0003 $ &  $ 1.399 \pm 0.02 $ &  K7V & $ 36.11 \pm  0.09 $&  Li UL & $\ge 4272.52 $ & $\ge 797.7 $ & $\ge 247.79 $ \\
HIP 110401 &  22 21 43.34 &  -17 09 40.90 &  $ 10.2378 \pm 0.0005 $ &  $ 1.208 \pm 0.013 $ &  K4V & $ 47.48 \pm  0.13 $&  Li UL & $\ge 4364.59 $ & $\ge 926.43 $ & $\ge 374.95 $ \\
HIP 110663 &  22 25 04.26 &  -13 59 53.92 &  $ 8.4368 \pm 0.0007 $ &  $ 0.77 \pm 0.005 $ &   G8/K0V & $ 43.29 \pm  0.05 $&  Li UL, Gyro, R'HK &  860.94 &  529.79-2168.83 &  392.82-2847.0 \\
HIP 111976 &  22 40 53.96 &  +64 54 26.04 &  $ 11.2184 \pm 0.0003 $ &  $ 1.548 \pm 0.1 $ &  K5 & $ 45.83 \pm  0.03 $&  Li UL & $\ge 4206.41 $ & $\ge 705.19 $ & $\ge 126.14 $ \\
HIP 113159 &  22 54 54.77 &  -22 22 08.23 &  $ 6.6536 \pm 0.0004 $ &  $ 0.469 \pm 0.005 $ &   F7V & $ 43.5 \pm  0.05 $&  Li, R'HK, CMD &  600.81 &  533.5-751.18 &  358.69-1086.46 \\
HIP 115280 &  23 20 53.26 &  +38 10 56.36 &  $ 5.6536 \pm 0.0004 $ &  $ 0.468 \pm 0.004 $ &   F5V & $ 42.05 \pm  0.06 $&  Li UL, R'HK, CMD &  1499.09 &  1124.39-1920.51 &  825.97-2277.7 \\
HIP 115411 &  23 22 43.20 &  -19 41 25.93 &  $ 8.5252 \pm 0.0003 $ &  $ 0.7 \pm 0.005 $ &   G6V & $ 42.88 \pm  0.04 $&  Li UL, R'HK &  9476.12 &  6862.42-11765.94 &  4396.19-12807.11 \\
HIP 116973 &  23 42 43.81 &  -19 52 57.32 &  $ 9.719 \pm 0.0004 $ &  $ 1.085 \pm 0.008 $ &   K3V & $ 41.93 \pm  0.06 $&  Li UL & $\ge 4468.03 $ & $\ge 1071.06 $ & $\ge 525.19 $ \\
HIP 117159 &  23 45 09.93 &  +29 33 42.72 &  $ 8.147 \pm 0.0002 $ &  $ 0.859 \pm 0.006 $ & K0 & $ 29.75 \pm  0.13 $&  Li UL & $\ge 4766.76 $ & $\ge 1475.94 $ & $\ge 801.54 $ \\
HIP 117235 &  23 46 17.95 &  +10 16 41.79 &  $ 11.1245 \pm 0.0002 $ &  $ 1.376 \pm 0.133 $ &   K7V & $ 47.8 \pm  0.08 $&  Li UL & $\ge 4278.54 $ & $\ge 806.1 $ & $\ge 233.62 $ \\
HIP 117410 &  23 48 25.69 &  -12 59 14.85 &  $ 9.3062 \pm 0.0019 $ &  $ 1.244 \pm 0.014 $ &   K5Vke & $ 28.57 \pm  0.22 $&  Li UL, Gyro &  635.14 &  394.14-1127.47 &  309.22-12215.08 \\
HIP 117795 &  23 53 19.84 &  +59 56 42.41 &  $ 10.0403 \pm 0.0002 $ &  $ 1.21 \pm 0.015 $ &  K8 & $ 26.72 \pm  0.01 $&  Li UL & $\ge 4346.62 $ & $\ge 901.3 $ & $\ge 350.04 $ \\
HIP 118086 &  23 57 14.38 &  -16 30 27.36 &  $ 10.348 \pm 0.0003 $ &  $ 1.3 \pm 0.015 $ &  K7V & $ 36.7 \pm  0.03 $&  Li UL & $\ge 4291.23 $ & $\ge 823.87 $ & $\ge 273.54 $ \\
HIP 118310 &  23 59 47.78 &  +06 39 50.95 &  $ 8.4993 \pm 0.0002 $ &  $ 1.187 \pm 0.052 $ &   K4V & $ 23.03 \pm  0.01 $&  Li UL & $\ge 4399.71 $ & $\ge 975.53 $ & $\ge 416.39 $
\label{tab:general}
\enddata

\end{deluxetable}
\end{longrotatetable}}

\clearpage

\subsection{Stars within the Lithium Dip}
There are several stars in our sample that are near the lithium dip, which is a range of $B-V$ ($\sim0.4-0.55$) where lithium depletes rapidly around the age of the Hyades. This rapid evolution means the feature is not well modeled \rev{(e.g. \citealt{2020ApJ...898...27S})}, we discuss the consistency of the ages for stars near the lithium dip below.

Many stars with $0.4<B-V<0.55$ have age posteriors from different tracers that are generally consistent, including the lithium posterior. For example, HIP 42343 has ages from lithium ($4964^{+4688}_{-2766}$ Myr), $R^{'}_{HK}$ ($4793^{+3345}_{-2103}$ Myr), and CMD position ($2105^{+1309}_{-1071}$ Myr) that give a combined posterior of $2791^{+933}_{-895}$ Myr. \rev{Similarly, the posteriors mainly agree for HIP 17118, HIP 12837, HIP 71899, and HIP 27021.}

HIP 47013 shows significant offset between posteriors. The lithium non-detection gives a $2\sigma$ lower limit of 1528 Myr, which is consistent with the $R^{'}_{HK}$ posterior of $4997^{+3340}_{-2175}$ Myr. However, the CMD posterior gives a significantly younger age of $1181^{+247}_{-203}$ Myr. \rev{This may be due to the small number of stars within the lithium dip in the $\sim500-1000$ Myr clusters used to calibrate \texttt{BAFFLES}. Because of this limitation, we recommend not relying on ages from lithium alone for stars within the lithium dip.} We see similar discrepancies between the posteriors for \rev{HIP 6943, HIP 113159, HIP 10050, HIP 71243, HIP 3293, HIP 6776, HIP 44162, HIP 2843, HIP 115280, and HIP 21066.}

\section{Discussion} \label{sec:discus}

From our full sample of 166 stars, \rev{102} age posteriors are based on lithium upper limits alone, while \rev{55} stars have an age posterior based on more than one tracer. \rev{31} of our targets ($\sim19\%$) have a median age younger than 1 Gyr, which is more than expected for a uniform star formation rate. We discuss the full age distribution of our sample below.

\subsection{Star Formation History of the Solar Neighborhood}

 \cite{1990Ap&SS.163..229R} and \cite{2023MNRAS.522.1643C} found that the local star formation rate has been near constant for the last 10 to 12 Gyr based on the luminosity distribution of white dwarfs. Other studies found slightly different histories: a maximum star formation rate ($\sim2$ times the typical star formation rate at other times) $\sim3$ Gyr ago \citep{2006A&A...459..783C} or two star forming events separated by $\sim5$ Gyr \citep{2018Natur.559..585N}.  We assume a uniform star formation history, which also matches the priors used in \texttt{BAFFLES} \citep{2020ApJ...898...27S} and \texttt{gyro-interp} \citep{2023ApJ...947L...3B}.

\subsection{Population Level Ages}

\rev{We investigate how significant it is that our sample contains more stars than expected with median ages younger than 1 Gyr. Given the small size of our sample ($<200$ stars), and the biased nature in which it was selected, we are not attempting here to constrain the local star formation history. Rather, we are examining whether the total age distribution we observe is broadly consistent with a uniform star formation rate in the solar neighborhood.} We construct a population-level age distribution by summing (and then normalizing) the age posteriors for the entire sample. This is equivalent to marginalizing over individual stars; in the limit where each age posterior is a delta-function, this method produces a histogram of stellar ages. With a large enough volume-limited sample, we would expect this population-level distribution to match the star formation of the solar neighborhood. \rev{If the star formation rate has been uniform from 0 to 13 Gyr, we expect 1/13 ($7.7\%$) of the summed age distribution  to be younger than 1 Gyr.}

\begin{figure*}[ht!]
\centering
 \includegraphics[scale = 0.35]{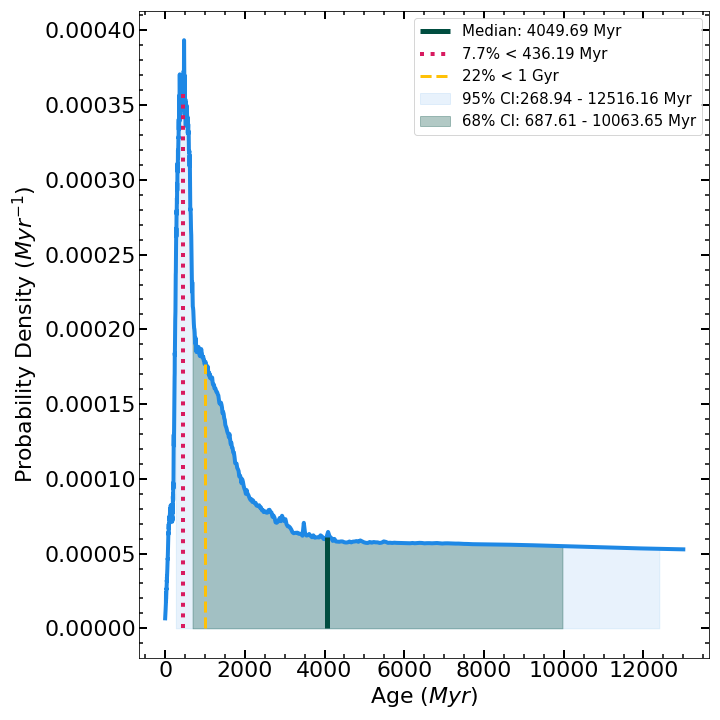}
 \includegraphics[scale = 0.35]{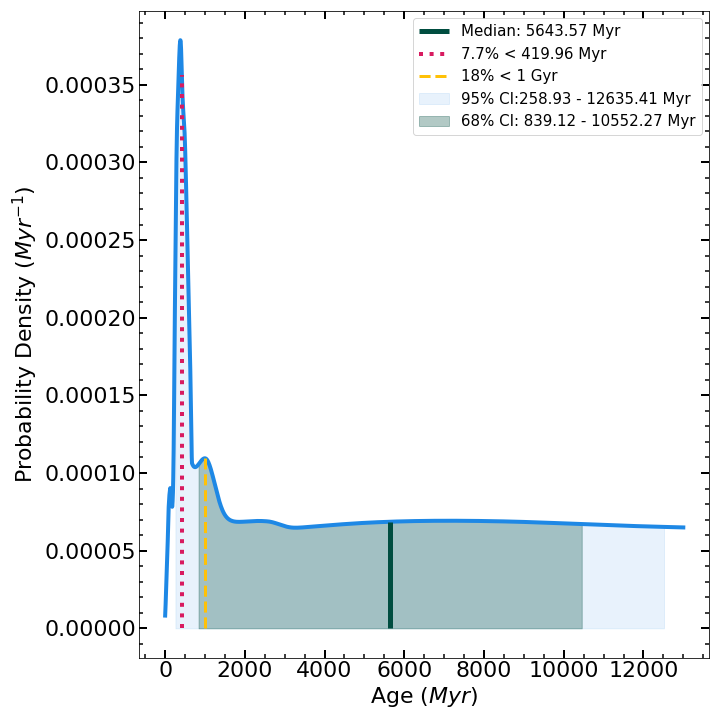}
 \caption{{\bf Left}: Sum of all the age posteriors for the full sample. The distribution of ages is peaked at young ages, but this is partially due to the inclusion of early-type stars with shorter main sequence lifetimes. {\bf Right}: Sum of age posteriors for only stars with $B-V > 0.595$, where there is still a peak at young ages, if less pronounced than for the full sample.}
 \label{global_post}
\end{figure*}

The left panel of Figure \ref{global_post} shows the population-level age distribution for our 166-star sample, which peaks at young ages ($\lesssim$2 Gyr). Part of this peak is due to early-type stars, which have short main sequence lifetimes and thus do not trace the full 13 Gyr star formation history. These stars make up $\sim19\%$ of the sample which is an over-representation relative to the initial mass function \citep{1955ApJ...121..161S, 2003PASP..115..763C, 2003ApJ...598.1076K}. This is likely due to the multiplicity fractions ($\sim50-70\%$) of these stars \citep{2023ASPC..534..275O} making them more likely to be accelerating. Removing stars with $B-V < 0.595$ (corresponding to a main sequence mass of $1.06 \ M_{\odot}$) \rev{leaves us with 135 stars and} produces the posterior in the right panel of Figure \ref{global_post}. The peak at young ages is still present, but less pronounced.

This general behavior is consistent with the types of age posteriors we produce here. For any given late-type star the posterior is either a well-constrained Gaussian-like peak or a lower age limit (corresponding to a lithium upper limit) with only the youngest ages ruled out. While $R^{'}_{HK}$ produces Gaussian-like posteriors at both young and old ages for stars earlier than K2, lithium and gyrochronology generally produce peaked posteriors for later spectral types if they are younger than $\sim$1 Gyr. Stars older than $\sim$1 Gyr mostly do not have detectable lithium (depending on spectral type), and either have rotation periods longer than a typical  $TESS$ sector or are less spotted. \rev{For the reduced sample of 135 stars, we expect $\sim10$ ($7.7\%$) stars to be younger than 1 Gyr. Following the Poisson distribution, we calculate a $65\%$ chance of between 8 and 13 stars being younger than 1 Gyr and a $96\%$ chance of between 5 and 17 stars being younger than 1 Gyr. Testing this prediction is complicated by the fact that we have somewhat broad age posteriors, rather than delta-function age measurements. Nevertheless, we can estimate the number of young stars in our sample by summing the probability in the posteriors for all stars. We calculate that $18\%$ of the age posterior for the reduced sample falls below 1 Gyr. This corresponds to $\sim24$ stars, which is significantly higher than expected. In fact, for a Poisson distribution with an expectation value of 10.4 stars, we only expect a value of 24 or higher $0.01\%$ of the time.}


\subsection{Testing Individual Age Tracers}

\begin{figure*}[ht!]
\centering
 \includegraphics[scale = 0.35]{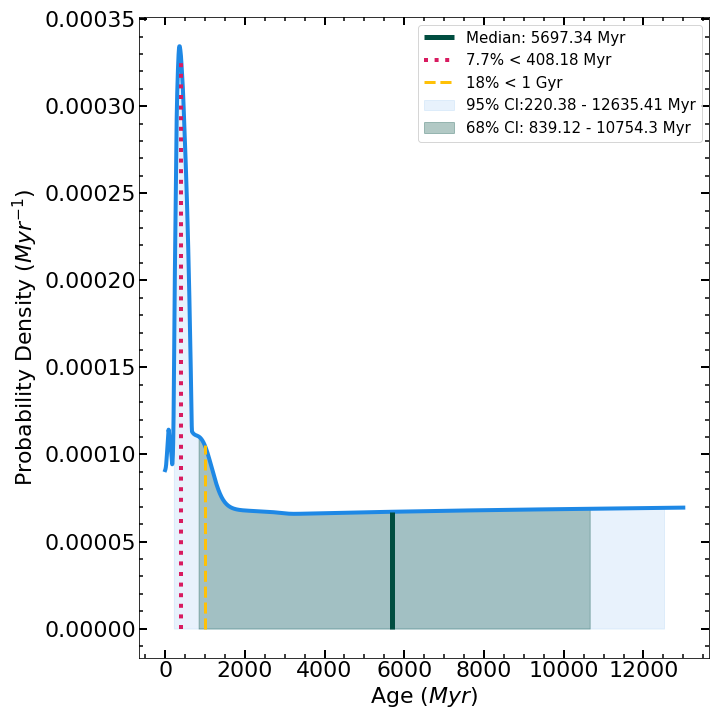}
 \includegraphics[scale = 0.35]
 {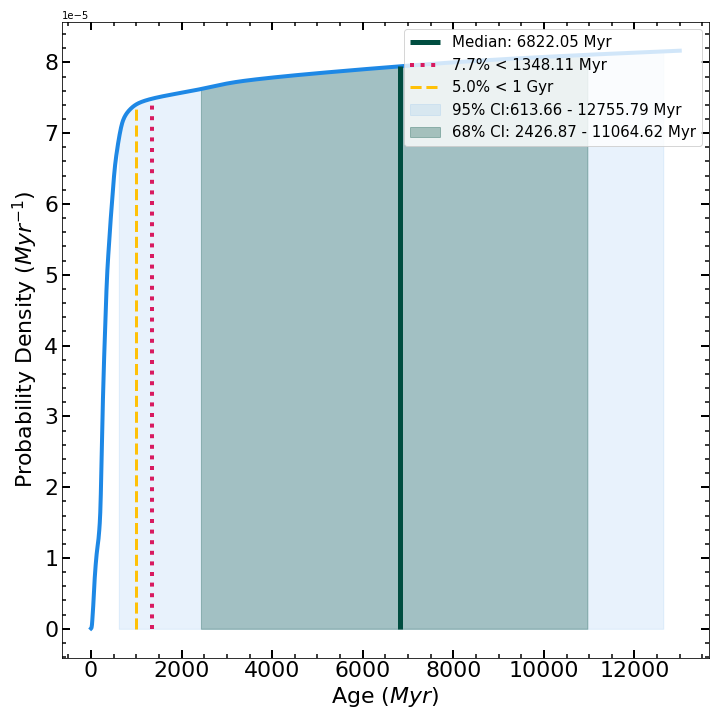}

 \caption{{\bf Left}: The combined age posterior from lithium detections and/or gyrochronology for the late-type stars in the sample. The peak at young ages is consistent with stars that are $<1$ Gyr being most likely to have detectable lithium or clear rotation signals. {\bf Right}: For stars that only have lithium upper limits, the age posterior is mostly flat, following the prior, with only the youngest ages excluded.}
 \label{global_post_late}
\end{figure*}

Figure \ref{global_post_late} further divides the late spectral type stars. The left panel only includes posteriors from lithium detections and/or rotation period detections. The right panel only includes posteriors derived from lithium upper limits. The posterior derived exclusively from lithium upper limits, as expected, is much less peaked and is essentially following the prior. The posterior derived from lithium detections and rotation rates peaks at ages younger than $\sim$1 Gyr. This peaked posterior is expected for age tracers that only return peaked posteriors for the youngest stars.

\rev{Nevertheless, this peak seems to include more probability at ages less than 1 Gyr than we would expect from a uniform star formation rate operating between 0 and 13 Gyr. We investigate the relative contributions of lithium and gyrochronology to this result in Figure \ref{LiGyroPost}. We consider ages from lithium, calcium, and gyrochronology separately. }

\rev{{\bf Lithium}: We find that the combined age posterior for lithium-only derived posteriors (114 posteriors) is consistent with a uniform star formation rate. From the Poisson distribution and an assumed uniform star formation rate history, there is an $82\%$ chance of  between $5\leq N_{stars} \leq 11$ being younger than 1 Gyr. We calculate $8.0\%$  of the posterior ($\sim9$ stars) falls below 1 Gyr, consistent with this prediction.} 

\rev{{\bf Calcium}: We derive age posteriors from $R^{'}_{HK}$ for 18 late-type stars. There are 7 additional late-type stars that have appropriate $B-V$ values to derive posteriors from $R^{'}_{HK}$ but have $R^{'}_{HK} \leq -5$, which is outside the calibration range of \texttt{BAFFLES}, but consistent with old ages. From the Poisson distribution, there is a $\sim60\%$ chance of measuring between 0 and 1 stars younger than 1 Gyr from a sample of 18 stars. $1\%$ of our 18 star $R^{'}_{HK}$ posterior falls below 1 Gyr, which corresponds to finding $0.18$ stars being less than 1 Gyr, which is again consistent with the prediction.} 

\rev{{\bf Gyrochronology}: We have  $TESS$ light curves for 129 stars, 106 of which have $B-V > 0.595$. We measure rotation periods for 20 of the late-type stars. From the gyrochronology population-level age distribution, $92\%$ of the probability is less than 1 Gyr, which corresponds to $\sim17\%$ of the total late-type gyrochronology sample of 106 stars, or about 18 young stars. Our ages from gyrochronology do not include upper limits because flat  $TESS$ light curves could indicate a long rotation period (longer than a typical  $TESS$ sector) or just a star without significant star spots. Overall, we would expect $\sim8$ stars out of 106 to have an age of 1 Gyr or younger. From the Poisson distribution with an expectation value of 8, there is a $\sim4\%$ chance of finding 13 or more stars below 1 Gyr, and a $\sim0.08\%$ chance of finding 18 or more. Thus, while the number of young stars identified from lithium and calcium alone are consistent with the predictions of a uniform star formation rate,  there appears to be $\sim5-10$ excess young stars identified by rotation. 

This result, that our combined age posterior suggests 18.0 stars are younger than 1 Gyr, does not change significantly even if we increase the size of the rotational period errors when computing age posteriors. Our period errors are between 0.11 days and 2.02 days, with a median of 0.55 days. Setting a minimum error bar to 1 day, 2 days, or 3 days still results in an estimate that 18.4, 17.8, and 16.8 stars are younger than 1 Gyr, respectively. As a result, we conclude that this excess of probability at young ages cannot be attributed solely to a systematic underestimate of the rotation period uncertainties.

\begin{figure*}[ht!]
\centering
 \includegraphics[scale = 0.35]{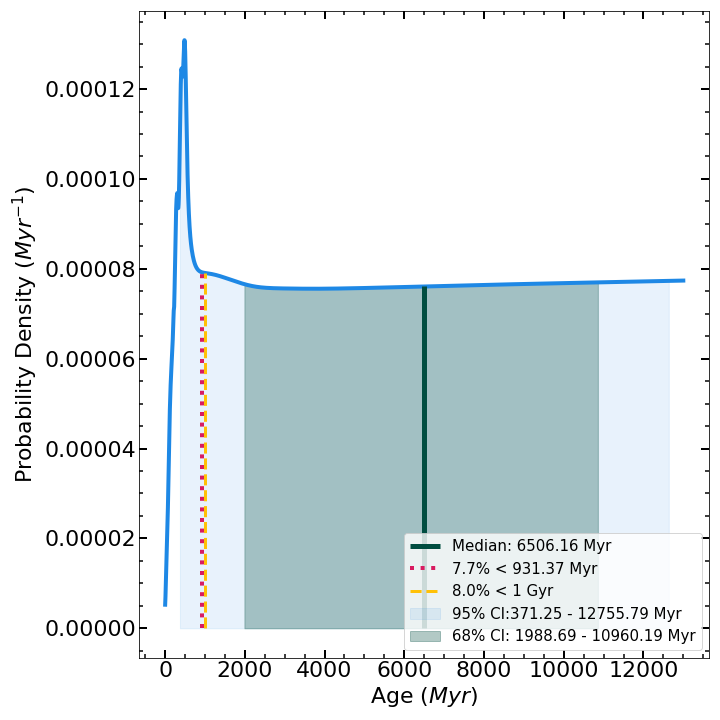}
  \includegraphics[scale = 0.35]{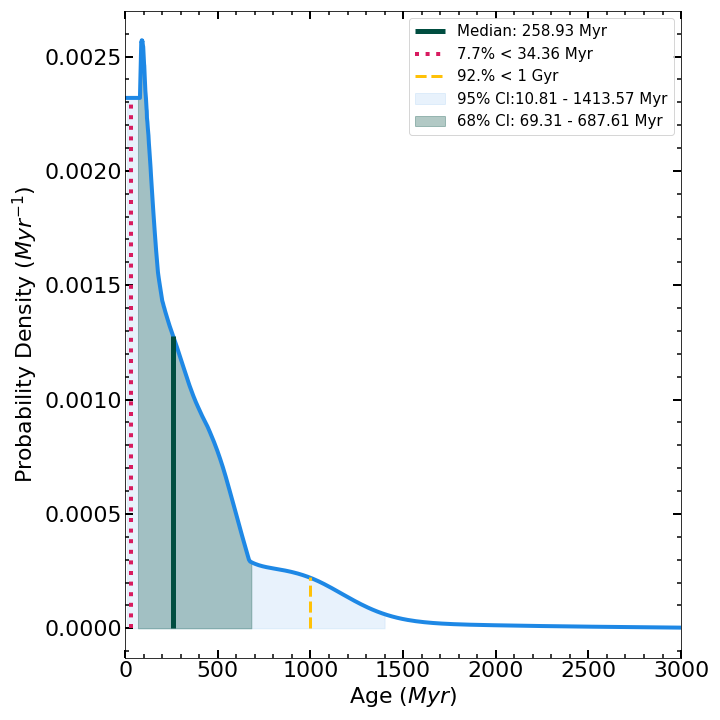}

 \caption{{\bf Left:} Ages derived from lithium alone yield an age distribution consistent with a constant star formation rate. $8\%$ of the combined distribution falls below 1 Gyr, which is consistent with the expected $7.7\%$. {\bf Right:} Ages derived from gyrochronology alone yield a significantly younger age distribution than expected. For the \rev{20} late-type stars with rotation periods $\lesssim15$ days, $92\%$ of the combined distribution of ages falls below 1 Gyr, which corresponds to \rev{$\sim17\%$} of the full gyrochronology sample.}
 \label{LiGyroPost}
\end{figure*}

There are several possible explanations for the higher-than-expected fraction of young stars in the gyrochronological sample. Since we select stars that are accelerating, there is a greater chance of selecting a short-period stellar binary \rev{(e.g. \citealt{2019AJ....158..226D})}. The rotational evolution of stars in close binaries could be influenced by tides resulting in shorter rotation periods than expected from age alone. Additionally,  $TESS$ pixels may be contaminated by multiple stars (either physical binaries or crowded fields). While we screen for contaminated pixels, additional components may be present in some light curves. Alternatively, this effect could be physical: if binaries become unbound over time, a sample of accelerating stars (which is more likely to contain binaries than the field) would be younger than a volume-limited sample.

While we screened our initial accelerating star sample for binaries from WDS, SB9, and archival imaging, such a search is not complete to all binary configurations. Indeed, by constructing our sample from accelerating stars not in these catalogs we expect undiscovered binaries to be overrepresented in our sample compared to an unbiased volume-limited sample. For example, three eclipsing binaries (HIP 16247, HIP 45731, and HIP 49577) discussed in Section \ref{sec:gyro} were discovered to be physical binaries only after our survey began. We are currently carrying out radial velocity monitoring of the full sample with the aim of identifying additional short-period stellar binaries, and will report the results in a future paper.}

\rev{\subsection{Companions}
Since the beginning of this survey in 2020, two brown dwarf companions have been discovered around two of the accelerating stars discussed in this work. \cite{2022ApJ...934L..18K} found a $\sim28  \ M_\textrm{Jup}$ companion to HIP 21152, a likely member of the Hyades Cluster ($\sim750$ Myr; \rev{\citealt{2018ApJ...856...23G})}. We measure a slightly older age $1116^{+406}_{-351}$ Myr based on a lithium upper limit and the star's CMD position. This star, however, is in the lithium dip as discussed in section \ref{sec:total_ages}. The CMD age $855^{+1348}_{-412}$, is consistent with the age of the Hyades. The lithium lower limit suggest an age a factor of two older than the Hyades ($>$1588 Myr), though this may be an overestimate due to how \texttt{BAFFLES} models the lithium dip.

\cite{2022AJ....164..152S} found a $\sim31 \  M_\textrm{Jup}$ companion to HIP 5319. With $B-V <0.45$, HIP 5319 is not a candidate for deriving an age from $R^{'}_{HK}$ using \texttt{BAFFLES}. \cite{2011A&A...530A.138C} report an age between $1.23^{+0.54}_{-0.6}$ Gyr by fitting isochrones from \cite{2008A&A...484..815B, 2009A&A...508..355B}. We measure an age of $843^{+387}_{-348}$ Myr from a lithium upper limit and the star's CMD position, which is consistent with the previously reported age. These detections underscore how effective astrometric selection can be at identifying stars hosting substellar companions.}

\clearpage

\section{Conclusions} \label{sec:con}

We identified a sample of 166 accelerating stars that may host substellar companions and do not have archival spectra. From APO/ARCES spectra of the full 166 star sample we measured lithium equivalent widths (or upper limits) and $R^{'}_{HK}$. We acquired TESS light curves for 129 stars and measured rotation rates for 23 out of the 129 stars. For the earlier spectral types, we also constrain the ages from color-magnitude diagram positions. Combining these methods we \rev{calculated median ages and $68\%$ and $95\%$ confidence intervals} for the full sample. We are continuing to collect APO/ARCES spectra of the full sample over at least 3 epochs per star to identify stellar binaries (which could be the source of the astrometric acceleration).

We find approximately twice as many young stars from gyrochronology than expected from a uniform star formation rate. We identify two possible explanations for this result. First, there could be unidentified short-period binaries in the sample: either short period binaries that spin each other up, or visual binaries that overlap in \textit{TESS} images. Second, if binaries tend to dissolve over time, then young stars may be more likely to be part of binaries (and possess astrometric accelerations) than older stars. We expect our multi-epoch radial velocity monitoring to identify additional binaries in the sample, and we will report on the results of this monitoring in a future paper.

Looking ahead, \textit{Gaia} DR4 is expected to be released in mid-2026. Unlike DR3, which only contains a single position and velocity measurement, DR4 will contain individual scans over several years. These individual scans will better differentiate between short-period stellar binaries and long-period substellar companions. Future \textit{TESS} data releases will provide light curves for more accelerating stars, allowing for additional gyrochronological age measurements. Additional light curves for accelerating stars that already have light curves will allow for monitoring of spot evolution over the course of a stellar cycle.

The youngest of the accelerating stars in our sample are high-priority targets for direct imaging since these substellar companions will be bright enough to be imaged. In addition to continued radial velocity monitoring, we have begun a direct imaging campaign targeting the youngest stars. Results from both studies will be presented in future work. Substellar companions found in this sample will become key benchmarks for substellar evolution models.

\begin{acknowledgments}
\section{Acknowledgments}
This research has made use of the Washington Double Star Catalog maintained at the U.S. Naval Observatory.  AEP, ELN, and JK are supported by NASA grant 80NSSC23K1008. AEP received funding from the William Webber Voyager Fellowship. This research has made use of the SIMBAD database,
operated at CDS, Strasbourg, France.

\rev{Based on observations obtained with the Apache Point Observatory 3.5-meter telescope, which is owned and operated by the Astrophysical Research Consortium. This work has made use of data from the European Space Agency (ESA) mission $Gaia$
(https://www.cosmos.esa.int/gaia), processed by the Gaia Data Processing and Analysis Consortium (DPAC, https://www.cosmos.esa.int/web/gaia/dpac/consortium). Funding for the DPAC has been
provided by national institutions, in particular the institutions participating in the $Gaia$ Multilateral Agreement.  This work includes data
collected by the $TESS$ mission. Funding for the $TESS$ mission is provided by the NASA's Science Mission Directorate.}

\rev{The $TESS$ data products used in this work are available in the MAST archive, specifically the ``$TESS$ Light Curves - All Sectors" repository \citep{10.17909/t9-nmc8-f686}, the  ``$TESS$ ``Fast" Light Curves - All Sectors" repository \citep{https://doi.org/10.17909/t9-st5g-3177}, the  ``$TESS$ Target Pixel Files - All Sectors" repository \citep{https://doi.org/10.17909/t9-yk4w-zc73}, and the ``$TESS$ ``FAST" Target Pixel Files - All Sectors" repository \citep{https://doi.org/10.17909/t9-tcn7-7g94}}

Thank you to Dr. Wladamir Lyra for his guidance and sharing his knowledge of age diagnostics.

\rev{We would like to thank the reviewer for their thoughtful and helpful suggestions, which have significantly improved this paper.}
%

\rev{\facilities{ARC 3.5m, $TESS$, $Gaia$, $Hipparcos$}}

\rev{\software{\texttt{Lightkurve} \citep{2018ascl.soft12013L}, \texttt{astropy} \citep{astropy:2013, astropy:2018, astropy:2022}, \texttt{SciPy} \citep{2020SciPy-NMeth}, \texttt{NumPy} \citep{harris2020array}, \texttt{emcee }\citep{2013PASP..125..306F}, \texttt{BAFFLES} \citep{2020ApJ...898...27S}, \texttt{gyro-interp} \citep{2023ApJ...947L...3B}}}



\end{acknowledgments}
\bibliography{main}{}
\bibliographystyle{aasjournal}


 \appendix 
\begin{figure*}[ht!]
\centering
 \includegraphics[scale = 0.4]{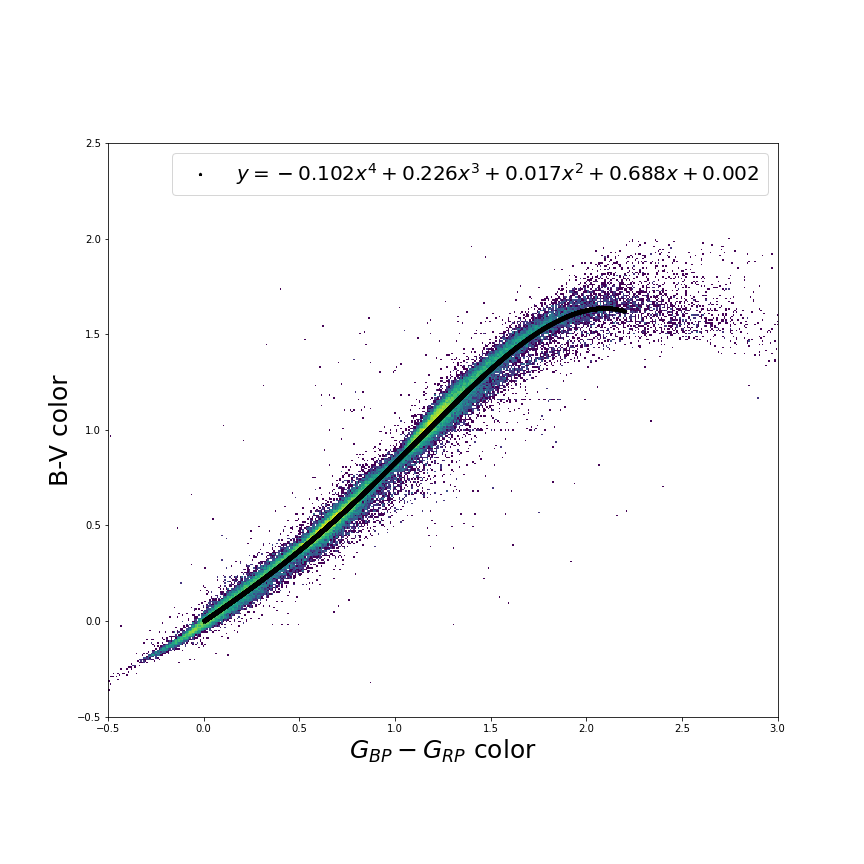}
 \caption{We transform \textit{Gaia} $G_{BP} - G_{RP}$ color to Johnson $B-V$ using a polynomial fit to a large sample of $Gaia-Hipparcos$ stars \citep{1997A&A...323L..61V, 2023A&A...674A...1G}. We use our derived transformation when \rev{calculating median ages $68\%$ and $95\%$ confidence intervals} from \texttt{BAFFLES} \citep{2020ApJ...898...27S} for stars without precise $B-V$ measurements.}
 
 \label{bmv}
\end{figure*}

{\setlength{\tabcolsep}{2pt}\begin{longtable}{|c|c|c|c|c|c|c|c|c|}

\caption{We present our lithium equivalent width measurements or upper limits for each star and its corresponding \texttt{BAFFLES} \rev{median age and confidence intervals} where appropriate. For stars with $B-V$ outside the \texttt{BAFFLES} range, we do not derive age posteriors.} \\

\hline
Target & B-V & Li EW & Age & $68\%$ CI& $95\%$ CI & $99.7\%$ LL & $95\%$ LL & $68\%$ LL\\
& & (m\r{A}) & (Myr) & (Myr) & (Myr) & (Myr) & (Myr) & (Myr)\\
\hline
\endfirsthead 
\hline
Target & B-V & Li EW & Age & $68\%$ CI& $95\%$ CI & $99.7\%$ LL & $95\%$ LL & $68\%$ LL\\
& & (m\r{A}) & (Myr) & (Myr) & (Myr) & (Myr) & (Myr) & (Myr)\\
\hline
\endhead
\hline
\endfoot
HIP 143 & $ 1.033 \pm 0.019 $ & $\le 7.64 $ &-- &-- &-- & $\ge 484.61 $ & $\ge 1034.76 $ & $\ge 4442.08 $ \\
HIP 669 & $ 0.624 \pm 0.015 $ & $ 31.45 \pm 1.89 $ & 4612.67 & 2164.81-9235.42 & 859.53-12375.61 &-- &-- &-- \\
HIP 1068 & $ 1.315 \pm 0.1 $ & $\le 68.37 $ &-- &-- &-- & $\ge 138.08 $ & $\ge 720.14 $ & $\ge 4217.04 $ \\
HIP 1412 & $ 1.414 \pm 0.005 $ & $\le 13.65 $ &-- &-- &-- & $\ge 254.02 $ & $\ge 803.55 $ & $\ge 4276.7 $ \\
HIP 1539 & $ 1.375 \pm 0.005 $ & $\le 12.17 $ &-- &-- &-- & $\ge 280.46 $ & $\ge 828.91 $ & $\ge 4294.84 $ \\
HIP 1771 & $ 1.489 \pm 0.1 $ & $\le 15.03 $ &-- &-- &-- & $\ge 164.0 $ & $\ge 744.83 $ & $\ge 4234.75 $ \\
HIP 2426 & $ 1.359 \pm 0.02 $ & $\le 10.93 $ &-- &-- &-- & $\ge 285.77 $ & $\ge 834.64 $ & $\ge 4298.94 $ \\
HIP 2843 & $ 0.466 \pm 0.007 $ & $\le 7.97 $ &-- &-- &-- & $\ge 716.16 $ & $\ge 1300.72 $ & $\ge 4632.99 $ \\
HIP 3008 & $ 1.424 \pm 0.02 $ & $\le 14.98 $ &-- &-- &-- & $\ge 246.95 $ & $\ge 797.04 $ & $\ge 4272.04 $ \\
HIP 3293 & $ 0.452 \pm 0.009 $ & $ 13.22 \pm 4.98 $ & 726.52 & 573.26-1064.21 & 224.61-1842.72 &-- &-- &-- \\
HIP 3633 & $ 0.814 \pm 0.061 $ & $\le 11.11 $ &-- &-- &-- & $\ge 571.54 $ & $\ge 1376.96 $ & $\ge 4795.91 $ \\
HIP 3724 & $ 1.4 \pm 0.02 $ & $\le 16.09 $ &-- &-- &-- & $\ge 255.04 $ & $\ge 804.65 $ & $\ge 4277.49 $ \\
HIP 4024 & $ 1.069 \pm 0.009 $ & $\le 4.33 $ &-- &-- &-- & $\ge 530.93 $ & $\ge 1075.56 $ & $\ge 4471.25 $ \\
HIP 5319 & $ 0.398 \pm 0.01 $ & $\le 35.97 $ &-- &-- &-- & $\ge 247.18 $ & $\ge 1301.88 $ & $\ge 4825.13 $ \\
HIP 5763 & $ 1.22 \pm 0.015 $ & $\le 8.14 $ &-- &-- &-- & $\ge 352.3 $ & $\ge 903.42 $ & $\ge 4348.13 $ \\
HIP 6776 & $ 0.453 \pm 0.007 $ & $\le 16.62 $ &-- &-- &-- & $\ge 573.96 $ & $\ge 1158.02 $ & $\ge 4530.53 $ \\
HIP 6796 & $ 1.157 \pm 0.074 $ & $\le 5.35 $ &-- &-- &-- & $\ge 409.71 $ & $\ge 976.89 $ & $\ge 4400.69 $ \\
HIP 6833 & $ 0.656 \pm 0.002 $ & $ 13.12 \pm 2.19 $ & 7124.14 & 4221.16-10604.78 & 1831.07-12609.77 &-- &-- &-- \\
HIP 6943 & $ 0.453 \pm 0.011 $ & $ 45.34 \pm 4.14 $ & 361.64 & 185.75-490.47 & 50.47-662.73 &-- &-- &-- \\
HIP 7287 & $ 1.2 \pm 0.02 $ & $\le 12.3 $ &-- &-- &-- & $\ge 332.72 $ & $\ge 884.25 $ & $\ge 4334.42 $ \\
HIP 8653 & $ 0.644 \pm 0.019 $ & $ 47.98 \pm 2.44 $ & 2101.63 & 818.22-7067.98 & 347.5-11896.89 &-- &-- &-- \\
HIP 9067 & $ 1.47 \pm 0.1 $ & $\le 6.96 $ &-- &-- &-- & $\ge 231.94 $ & $\ge 801.85 $ & $\ge 4275.7 $ \\
HIP 9989 & $ 1.248 \pm 0.002 $ & $\le 7.92 $ &-- &-- &-- & $\ge 341.68 $ & $\ge 892.08 $ & $\ge 4340.02 $ \\
HIP 10050 & $ 0.471 \pm 0.001 $ & $ 14.76 \pm 2.12 $ & 641.25 & 536.76-745.32 & 166.15-976.21 &-- &-- &-- \\
HIP 10532 & $ 0.88 \pm 0.033 $ & $\le 4.38 $ &-- &-- &-- & $\ge 647.91 $ & $\ge 1305.19 $ & $\ge 4648.61 $ \\
HIP 11090 & $ 0.289 \pm 0.006 $ & $\le 37.8 $ & -- & -- &-- & -- & -- & -- \\
HIP 11815 & $ 1.357 \pm 0.02 $ & $\le 21.0 $ &-- &-- &-- & $\ge 253.41 $ & $\ge 803.47 $ & $\ge 4276.65 $ \\
HIP 12837 & $ 0.517 \pm 0.008 $ & $ 82.5 \pm 4.08 $ & 2947.74 & 458.78-8919.31 & 118.81-12361.88 &-- &-- &-- \\
HIP 13681 & $ 1.062 \pm 0.015 $ & $\le 5.91 $ &-- &-- &-- & $\ge 509.53 $ & $\ge 1054.8 $ & $\ge 4456.4 $ \\
HIP 13782 & $ 0.238 \pm 0.003 $ & $\le 25.62 $ & -- & -- &-- & -- & -- & -- \\
HIP 15211 & $ 0.936 \pm 0.007 $ & $\le 3.61 $ &-- &-- &-- & $\ge 609.82 $ & $\ge 1187.49 $ & $\ge 4551.82 $ \\
HIP 16247 & $ 1.02 \pm 0.037 $ & $\le 9.35 $ &-- &-- &-- & $\ge 461.81 $ & $\ge 1015.84 $ & $\ge 4428.57 $ \\
HIP 16709 & $ 1.225 \pm 0.02 $ & $\le 11.19 $ &-- &-- &-- & $\ge 328.75 $ & $\ge 879.86 $ & $\ge 4331.28 $ \\
HIP 17118 & $ 0.478 \pm 0.023 $ & $ 40.49 \pm 10.51 $ & 3616.84 & 512.59-9572.67 & 172.65-12482.35 &-- &-- &-- \\
HIP 17213 & $ 1.1 \pm 0.015 $ & $\le 6.23 $ &-- &-- &-- & $\ge 453.14 $ & $\ge 1004.51 $ & $\ge 4420.43 $ \\
HIP 18527 & $ 0.899 \pm 0.021 $ & $\le 4.12 $ &-- &-- &-- & $\ge 633.73 $ & $\ge 1222.56 $ & $\ge 4577.49 $ \\
HIP 19739 & $ 1.424 \pm 0.02 $ & $\le 11.89 $ &-- &-- &-- & $\ge 257.67 $ & $\ge 807.4 $ & $\ge 4279.45 $ \\
HIP 19912 & $ 1.148 \pm 0.002 $ & $\le 4.65 $ &-- &-- &-- & $\ge 433.87 $ & $\ge 987.64 $ & $\ge 4408.38 $ \\
HIP 20419 & $ 1.183 \pm 0.005 $ & $\le 13.39 $ &-- &-- &-- & $\ge 334.12 $ & $\ge 885.61 $ & $\ge 4335.39 $ \\
HIP 20648 & $ 0.049 \pm 0.007 $ & $\le 3.04 $ & -- & -- &-- & -- & -- & -- \\
HIP 21066 & $ 0.472 \pm 0.013 $ & $\le 31.26 $ &-- &-- &-- & $\ge 391.93 $ & $\ge 1076.4 $ & $\ge 4478.33 $ \\
HIP 21152 & $ 0.42 \pm 0.014 $ & $\le 13.2 $ &-- &-- &-- & $\ge 566.24 $ & $\ge 1588.14 $ & $\ge 5230.07 $ \\
HIP 22361 & $ 0.283 \pm 0.004 $ & $\le 26.75 $ & -- & -- &-- & -- & -- & -- \\
HIP 23208 & $ 1.18 \pm 0.006 $ & $\le 7.62 $ &-- &-- &-- & $\ge 376.02 $ & $\ge 927.89 $ & $\ge 4365.63 $ \\
HIP 24017 & $ 0.448 \pm 0.004 $ & $\le 9.48 $ &-- &-- &-- & $\ge 669.48 $ & $\ge 1254.06 $ & $\ge 4599.58 $ \\
HIP 24177 & $ 1.461 \pm 0.005 $ & $\le 33.19 $ &-- &-- &-- & $\ge 159.14 $ & $\ge 719.62 $ & $\ge 4216.67 $ \\
HIP 24292 & $ 1.23 \pm 0.015 $ & $\le 6.29 $ &-- &-- &-- & $\ge 365.54 $ & $\ge 916.6 $ & $\ge 4357.55 $ \\
HIP 24457 & $ 1.118 \pm 0.005 $ & $\le 3.42 $ &-- &-- &-- & $\ge 482.73 $ & $\ge 1038.96 $ & $\ge 4445.1 $ \\
HIP 25606 & $ 0.807 \pm 0.027 $ & $\le 3.03 $ &-- &-- &-- & $\ge 1021.51 $ & $\ge 2111.49 $ & $\ge 5405.2 $ \\
HIP 27021 & $ 0.531 \pm 0.006 $ & $ 44.96 \pm 6.8 $ & 6549.83 & 2786.96-10753.71 & 830.61-12664.69 &-- &-- &-- \\
HIP 28823 & $ 0.358 \pm 0.005 $ & $\le 38.81 $ &-- &-- &-- & $\ge 199.04 $ & $\ge 1161.78 $ & $\ge 4649.16 $ \\
HIP 29208 & $ 0.895 \pm 0.022 $ & $\le 6.74 $ &-- &-- &-- & $\ge 568.66 $ & $\ge 1152.02 $ & $\ge 4526.8 $ \\
HIP 33282 & $ 1.344 \pm 0.035 $ & $\le 13.59 $ &-- &-- &-- & $\ge 278.21 $ & $\ge 827.95 $ & $\ge 4294.15 $ \\
HIP 33368 & $ 1.346 \pm 0.1 $ & $\le 8.69 $ &-- &-- &-- & $\ge 288.0 $ & $\ge 847.34 $ & $\ge 4308.04 $ \\
HIP 34804 & $ 1.591 \pm 0.082 $ & $\le 8.55 $ &-- &-- &-- & $\ge 157.52 $ & $\ge 730.29 $ & $\ge 4224.76 $ \\
HIP 36081 & $ 0.62 \pm 0.005 $ & $ 49.7 \pm 1.89 $ & 2660.91 & 1006.18-8005.41 & 403.71-12139.12 &-- &-- &-- \\
HIP 37267 & $ 1.277 \pm 0.022 $ & $\le 8.48 $ &-- &-- &-- & $\ge 325.41 $ & $\ge 875.61 $ & $\ge 4328.24 $ \\
HIP 40518 & $ 1.134 \pm 0.004 $ & $\le 5.26 $ &-- &-- &-- & $\ge 430.98 $ & $\ge 985.1 $ & $\ge 4406.56 $ \\
HIP 41274 & $ 0.952 \pm 0.011 $ & $ 7.47 \pm 1.22 $ & 505.8 & 439.44-602.43 & 335.92-820.22 &-- &-- &-- \\
HIP 41277 & $ 1.03 \pm 0.015 $ & $ 125.14 \pm 5.01 $ & 173.84 & 96.87-252.49 & 31.29-327.98 &-- &-- &-- \\
HIP 41319 & $ 0.448 \pm 0.005 $ & $ 29.11 \pm 2.49 $ & 489.33 & 249.94-579.52 & 68.41-782.82 &-- &-- &-- \\
HIP 42231 & $ 1.241 \pm 0.015 $ & $\le 16.83 $ &-- &-- &-- & $\ge 296.87 $ & $\ge 847.87 $ & $\ge 4308.4 $ \\
HIP 42343 & $ 0.516 \pm 0.001 $ & $ 35.12 \pm 2.7 $ & 4964.37 & 2198.09-9633.86 & 776.39-12466.6 &-- &-- &-- \\
HIP 42507 & $ 1.361 \pm 0.001 $ & $\le 30.52 $ &-- &-- &-- & $\ge 226.84 $ & $\ge 778.95 $ & $\ge 4259.1 $ \\
HIP 43745 & $ 1.423 \pm 0.02 $ & $\le 12.37 $ &-- &-- &-- & $\ge 256.35 $ & $\ge 806.11 $ & $\ge 4278.53 $ \\
HIP 44162 & $ 0.461 \pm 0.005 $ & $\le 6.86 $ &-- &-- &-- & $\ge 735.01 $ & $\ge 1316.0 $ & $\ge 4643.77 $ \\
HIP 44953 & $ 0.767 \pm 0.002 $ & $\le 3.52 $ &-- &-- &-- & $\ge 1323.58 $ & $\ge 2846.19 $ & $\ge 5993.75 $ \\
HIP 45621 & $ 0.869 \pm 0.013 $ & $ 28.92 \pm 4.34 $ & 409.4 & 336.01-508.61 & 157.11-770.0 &-- &-- &-- \\
HIP 45731 & $ 1.586 \pm 0.02 $ & $\le 23.11 $ &-- &-- &-- & $\ge 91.63 $ & $\ge 650.87 $ & $\ge 4167.6 $ \\
HIP 46385 & $ 1.586 \pm 0.1 $ & $\le 30.6 $ &-- &-- &-- & $\ge 92.44 $ & $\ge 661.4 $ & $\ge 4175.04 $ \\
HIP 47013 & $ 0.53 \pm 0.003 $ & $\le 9.22 $ &-- &-- &-- & $\ge 215.33 $ & $\ge 1527.67 $ & $\ge 5866.68 $ \\
HIP 47261 & $ 1.21 \pm 0.015 $ & $\le 8.15 $ &-- &-- &-- & $\ge 356.79 $ & $\ge 908.11 $ & $\ge 4351.49 $ \\
HIP 47300 & $ 0.223 \pm 0.004 $ & $\le 59.79 $ & -- & -- &-- & -- & -- & -- \\
HIP 47387 & $ 1.07 \pm 0.047 $ & $\le 5.05 $ &-- &-- &-- & $\ge 489.6 $ & $\ge 1046.57 $ & $\ge 4450.55 $ \\
HIP 47539 & $ 1.353 \pm 0.011 $ & $\le 15.01 $ &-- &-- &-- & $\ge 270.07 $ & $\ge 819.51 $ & $\ge 4288.12 $ \\
HIP 47741 & $ 1.542 \pm 0.039 $ & $\le 15.13 $ &-- &-- &-- & $\ge 130.85 $ & $\ge 698.97 $ & $\ge 4202.01 $ \\
HIP 48165 & $ 0.925 \pm 0.015 $ & $\le 3.19 $ &-- &-- &-- & $\ge 638.59 $ & $\ge 1219.3 $ & $\ge 4574.64 $ \\
HIP 48629 & $ 1.581 \pm 0.197 $ & $\le 9.94 $ &-- &-- &-- & $\ge 152.31 $ & $\ge 744.26 $ & $\ge 4234.47 $ \\
HIP 49046 & $ 1.372 \pm 0.015 $ & $\le 12.76 $ &-- &-- &-- & $\ge 276.52 $ & $\ge 825.31 $ & $\ge 4292.26 $ \\
HIP 49526 & $ 1.492 \pm 0.02 $ & $\le 24.48 $ &-- &-- &-- & $\ge 147.69 $ & $\ge 710.86 $ & $\ge 4210.41 $ \\
HIP 49577 & $ 1.331 \pm 0.014 $ & $\le 6.65 $ &-- &-- &-- & $\ge 318.63 $ & $\ge 867.5 $ & $\ge 4322.44 $ \\
HIP 49910 & $ 1.391 \pm 0.023 $ & $\le 12.9 $ &-- &-- &-- & $\ge 268.72 $ & $\ge 817.89 $ & $\ge 4286.95 $ \\
HIP 51208 & $ 1.2 \pm 0.015 $ & $\le 7.14 $ &-- &-- &-- & $\ge 371.01 $ & $\ge 922.66 $ & $\ge 4361.89 $ \\
HIP 52339 & $ 1.42 \pm 0.015 $ & $\le 11.84 $ &-- &-- &-- & $\ge 260.19 $ & $\ge 809.62 $ & $\ge 4281.04 $ \\
HIP 53175 & $ 1.293 \pm 0.005 $ & $\le 11.01 $ &-- &-- &-- & $\ge 304.58 $ & $\ge 854.28 $ & $\ge 4312.99 $ \\
HIP 53869 & $ 0.868 \pm 0.045 $ & $\le 5.27 $ &-- &-- &-- & $\ge 628.48 $ & $\ge 1339.17 $ & $\ge 4698.29 $ \\
HIP 54094 & $ 1.1 \pm 0.02 $ & $\le 4.44 $ &-- &-- &-- & $\ge 481.2 $ & $\ge 1033.83 $ & $\ge 4441.41 $ \\
HIP 54199 & $ 0.874 \pm 0.016 $ & $\le 34.44 $ &-- &-- &-- & $\ge 383.88 $ & $\ge 952.11 $ & $\ge 4383.16 $ \\
HIP 55192 & $ 0.776 \pm 0.015 $ & $ 75.4 \pm 1.89 $ & 308.96 & 218.51-539.13 & 66.28-1459.69 &-- &-- &-- \\
HIP 55409 & $ 0.658 \pm 0.003 $ & $\le 5.57 $ &-- &-- &-- & $\ge 892.89 $ & $\ge 3214.05 $ & $\ge 7741.72 $ \\
HIP 57361 & $ 1.482 \pm 0.015 $ & $\le 70.46 $ &-- &-- &-- & $\ge 85.43 $ & $\ge 646.46 $ & $\ge 4164.34 $ \\
HIP 57571 & $ 1.369 \pm 0.02 $ & $\le 11.45 $ &-- &-- &-- & $\ge 281.31 $ & $\ge 830.14 $ & $\ge 4295.72 $ \\
HIP 57645 & $ 0.895 \pm 0.015 $ & $\le 3.69 $ &-- &-- &-- & $\ge 659.12 $ & $\ge 1242.26 $ & $\ge 4591.27 $ \\
HIP 57984 & $ 1.316 \pm 0.015 $ & $\le 10.03 $ &-- &-- &-- & $\ge 302.07 $ & $\ge 851.43 $ & $\ge 4310.95 $ \\
HIP 59000 & $ 1.336 \pm 0.014 $ & $\le 10.5 $ &-- &-- &-- & $\ge 292.85 $ & $\ge 842.01 $ & $\ge 4304.21 $ \\
HIP 59198 & $ 1.409 \pm 0.015 $ & $\le 11.98 $ &-- &-- &-- & $\ge 264.26 $ & $\ge 813.52 $ & $\ge 4283.83 $ \\
HIP 59199 & $ 0.334 \pm 0.015 $ & $ 47.3 \pm 3.74 $ & -- & -- & -- & -- & -- & -- \\
HIP 59247 & $ 1.29 \pm 0.015 $ & $\le 10.15 $ &-- &-- &-- & $\ge 310.2 $ & $\ge 860.01 $ & $\ge 4317.08 $ \\
HIP 59767 & $ 0.346 \pm 0.011 $ & $ 32.5 \pm 3.95 $ & -- & -- & -- & -- & -- & -- \\
HIP 59905 & $ 1.125 \pm 0.015 $ & $\le 5.68 $ &-- &-- &-- & $\ge 430.87 $ & $\ge 985.26 $ & $\ge 4406.67 $ \\
HIP 59953 & $ 1.602 \pm 0.1 $ & $\le 25.28 $ &-- &-- &-- & $\ge 96.84 $ & $\ge 665.13 $ & $\ge 4177.73 $ \\
HIP 60829 & $ 0.71 \pm 0.012 $ & $\le 3.99 $ &-- &-- &-- & $\ge 1515.85 $ & $\ge 3828.19 $ & $\ge 7411.7 $ \\
HIP 61947 & $ 1.582 \pm 0.082 $ & $\le 21.91 $ &-- &-- &-- & $\ge 104.87 $ & $\ge 673.77 $ & $\ge 4183.93 $ \\
HIP 62325 & $ 0.989 \pm 0.007 $ & $\le 1.63 $ &-- &-- &-- & $\ge 585.8 $ & $\ge 1156.33 $ & $\ge 4529.3 $ \\
HIP 62627 & $ 0.995 \pm 0.003 $ & $\le 7.1 $ &-- &-- &-- & $\ge 480.59 $ & $\ge 1042.53 $ & $\ge 4447.71 $ \\
HIP 63419 & $ 0.779 \pm 0.021 $ & $ 7.65 \pm 2.67 $ & 2340.25 & 1276.27-5485.81 & 757.54-10722.15 &-- &-- &-- \\
HIP 63421 & $ 1.354 \pm 0.01 $ & $\le 16.29 $ &-- &-- &-- & $\ge 265.57 $ & $\ge 815.16 $ & $\ge 4285.0 $ \\
HIP 63661 & $ 1.468 \pm 0.1 $ & $\le 17.8 $ &-- &-- &-- & $\ge 166.58 $ & $\ge 747.11 $ & $\ge 4236.36 $ \\
HIP 65120 & $ 1.49 \pm 0.02 $ & $\le 26.81 $ &-- &-- &-- & $\ge 142.13 $ & $\ge 706.06 $ & $\ge 4206.97 $ \\
HIP 65602 & $ 0.911 \pm 0.002 $ & $\le 3.47 $ &-- &-- &-- & $\ge 641.92 $ & $\ge 1219.41 $ & $\ge 4574.65 $ \\
HIP 65651 & $ 1.203 \pm 0.02 $ & $\le 17.22 $ &-- &-- &-- & $\ge 309.3 $ & $\ge 860.85 $ & $\ge 4317.68 $ \\
HIP 65706 & $ 1.577 \pm 0.082 $ & $\le 10.88 $ &-- &-- &-- & $\ge 144.27 $ & $\ge 717.64 $ & $\ge 4215.51 $ \\
HIP 66262 & $ 1.165 \pm 0.067 $ & $\le 6.35 $ &-- &-- &-- & $\ge 393.3 $ & $\ge 956.97 $ & $\ge 4386.44 $ \\
HIP 66315 & $ 1.039 \pm 0.121 $ & $\le 12.91 $ &-- &-- &-- & $\ge 395.09 $ & $\ge 985.51 $ & $\ge 4412.74 $ \\
HIP 66587 & $ 1.419 \pm 0.009 $ & $\le 15.03 $ &-- &-- &-- & $\ge 249.51 $ & $\ge 799.18 $ & $\ge 4273.57 $ \\
HIP 66828 & $ 1.33 \pm 0.02 $ & $\le 8.89 $ &-- &-- &-- & $\ge 303.69 $ & $\ge 852.86 $ & $\ge 4311.97 $ \\
HIP 69311 & $ 1.172 \pm 0.07 $ & $\le 9.39 $ &-- &-- &-- & $\ge 360.75 $ & $\ge 923.47 $ & $\ge 4362.47 $ \\
HIP 69333 & $ 1.173 \pm 0.004 $ & $\le 4.28 $ &-- &-- &-- & $\ge 426.36 $ & $\ge 979.3 $ & $\ge 4402.4 $ \\
HIP 69860 & $ 1.629 \pm 0.1 $ & $\le 25.8 $ &-- &-- &-- & $\ge 92.18 $ & $\ge 656.97 $ & $\ge 4171.9 $ \\
HIP 69963 & $ 1.39 \pm 0.02 $ & $\le 34.7 $ &-- &-- &-- & $\ge 217.68 $ & $\ge 769.92 $ & $\ge 4252.64 $ \\
HIP 70472 & $ 1.257 \pm 0.01 $ & $\le 5.02 $ &-- &-- &-- & $\ge 368.06 $ & $\ge 918.5 $ & $\ge 4358.91 $ \\
HIP 71243 & $ 0.45 \pm 0.003 $ & $\le 12.35 $ &-- &-- &-- & $\ge 624.45 $ & $\ge 1202.77 $ & $\ge 4562.54 $ \\
HIP 71515 & $ 0.95 \pm 0.002 $ & $\le 4.52 $ &-- &-- &-- & $\ge 566.76 $ & $\ge 1138.55 $ & $\ge 4516.62 $ \\
HIP 71899 & $ 0.533 \pm 0.01 $ & $ 90.22 \pm 3.36 $ & 3446.91 & 535.05-9289.75 & 124.97-12428.99 &-- &-- &-- \\
HIP 71957 & $ 0.385 \pm 0.006 $ & $ 62.96 \pm 6.43 $ & 742.99 & 198.67-2858.42 & 57.32-9369.38 &-- &-- &-- \\
HIP 73121 & $ 1.035 \pm 0.009 $ & $\le 3.93 $ &-- &-- &-- & $\ge 544.57 $ & $\ge 1093.34 $ & $\ge 4483.97 $ \\
HIP 73183 & $ 1.555 \pm 0.197 $ & $\le 10.92 $ &-- &-- &-- & $\ge 152.29 $ & $\ge 747.08 $ & $\ge 4236.44 $ \\
HIP 73252 & $ 1.565 \pm 0.082 $ & $\le 14.68 $ &-- &-- &-- & $\ge 129.34 $ & $\ge 703.31 $ & $\ge 4205.13 $ \\
HIP 73787 & $ 1.21 \pm 0.008 $ & $\le 7.17 $ &-- &-- &-- & $\ge 366.04 $ & $\ge 917.25 $ & $\ge 4358.02 $ \\
HIP 74926 & $ 1.214 \pm 0.009 $ & $\le 8.04 $ &-- &-- &-- & $\ge 356.02 $ & $\ge 907.11 $ & $\ge 4350.77 $ \\
HIP 79203 & $ 0.447 \pm 0.007 $ & $ 52.15 \pm 2.24 $ & 316.16 & 170.59-447.18 & 48.25-603.54 &-- &-- &-- \\
HIP 79593 & $ 1.584 \pm 0.01 $ & $\le 64.58 $ &-- &-- &-- & $\ge 62.8 $ & $\ge 619.13 $ & $\ge 4144.8 $ \\
HIP 81655 & $ 1.605 \pm 0.1 $ & $\le 16.7 $ &-- &-- &-- & $\ge 114.09 $ & $\ge 684.91 $ & $\ge 4191.94 $ \\
HIP 82260 & $ 0.767 \pm 0.002 $ & $\le 2.66 $ &-- &-- &-- & $\ge 1383.07 $ & $\ge 2980.3 $ & $\ge 6115.02 $ \\
HIP 83676 & $ 1.015 \pm 0.015 $ & $\le 5.23 $ &-- &-- &-- & $\ge 513.21 $ & $\ge 1071.18 $ & $\ge 4468.19 $ \\
HIP 83929 & $ 1.597 \pm 0.197 $ & $\le 7.89 $ &-- &-- &-- & $\ge 164.45 $ & $\ge 754.63 $ & $\ge 4242.01 $ \\
HIP 87370 & $ 0.656 \pm 0.015 $ & $ 5.45 \pm 1.73 $ & 9440.7 & 5779.69-11965.31 & 2459.94-12855.22 &-- &-- &-- \\
HIP 88962 & $ 1.372 \pm 0.004 $ & $\le 7.43 $ &-- &-- &-- & $\ge 302.95 $ & $\ge 850.97 $ & $\ge 4310.61 $ \\
HIP 89320 & $ 1.586 \pm 0.197 $ & $\le 17.74 $ &-- &-- &-- & $\ge 119.63 $ & $\ge 709.43 $ & $\ge 4209.43 $ \\
HIP 89449 & $ 1.323 \pm 0.019 $ & $\le 13.44 $ &-- &-- &-- & $\ge 284.08 $ & $\ge 833.75 $ & $\ge 4298.3 $ \\
HIP 90306 & $ 1.442 \pm 0.02 $ & $\le 20.2 $ &-- &-- &-- & $\ge 220.21 $ & $\ge 771.96 $ & $\ge 4254.11 $ \\
HIP 93072 & $ 1.437 \pm 0.004 $ & $\le 9.38 $ &-- &-- &-- & $\ge 266.15 $ & $\ge 815.5 $ & $\ge 4285.25 $ \\
HIP 98007 & $ 0.978 \pm 0.003 $ & $\le 4.06 $ &-- &-- &-- & $\ge 561.44 $ & $\ge 1130.48 $ & $\ge 4510.78 $ \\
HIP 99969 & $ 1.14 \pm 0.015 $ & $\le 6.74 $ &-- &-- &-- & $\ge 405.74 $ & $\ge 959.42 $ & $\ge 4388.19 $ \\
HIP 100133 & $ 1.43 \pm 0.012 $ & $\le 10.69 $ &-- &-- &-- & $\ge 261.29 $ & $\ge 810.76 $ & $\ge 4281.86 $ \\
HIP 101852 & $ 0.597 \pm 0.012 $ & $\le 5.55 $ &-- &-- &-- & $\ge 413.67 $ & $\ge 2147.3 $ & $\ge 6660.85 $ \\
HIP 102119 & $ 1.121 \pm 0.009 $ & $\le 15.56 $ &-- &-- &-- & $\ge 349.81 $ & $\ge 902.7 $ & $\ge 4347.61 $ \\
HIP 104687 & $ 0.654 \pm 0.041 $ & $ 63.32 \pm 3.17 $ & 1661.55 & 516.08-6885.97 & 216.02-11896.04 &-- &-- &-- \\
HIP 105504 & $ 1.255 \pm 0.015 $ & $\le 7.34 $ &-- &-- &-- & $\ge 343.5 $ & $\ge 893.98 $ & $\ge 4341.38 $ \\
HIP 107317 & $ 1.522 \pm 0.021 $ & $\le 19.01 $ &-- &-- &-- & $\ge 131.93 $ & $\ge 696.16 $ & $\ge 4199.9 $ \\
HIP 108036 & $ 0.378 \pm 0.011 $ & $ 36.67 \pm 8.24 $ & 1531.65 & 629.16-4486.11 & 186.82-10568.9 &-- &-- &-- \\
HIP 109807 & $ 1.252 \pm 0.005 $ & $\le 8.56 $ &-- &-- &-- & $\ge 334.91 $ & $\ge 885.27 $ & $\ge 4335.15 $ \\
HIP 110106 & $ 1.399 \pm 0.02 $ & $\le 18.94 $ &-- &-- &-- & $\ge 247.79 $ & $\ge 797.7 $ & $\ge 4272.52 $ \\
HIP 110401 & $ 1.208 \pm 0.013 $ & $\le 6.42 $ &-- &-- &-- & $\ge 374.95 $ & $\ge 926.43 $ & $\ge 4364.59 $ \\
HIP 110663 & $ 0.77 \pm 0.005 $ & $\le 30.38 $ &-- &-- &-- & $\ge 449.42 $ & $\ge 1135.54 $ & $\ge 4534.24 $ \\
HIP 111976 & $ 1.548 \pm 0.1 $ & $\le 17.49 $ &-- &-- &-- & $\ge 126.14 $ & $\ge 705.19 $ & $\ge 4206.41 $ \\
HIP 113159 & $ 0.469 \pm 0.005 $ & $ 22.57 \pm 2.34 $ & 563.71 & 364.7-656.67 & 95.73-869.36 &-- &-- &-- \\
HIP 115280 & $ 0.468 \pm 0.004 $ & $\le 13.37 $ &-- &-- &-- & $\ge 630.31 $ & $\ge 1208.72 $ & $\ge 4566.8 $ \\
HIP 115411 & $ 0.7 \pm 0.005 $ & $\le 3.79 $ &-- &-- &-- & $\ge 1551.06 $ & $\ge 4259.39 $ & $\ge 8276.42 $ \\
HIP 116973 & $ 1.085 \pm 0.008 $ & $\le 3.36 $ &-- &-- &-- & $\ge 525.19 $ & $\ge 1071.06 $ & $\ge 4468.03 $ \\
HIP 117159 & $ 0.859 \pm 0.006 $ & $\le 2.4 $ &-- &-- &-- & $\ge 801.54 $ & $\ge 1475.94 $ & $\ge 4766.76 $ \\
HIP 117235 & $ 1.376 \pm 0.133 $ & $\le 13.91 $ &-- &-- &-- & $\ge 233.62 $ & $\ge 806.1 $ & $\ge 4278.54 $ \\
HIP 117410 & $ 1.244 \pm 0.014 $ & $\le 6.27 $ &-- &-- &-- & $\ge 359.04 $ & $\ge 909.77 $ & $\ge 4352.67 $ \\
HIP 117795 & $ 1.21 \pm 0.015 $ & $\le 8.98 $ &-- &-- &-- & $\ge 350.04 $ & $\ge 901.3 $ & $\ge 4346.62 $ \\
HIP 118086 & $ 1.3 \pm 0.015 $ & $\le 18.5 $ &-- &-- &-- & $\ge 273.54 $ & $\ge 823.87 $ & $\ge 4291.23 $ \\
HIP 118310 & $ 1.187 \pm 0.052 $ & $\le 4.14 $ &-- &-- &-- & $\ge 416.39 $ & $\ge 975.53 $ & $\ge 4399.71 $
\label{tab:lithium}
\end{longtable}}

\begin{longtable}{|c|c|c|c|c|c|c|c|}
\caption{Each target star and its adopted $B-V$, measured S index and measured $log(R^{'}_{HK})$ are listed. $R^{'}_{HK}$ is only calibrated as an age metric for F5-K2 stars for which report a median age, $68\%$ and $98\%$ confidence intervals.}\\

\hline
Target & Adopted  & S & $log(R^{'}_{HK})$ & Age  & $68\%$ CI& $95\%$ CI \\
 & B-V& & & (Myr)  & (Myr) & (Myr)  \\
\hline
\endfirsthead 
\hline
Target & Adopted  & S & $log(R^{'}_{HK})$ & Age  & $68\%$ CI& $95\%$ CI \\
 & B-V& & & (Myr)  & (Myr) & (Myr)  \\
\hline
\endhead
\hline
\endfoot
\hline
\endlastfoot
HIP 143 & $ 1.033 \pm 0.019 $ & 0.41 & -4.58 &-- &-- &-- \\
HIP 669 & $ 0.624 \pm 0.015 $ & 0.11 & -4.88 & 6468.35 & 3805.7-9673.65 & 1811.36-12362.31 \\
HIP 1068 & $ 1.315 \pm 0.1 $ & 0.41 & -4.3 &-- &-- &-- \\
HIP 1412 & $ 1.414 \pm 0.005 $ & 1.38 & -3.7 &-- &-- &-- \\
HIP 1539 & $ 1.375 \pm 0.005 $ & 1.25 & -3.77 &-- &-- &-- \\
HIP 1771 & $ 1.489 \pm 0.1 $ & 1.4 & -3.66 &-- &-- &-- \\
HIP 2426 & $ 1.359 \pm 0.02 $ & 0.85 & -3.95 &-- &-- &-- \\
HIP 2843 & $ 0.466 \pm 0.007 $ & 0.1 & -4.82 & 5038.2 & 2848.33-8376.32 & 1319.5-12018.59 \\
HIP 3008 & $ 1.424 \pm 0.02 $ & 1.57 & -3.64 &-- &-- &-- \\
HIP 3293 & $ 0.452 \pm 0.009 $ & 0.06 & -4.8 & 4744.41 & 2658.9-8090.39 & 1223.88-11927.21 \\
HIP 3633 & $ 0.814 \pm 0.061 $ & 0.28 & -4.93 & 7627.38 & 4629.95-10777.13 & 2245.46-12580.04 \\
HIP 3724 & $ 1.4 \pm 0.02 $ & 1.95 & -3.56 &-- &-- &-- \\
HIP 4024 & $ 1.069 \pm 0.009 $ & 0.39 & -4.55 &-- &-- &-- \\
HIP 5319 & $ 0.398 \pm 0.01 $ & 0.13 & -4.83 &-- &-- &-- \\
HIP 5763 & $ 1.22 \pm 0.015 $ & 0.67 & -4.16 &-- &-- &-- \\
HIP 6776 & $ 0.453 \pm 0.007 $ & 0.25 & -4.89 & 6703.86 & 3969.45-9900.36 & 1896.83-12406.95 \\
HIP 6796 & $ 1.157 \pm 0.074 $ & 0.58 & -4.28 &-- &-- &-- \\
HIP 6833 & $ 0.656 \pm 0.002 $ & 0.12 & -4.91 & 7056.53 & 4218.1-10241.09 & 2027.37-12469.75 \\
HIP 6943 & $ 0.453 \pm 0.011 $ & 0.06 & -4.8 & 4719.62 & 2643.05-8065.82 & 1215.9-11919.12 \\
HIP 7287 & $ 1.2 \pm 0.02 $ & 0.44 & -4.36 &-- &-- &-- \\
HIP 8653 & $ 0.644 \pm 0.019 $ & 0.22 & -4.91 & 7289.8 & 4385.03-10462.51 & 2115.46-12502.65 \\
HIP 9067 & $ 1.47 \pm 0.1 $ & 0.74 & -3.94 &-- &-- &-- \\
HIP 9989 & $ 1.248 \pm 0.002 $ & 0.78 & -4.07 &-- &-- &-- \\
HIP 10050 & $ 0.471 \pm 0.001 $ & 0.05 & -4.8 & 4649.09 & 2597.95-7995.11 & 1193.23-11895.76 \\
HIP 10532 & $ 0.88 \pm 0.033 $ & 0.19 & -5.0 & 8917.93 & 5673.64-11623.81 & 2817.71-12794.38 \\
HIP 11090 & $ 0.289 \pm 0.006 $ & 0.08 & -4.82 &-- &-- &-- \\
HIP 11815 & $ 1.357 \pm 0.02 $ & 2.34 & -3.51 &-- &-- &-- \\
HIP 12837 & $ 0.517 \pm 0.008 $ & 0.13 & -4.84 & 5572.64 & 3199.05-8862.9 & 1498.1-12163.83 \\
HIP 13681 & $ 1.062 \pm 0.015 $ & 0.55 & -4.42 &-- &-- &-- \\
HIP 13782 & $ 0.238 \pm 0.003 $ & 0.11 & -4.83 &-- &-- &-- \\
HIP 15211 & $ 0.936 \pm 0.007 $ & 0.21 & -4.95 &-- &-- &-- \\
HIP 16247 & $ 1.02 \pm 0.037 $ & 1.24 & -4.13 &-- &-- &-- \\
HIP 16709 & $ 1.225 \pm 0.02 $ & 0.46 & -4.32 &-- &-- &-- \\
HIP 17118 & $ 0.478 \pm 0.023 $ & 0.14 & -4.84 & 5506.82 & 3155.38-8805.1 & 1475.75-12147.31 \\
HIP 17213 & $ 1.1 \pm 0.015 $ & 0.8 & -4.21 &-- &-- &-- \\
HIP 18527 & $ 0.899 \pm 0.021 $ & 0.45 & -4.72 & 3335.89 & 1782.91-6529.46 & 789.98-11322.22 \\
HIP 19739 & $ 1.424 \pm 0.02 $ & 1.01 & -3.83 &-- &-- &-- \\
HIP 19912 & $ 1.148 \pm 0.002 $ & 0.39 & -4.47 &-- &-- &-- \\
HIP 20419 & $ 1.183 \pm 0.005 $ & 0.47 & -4.35 &-- &-- &-- \\
HIP 20648 & $ 0.049 \pm 0.007 $ & 0.1 & -4.76 &-- &-- &-- \\
HIP 21066 & $ 0.472 \pm 0.013 $ & 0.13 & -4.83 & 5415.69 & 3095.17-8724.59 & 1445.0-12123.82 \\
HIP 21152 & $ 0.42 \pm 0.014 $ & 0.12 & -4.82 &-- &-- &-- \\
HIP 22361 & $ 0.283 \pm 0.004 $ & 0.13 & -4.83 &-- &-- &-- \\
HIP 23208 & $ 1.18 \pm 0.006 $ & 1.03 & -4.01 &-- &-- &-- \\
HIP 24017 & $ 0.448 \pm 0.004 $ & 0.11 & -4.82 &-- &-- &-- \\
HIP 24177 & $ 1.461 \pm 0.005 $ & 0.65 & -4.0 &-- &-- &-- \\
HIP 24292 & $ 1.23 \pm 0.015 $ & 0.7 & -4.13 &-- &-- &-- \\
HIP 24457 & $ 1.118 \pm 0.005 $ & 0.45 & -4.44 &-- &-- &-- \\
HIP 25606 & $ 0.807 \pm 0.027 $ & 0.07 & -5.1 &-- &-- &-- \\
HIP 27021 & $ 0.531 \pm 0.006 $ & 0.05 & -4.81 & 4955.97 & 2795.06-8297.47 & 1292.55-11993.99 \\
HIP 28823 & $ 0.358 \pm 0.005 $ & 0.11 & -4.83 &-- &-- &-- \\
HIP 29208 & $ 0.895 \pm 0.022 $ & 0.11 & -5.13 &-- &-- &-- \\
HIP 33282 & $ 1.344 \pm 0.035 $ & 1.12 & -3.83 &-- &-- &-- \\
HIP 33368 & $ 1.346 \pm 0.1 $ & 1.19 & -3.81 &-- &-- &-- \\
HIP 34804 & $ 1.591 \pm 0.082 $ & 0.38 & -4.18 &-- &-- &-- \\
HIP 36081 & $ 0.62 \pm 0.005 $ & 0.06 & -4.86 & 6087.61 & 3544.88-9317.97 & 1676.01-12283.7 \\
HIP 37267 & $ 1.277 \pm 0.022 $ & 0.72 & -4.08 &-- &-- &-- \\
HIP 40518 & $ 1.134 \pm 0.004 $ & 0.48 & -4.39 &-- &-- &-- \\
HIP 41274 & $ 0.952 \pm 0.011 $ & 0.61 & -4.53 &-- &-- &-- \\
HIP 41277 & $ 1.03 \pm 0.015 $ & 1.68 & -3.98 &-- &-- &-- \\
HIP 41319 & $ 0.448 \pm 0.005 $ & 0.22 & -4.88 &-- &-- &-- \\
HIP 42231 & $ 1.241 \pm 0.015 $ & 0.53 & -4.24 &-- &-- &-- \\
HIP 42343 & $ 0.516 \pm 0.001 $ & 0.05 & -4.8 & 4792.54 & 2689.78-8138.18 & 1239.43-11942.73 \\
HIP 42507 & $ 1.361 \pm 0.001 $ & 1.48 & -3.7 &-- &-- &-- \\
HIP 43745 & $ 1.423 \pm 0.02 $ & 1.28 & -3.73 &-- &-- &-- \\
HIP 44162 & $ 0.461 \pm 0.005 $ & 0.1 & -4.82 & 5091.45 & 2882.9-8426.86 & 1337.04-12034.09 \\
HIP 44953 & $ 0.767 \pm 0.002 $ & 0.12 & -5.01 &-- &-- &-- \\
HIP 45621 & $ 0.869 \pm 0.013 $ & 0.36 & -4.83 & 5312.27 & 3027.16-8631.79 & 1410.33-12096.23 \\
HIP 45731 & $ 1.586 \pm 0.02 $ & 4.53 & -3.11 &-- &-- &-- \\
HIP 46385 & $ 1.586 \pm 0.1 $ & 1.07 & -3.73 &-- &-- &-- \\
HIP 47013 & $ 0.53 \pm 0.003 $ & 0.06 & -4.81 & 4997.36 & 2821.85-8337.18 & 1306.1-12006.47 \\
HIP 47261 & $ 1.21 \pm 0.015 $ & 1.16 & -3.93 &-- &-- &-- \\
HIP 47300 & $ 0.223 \pm 0.004 $ & 0.08 & -4.83 &-- &-- &-- \\
HIP 47387 & $ 1.07 \pm 0.047 $ & 0.49 & -4.46 &-- &-- &-- \\
HIP 47539 & $ 1.353 \pm 0.011 $ & 0.95 & -3.9 &-- &-- &-- \\
HIP 47741 & $ 1.542 \pm 0.039 $ & 0.61 & -3.99 &-- &-- &-- \\
HIP 48165 & $ 0.925 \pm 0.015 $ & 0.16 & -5.03 &-- &-- &-- \\
HIP 48629 & $ 1.581 \pm 0.197 $ & 1.11 & -3.72 &-- &-- &-- \\
HIP 49046 & $ 1.372 \pm 0.015 $ & 1.26 & -3.77 &-- &-- &-- \\
HIP 49526 & $ 1.492 \pm 0.02 $ & 0.97 & -3.81 &-- &-- &-- \\
HIP 49577 & $ 1.331 \pm 0.014 $ & 0.71 & -4.05 &-- &-- &-- \\
HIP 49910 & $ 1.391 \pm 0.023 $ & 1.09 & -3.82 &-- &-- &-- \\
HIP 51208 & $ 1.2 \pm 0.015 $ & 0.59 & -4.23 &-- &-- &-- \\
HIP 52339 & $ 1.42 \pm 0.015 $ & 1.32 & -3.72 &-- &-- &-- \\
HIP 53175 & $ 1.293 \pm 0.005 $ & 1.41 & -3.77 &-- &-- &-- \\
HIP 53869 & $ 0.868 \pm 0.045 $ & 0.36 & -4.83 & 5401.81 & 3086.04-8712.37 & 1440.37-12120.17 \\
HIP 54094 & $ 1.1 \pm 0.02 $ & 0.16 & -4.9 &-- &-- &-- \\
HIP 54199 & $ 0.874 \pm 0.016 $ & 0.13 & -5.09 &-- &-- &-- \\
HIP 55192 & $ 0.776 \pm 0.015 $ & 0.1 & -5.03 &-- &-- &-- \\
HIP 55409 & $ 0.658 \pm 0.003 $ & 0.12 & -4.91 & 7119.41 & 4262.9-10301.28 & 2050.96-12479.67 \\
HIP 57361 & $ 1.482 \pm 0.015 $ & 0.64 & -4.0 &-- &-- &-- \\
HIP 57571 & $ 1.369 \pm 0.02 $ & 0.73 & -4.0 &-- &-- &-- \\
HIP 57645 & $ 0.895 \pm 0.015 $ & 0.29 & -4.87 & 6194.96 & 3617.97-9416.26 & 1713.82-12306.8 \\
HIP 57984 & $ 1.316 \pm 0.015 $ & 1.08 & -3.87 &-- &-- &-- \\
HIP 59000 & $ 1.336 \pm 0.014 $ & 1.96 & -3.6 &-- &-- &-- \\
HIP 59198 & $ 1.409 \pm 0.015 $ & 0.93 & -3.87 &-- &-- &-- \\
HIP 59199 & $ 0.334 \pm 0.015 $ & 0.1 & -4.82 &-- &-- &-- \\
HIP 59247 & $ 1.29 \pm 0.015 $ & 0.81 & -4.01 &-- &-- &-- \\
HIP 59767 & $ 0.346 \pm 0.011 $ & 0.12 & -4.83 &-- &-- &-- \\
HIP 59905 & $ 1.125 \pm 0.015 $ & 0.58 & -4.32 &-- &-- &-- \\
HIP 59953 & $ 1.602 \pm 0.1 $ & 0.97 & -3.77 &-- &-- &-- \\
HIP 60829 & $ 0.71 \pm 0.012 $ & 0.07 & -4.96 & 8269.16 & 5128.5-11264.78 & 2514.96-12724.17 \\
HIP 61947 & $ 1.582 \pm 0.082 $ & 1.12 & -3.71 &-- &-- &-- \\
HIP 62325 & $ 0.989 \pm 0.007 $ & 0.08 & -5.23 &-- &-- &-- \\
HIP 62627 & $ 0.995 \pm 0.003 $ & 0.2 & -4.91 &-- &-- &-- \\
HIP 63419 & $ 0.779 \pm 0.021 $ & 0.36 & -4.9 & 6919.61 & 4121.07-10109.29 & 1976.29-12446.08 \\
HIP 63421 & $ 1.354 \pm 0.01 $ & 0.93 & -3.91 &-- &-- &-- \\
HIP 63661 & $ 1.468 \pm 0.1 $ & 0.95 & -3.84 &-- &-- &-- \\
HIP 65120 & $ 1.49 \pm 0.02 $ & 0.99 & -3.81 &-- &-- &-- \\
HIP 65602 & $ 0.911 \pm 0.002 $ & 0.24 & -4.92 &-- &-- &-- \\
HIP 65651 & $ 1.203 \pm 0.02 $ & 1.16 & -3.93 &-- &-- &-- \\
HIP 65706 & $ 1.577 \pm 0.082 $ & 0.96 & -3.78 &-- &-- &-- \\
HIP 66262 & $ 1.165 \pm 0.067 $ & 0.48 & -4.35 &-- &-- &-- \\
HIP 66315 & $ 1.039 \pm 0.121 $ & 0.33 & -4.66 &-- &-- &-- \\
HIP 66587 & $ 1.419 \pm 0.009 $ & 0.81 & -3.93 &-- &-- &-- \\
HIP 66828 & $ 1.33 \pm 0.02 $ & 1.17 & -3.83 &-- &-- &-- \\
HIP 69311 & $ 1.172 \pm 0.07 $ & 1.15 & -3.97 &-- &-- &-- \\
HIP 69333 & $ 1.173 \pm 0.004 $ & 0.49 & -4.34 &-- &-- &-- \\
HIP 69860 & $ 1.629 \pm 0.1 $ & 1.22 & -3.66 &-- &-- &-- \\
HIP 69963 & $ 1.39 \pm 0.02 $ & 0.57 & -4.1 &-- &-- &-- \\
HIP 70472 & $ 1.257 \pm 0.01 $ & 0.65 & -4.14 &-- &-- &-- \\
HIP 71243 & $ 0.45 \pm 0.003 $ & 0.1 & -4.81 & 5027.51 & 2841.38-8366.19 & 1315.99-12015.47 \\
HIP 71515 & $ 0.95 \pm 0.002 $ & 0.33 & -4.77 &-- &-- &-- \\
HIP 71899 & $ 0.533 \pm 0.01 $ & 0.16 & -4.86 & 5942.96 & 3446.92-9187.67 & 1625.41-12251.95 \\
HIP 71957 & $ 0.385 \pm 0.006 $ & 0.11 & -4.82 &-- &-- &-- \\
HIP 73121 & $ 1.035 \pm 0.009 $ & 0.19 & -4.89 &-- &-- &-- \\
HIP 73183 & $ 1.555 \pm 0.197 $ & 1.13 & -3.72 &-- &-- &-- \\
HIP 73252 & $ 1.565 \pm 0.082 $ & 1.26 & -3.67 &-- &-- &-- \\
HIP 73787 & $ 1.21 \pm 0.008 $ & 0.76 & -4.11 &-- &-- &-- \\
HIP 74926 & $ 1.214 \pm 0.009 $ & 0.57 & -4.23 &-- &-- &-- \\
HIP 79203 & $ 0.447 \pm 0.007 $ & 0.07 & -4.8 &-- &-- &-- \\
HIP 79593 & $ 1.584 \pm 0.01 $ & 0.19 & -4.48 &-- &-- &-- \\
HIP 81655 & $ 1.605 \pm 0.1 $ & 0.6 & -3.98 &-- &-- &-- \\
HIP 82260 & $ 0.767 \pm 0.002 $ & 0.08 & -5.03 &-- &-- &-- \\
HIP 83676 & $ 1.015 \pm 0.015 $ & 0.37 & -4.65 &-- &-- &-- \\
HIP 83929 & $ 1.597 \pm 0.197 $ & 0.73 & -3.9 &-- &-- &-- \\
HIP 87370 & $ 0.656 \pm 0.015 $ & 0.07 & -4.9 & 6908.91 & 4113.47-10099.0 & 1972.33-12444.19 \\
HIP 88962 & $ 1.372 \pm 0.004 $ & 0.91 & -3.91 &-- &-- &-- \\
HIP 89320 & $ 1.586 \pm 0.197 $ & 1.31 & -3.65 &-- &-- &-- \\
HIP 89449 & $ 1.323 \pm 0.019 $ & 0.82 & -3.99 &-- &-- &-- \\
HIP 90306 & $ 1.442 \pm 0.02 $ & 0.78 & -3.93 &-- &-- &-- \\
HIP 93072 & $ 1.437 \pm 0.004 $ & 0.98 & -3.84 &-- &-- &-- \\
HIP 98007 & $ 0.978 \pm 0.003 $ & 0.17 & -4.99 &-- &-- &-- \\
HIP 99969 & $ 1.14 \pm 0.015 $ & 0.62 & -4.27 &-- &-- &-- \\
HIP 100133 & $ 1.43 \pm 0.012 $ & 0.82 & -3.92 &-- &-- &-- \\
HIP 101852 & $ 0.597 \pm 0.012 $ & 0.16 & -4.88 & 6452.83 & 3795.0-9658.97 & 1805.77-12359.28 \\
HIP 102119 & $ 1.121 \pm 0.009 $ & 1.9 & -3.81 &-- &-- &-- \\
HIP 104687 & $ 0.654 \pm 0.041 $ & 0.25 & -4.92 & 7504.1 & 4539.92-10663.71 & 2197.54-12545.56 \\
HIP 105504 & $ 1.255 \pm 0.015 $ & 1.39 & -3.81 &-- &-- &-- \\
HIP 107317 & $ 1.522 \pm 0.021 $ & 1.73 & -3.55 &-- &-- &-- \\
HIP 108036 & $ 0.378 \pm 0.011 $ & 0.14 & -4.84 &-- &-- &-- \\
HIP 109807 & $ 1.252 \pm 0.005 $ & 1.21 & -3.87 &-- &-- &-- \\
HIP 110106 & $ 1.399 \pm 0.02 $ & 1.26 & -3.75 &-- &-- &-- \\
HIP 110401 & $ 1.208 \pm 0.013 $ & 0.59 & -4.23 &-- &-- &-- \\
HIP 110663 & $ 0.77 \pm 0.005 $ & 0.35 & -4.91 & 7152.13 & 4286.34-10332.43 & 2063.3-12484.35 \\
HIP 111976 & $ 1.548 \pm 0.1 $ & 1.07 & -3.74 &-- &-- &-- \\
HIP 113159 & $ 0.469 \pm 0.005 $ & 0.08 & -4.81 & 4922.09 & 2773.17-8264.65 & 1281.48-11983.6 \\
HIP 115280 & $ 0.468 \pm 0.004 $ & 0.13 & -4.83 & 5405.28 & 3088.32-8715.42 & 1441.52-12121.08 \\
HIP 115411 & $ 0.7 \pm 0.005 $ & 0.09 & -4.95 & 8010.07 & 4920.15-11092.03 & 2401.41-12683.1 \\
HIP 116973 & $ 1.085 \pm 0.008 $ & 0.55 & -4.39 &-- &-- &-- \\
HIP 117159 & $ 0.859 \pm 0.006 $ & 0.1 & -5.13 &-- &-- &-- \\
HIP 117235 & $ 1.376 \pm 0.133 $ & 0.12 & -4.78 &-- &-- &-- \\
HIP 117410 & $ 1.244 \pm 0.014 $ & 2.99 & -3.49 &-- &-- &-- \\
HIP 117795 & $ 1.21 \pm 0.015 $ & 0.55 & -4.26 &-- &-- &-- \\
HIP 118086 & $ 1.3 \pm 0.015 $ & 0.82 & -4.0 &-- &-- &-- \\
HIP 118310 & $ 1.187 \pm 0.052 $ & 0.76 & -4.14 &-- &-- &-- 
\label{tab:rhk}
\end{longtable}

\begin{longtable}{|c|c|c|c|c|c|c|}

\caption{Rotation period measurements for 23 of our 166 accelerating stars. An additional 106 stars have a TESS light curve but display either no variability signal or a signal that is not due to rotation, and 37 stars do not have a TESS light curve. \rev{Median ages and $68\%$ and $95\%$ confidence intervals} are given for stars with a rotation period measurement that are also within the \texttt{gyro-interp} $T_{eff}$ range.}\\

\hline
Target & $T_{eff}$  & $P_{rot}$  & Age & $68\%$ CI & $95\%$ CI  \\
& (K)& (d) &  (Myr) & (Myr) & (Myr) \\
\hline
\endfirsthead
\hline
Target & $T_{eff}$  & $P_{rot}$  & Age & $68\%$ CI & $95\%$ CI  \\
& (K)&  (d) &  (Myr) & (Myr) & (Myr) \\
\hline
\endhead
\hline
\endfoot
\hline
\endlastfoot
HIP 143 & $ 4776 \pm 200 $ & no LC &-- &-- &-- \\
HIP 669 & $ 5955 \pm 343 $ & no signal &-- &-- &-- \\
HIP 1068 & $ 3401 \pm 51 $ & no signal &-- &-- &-- \\
HIP 1412 & $ 3807 \pm 82 $ &$ 5.1 \pm 0.44 $ & 134.6 & 41.14 - 385.6 & 6.73 - 568.83 \\
HIP 1539 & $ 3956 \pm 153 $ & no LC &-- &-- &-- \\
HIP 1771 & $ 4055 \pm 154 $ & no signal &-- &-- &-- \\
HIP 2426 & $ 4104 \pm 151 $ & no signal &-- &-- &-- \\
HIP 2843 & $ 6465 \pm 335 $ & no signal &-- &-- &-- \\
HIP 3008 & $ 3891 \pm 121 $ &$ 5.11 \pm 0.36 $ & 128.37 & 39.61 - 378.36 & 6.54 - 563.46 \\
HIP 3293 & $ 6442 \pm 344 $ & no signal &-- &-- &-- \\
HIP 3633 & $ 4831 \pm 211 $ & no signal &-- &-- &-- \\
HIP 3724 & $ 3873 \pm 104 $ & no LC &-- &-- &-- \\
HIP 4024 & $ 4692 \pm 184 $ & no signal &-- &-- &-- \\
HIP 5319 & $ 6693 \pm 324 $ & no signal &-- &-- &-- \\
HIP 5763 & $ 4370 \pm 144 $ & no signal &-- &-- &-- \\
HIP 6776 & $ 6496 \pm 324 $ & no signal &-- &-- &-- \\
HIP 6796 & $ 4590 \pm 174 $ & no signal &-- &-- &-- \\
HIP 6833 & $ 5765 \pm 305 $ & no signal &-- &-- &-- \\
HIP 6943 & $ 6464 \pm 336 $ & no signal &-- &-- &-- \\
HIP 7287 & $ 4242 \pm 123 $ & no signal &-- &-- &-- \\
HIP 8653 & $ 5755 \pm 306 $ &$ 4.11 \pm 0.34 $ & 220.38 & 76.93 - 423.96 & 11.88 - 1158.34 \\
HIP 9067 & $ 4117 \pm 150 $ & no signal &-- &-- &-- \\
HIP 9989 & $ 4279 \pm 126 $ & no signal &-- &-- &-- \\
HIP 10050 & $ 6461 \pm 337 $ & no signal &-- &-- &-- \\
HIP 10532 & $ 5149 \pm 198 $ & no signal &-- &-- &-- \\
HIP 11090 & $ 7127 \pm 426 $ & not rotation &-- &-- &-- \\
HIP 11815 & $ 3899 \pm 129 $ &2 periods & nan & nan & nan \\
HIP 12837 & $ 6293 \pm 311 $ & no signal &-- &-- &-- \\
HIP 13681 & $ 4709 \pm 188 $ & no signal &-- &-- &-- \\
HIP 13782 & $ 7615 \pm 441 $ & no signal &-- &-- &-- \\
HIP 15211 & $ 4867 \pm 211 $ & no signal &-- &-- &-- \\
HIP 16247 & $ 4453 \pm 160 $ &EB &-- &-- &-- \\
HIP 16709 & $ 4104 \pm 151 $ & no signal &-- &-- &-- \\
HIP 17118 & $ 6406 \pm 343 $ & no LC &-- &-- &-- \\
HIP 17213 & $ 4650 \pm 177 $ & no LC &-- &-- &-- \\
HIP 18527 & $ 5139 \pm 201 $ &$ 9.59 \pm 1.78 $ & 734.8 & 412.07 - 1053.55 & 153.71 - 1335.39 \\
HIP 19739 & $ 3818 \pm 88 $ &$ 7.84 \pm 1.06 $ & 254.07 & 129.59 - 453.05 & 71.31 - 590.82 \\
HIP 19912 & $ 4550 \pm 170 $ & no signal &-- &-- &-- \\
HIP 20419 & $ 4438 \pm 158 $ &$ 10.15 \pm 2.02 $ & 674.7 & 284.69 - 1180.52 & 132.07 - 1568.97 \\
HIP 20648 & $ 8718 \pm 670 $ & no signal &-- &-- &-- \\
HIP 21066 & $ 6411 \pm 344 $ & not rotation &-- &-- &-- \\
HIP 21152 & $ 6629 \pm 311 $ & no signal &-- &-- &-- \\
HIP 22361 & $ 7155 \pm 428 $ & not rotation &-- &-- &-- \\
HIP 23208 & $ 4268 \pm 125 $ & no LC &-- &-- &-- \\
HIP 24017 & $ 6579 \pm 307 $ & no signal &-- &-- &-- \\
HIP 24177 & $ 3687 \pm 87 $ & no signal &-- &-- &-- \\
HIP 24292 & $ 4342 \pm 137 $ & no signal &-- &-- &-- \\
HIP 24457 & $ 4627 \pm 177 $ & no signal &-- &-- &-- \\
HIP 25606 & $ 5109 \pm 184 $ & no signal &-- &-- &-- \\
HIP 27021 & $ 6123 \pm 315 $ & no signal &-- &-- &-- \\
HIP 28823 & $ 6887 \pm 322 $ & no signal &-- &-- &-- \\
HIP 29208 & $ 5112 \pm 186 $ & no LC &-- &-- &-- \\
HIP 33282 & $ 4200 \pm 137 $ &$ 9.87 \pm 1.77 $ & 507.65 & 220.38 - 1073.73 & 127.15 - 1510.58 \\
HIP 33368 & $ 4299 \pm 128 $ & contaminated pixel &-- &-- &-- \\
HIP 34804 & $ 4517 \pm 166 $ & no LC &-- &-- &-- \\
HIP 36081 & $ 5957 \pm 343 $ & no signal &-- &-- &-- \\
HIP 37267 & $ 4329 \pm 135 $ & no signal &-- &-- &-- \\
HIP 40518 & $ 4435 \pm 157 $ & no LC &-- &-- &-- \\
HIP 41274 & $ 5058 \pm 173 $ & not rotation &-- &-- &-- \\
HIP 41277 & $ 4550 \pm 170 $ &$ 3.7 \pm 0.24 $ & 95.67 & 30.09 - 315.98 & 5.16 - 532.3 \\
HIP 41319 & $ 6511 \pm 319 $ & no signal &-- &-- &-- \\
HIP 42231 & $ 4200 \pm 137 $ & no LC &-- &-- &-- \\
HIP 42343 & $ 6306 \pm 316 $ & no signal &-- &-- &-- \\
HIP 42507 & $ 3955 \pm 153 $ & no signal  & -- &-- &-- \\
HIP 43745 & $ 3936 \pm 148 $ &$ 5.82 \pm 0.56 $ & 158.14 & 55.2 - 396.73 & 8.77 - 574.25 \\
HIP 44162 & $ 6499 \pm 323 $ & no signal &-- &-- &-- \\
HIP 44953 & $ 5322 \pm 239 $ & no signal &-- &-- &-- \\
HIP 45621 & $ 5150 \pm 197 $ &$ 5.48 \pm 0.51 $ & 130.82 & 39.99 - 287.4 & 6.6 - 419.96 \\
HIP 45731 & $ 3709 \pm 87 $ &EB &-- &-- &-- \\
HIP 46385 & $ 3529 \pm 57 $ & $ 10.33 \pm 1.69 $ &-- &-- &-- \\
HIP 47013 & $ 6286 \pm 309 $ & no signal &-- &-- &-- \\
HIP 47261 & $ 4334 \pm 136 $ & contaminated pixel &-- &-- &-- \\
HIP 47300 & $ 7661 \pm 466 $ & no signal &-- &-- &-- \\
HIP 47387 & $ 4657 \pm 178 $ & no LC &-- &-- &-- \\
HIP 47539 & $ 3954 \pm 153 $ & no LC &-- &-- &-- \\
HIP 47741 & $ 3363 \pm 57 $ & no signal &-- &-- &-- \\
HIP 48165 & $ 4854 \pm 214 $ & no signal &-- &-- &-- \\
HIP 48629 & $ 3961 \pm 155 $ & no signal &-- &-- &-- \\
HIP 49046 & $ 3835 \pm 91 $ & no signal &-- &-- &-- \\
HIP 49526 & $ 3713 \pm 87 $ & no signal &-- &-- &-- \\
HIP 49577 & $ 4133 \pm 151 $ &EB &-- &-- &-- \\
HIP 49910 & $ 3902 \pm 132 $ & no signal &-- &-- &-- \\
HIP 51208 & $ 4384 \pm 146 $ & no LC &-- &-- &-- \\
HIP 52339 & $ 3859 \pm 96 $ & no signal &-- &-- &-- \\
HIP 53175 & $ 4211 \pm 126 $ & not rotation &-- &-- &-- \\
HIP 53869 & $ 4937 \pm 197 $ &$ 5.9 \pm 0.51 $ & 102.24 & 32.77 - 263.89 & 5.56 - 436.19 \\
HIP 54094 & $ 4839 \pm 212 $ & no signal &-- &-- &-- \\
HIP 54199 & $ 5114 \pm 187 $ & no LC &-- &-- &-- \\
HIP 55192 & $ 5301 \pm 229 $ & no signal &-- &-- &-- \\
HIP 55409 & $ 5735 \pm 307 $ & no signal &-- &-- &-- \\
HIP 57361 & $ 3516 \pm 54 $ & no signal &-- &-- &-- \\
HIP 57571 & $ 3933 \pm 147 $ & no signal &-- &-- &-- \\
HIP 57645 & $ 5093 \pm 174 $ & no signal &-- &-- &-- \\
HIP 57984 & $ 4214 \pm 123 $ & no signal &-- &-- &-- \\
HIP 59000 & $ 4101 \pm 151 $ &$ 6.56 \pm 0.87 $ & 185.8 & 82.99 - 412.07 & 12.82 - 574.25 \\
HIP 59198 & $ 3831 \pm 89 $ & no LC &-- &-- &-- \\
HIP 59199 & $ 6799 \pm 318 $ & not rotation &-- &-- &-- \\
HIP 59247 & $ 4166 \pm 153 $ & no signal &-- &-- &-- \\
HIP 59767 & $ 6943 \pm 327 $ & not rotation &-- &-- &-- \\
HIP 59905 & $ 4551 \pm 170 $ & no signal &-- &-- &-- \\
HIP 59953 & $ 3801 \pm 79 $ & no signal &-- &-- &-- \\
HIP 60829 & $ 5636 \pm 323 $ & no signal &-- &-- &-- \\
HIP 61947 & $ 3853 \pm 96 $ & no signal &-- &-- &-- \\
HIP 62325 & $ 4937 \pm 197 $ & no LC &-- &-- &-- \\
HIP 62627 & $ 4799 \pm 205 $ & no signal &-- &-- &-- \\
HIP 63419 & $ 5278 \pm 216 $ &$ 9.84 \pm 1.5 $ & 847.11 & 563.46 - 1115.23 & 251.67 - 1373.92 \\
HIP 63421 & $ 3894 \pm 124 $ & no signal &-- &-- &-- \\
HIP 63661 & $ 4120 \pm 150 $ & no signal &-- &-- &-- \\
HIP 65120 & $ 3760 \pm 78 $ & no LC &-- &-- &-- \\
HIP 65602 & $ 5060 \pm 172 $ & no signal &-- &-- &-- \\
HIP 65651 & $ 4355 \pm 140 $ &$ 4.86 \pm 0.4 $ & 106.19 & 33.08 - 337.67 & 5.62 - 542.49 \\
HIP 65706 & $ 3964 \pm 156 $ & no signal &-- &-- &-- \\
HIP 66262 & $ 4470 \pm 161 $ & no signal &-- &-- &-- \\
HIP 66315 & $ 4580 \pm 173 $ & no signal &-- &-- &-- \\
HIP 66587 & $ 3824 \pm 89 $ & no LC &-- &-- &-- \\
HIP 66828 & $ 4186 \pm 150 $ &$ 4.39 \pm 0.28 $ & 107.2 & 33.08 - 344.13 & 5.62 - 547.66 \\
HIP 69311 & $ 4399 \pm 150 $ &$ 6.45 \pm 0.62 $ & 129.59 & 43.96 - 357.44 & 7.19 - 547.66 \\
HIP 69333 & $ 4416 \pm 153 $ & no signal &-- &-- &-- \\
HIP 69860 & $ 3729 \pm 84 $ & no signal &-- &-- &-- \\
HIP 69963 & $ 3762 \pm 77 $ & no signal &-- &-- &-- \\
HIP 70472 & $ 4126 \pm 150 $ & no LC &-- &-- &-- \\
HIP 71243 & $ 6550 \pm 310 $ & no signal &-- &-- &-- \\
HIP 71515 & $ 4999 \pm 185 $ & no LC &-- &-- &-- \\
HIP 71899 & $ 6268 \pm 311 $ & $ 2.53 \pm 0.11 $ &-- &-- &-- \\
HIP 71957 & $ 6497 \pm 324 $ & no signal &-- &-- &-- \\
HIP 73121 & $ 4924 \pm 200 $ & no LC &-- &-- &-- \\
HIP 73183 & $ 4094 \pm 152 $ & no signal &-- &-- &-- \\
HIP 73252 & $ 4045 \pm 154 $ & no signal &-- &-- &-- \\
HIP 73787 & $ 4294 \pm 128 $ & no LC &-- &-- &-- \\
HIP 74926 & $ 4266 \pm 125 $ & no LC &-- &-- &-- \\
HIP 79203 & $ 6599 \pm 305 $ & no LC &-- &-- &-- \\
HIP 79593 & $ 4260 \pm 125 $ & no LC &-- &-- &-- \\
HIP 81655 & $ 3551 \pm 57 $ & no signal &-- &-- &-- \\
HIP 82260 & $ 5473 \pm 331 $ & no LC &-- &-- &-- \\
HIP 83676 & $ 4871 \pm 211 $ & no LC &-- &-- &-- \\
HIP 83929 & $ 4177 \pm 154 $ &$ 6.19 \pm 0.63 $ & 162.7 & 60.69 - 392.99 & 9.55 - 568.83 \\
HIP 87370 & $ 5831 \pm 319 $ & no LC &-- &-- &-- \\
HIP 88962 & $ 3949 \pm 151 $ & no LC &-- &-- &-- \\
HIP 89320 & $ 4206 \pm 131 $ & contaminated pixel &-- &-- &-- \\
HIP 89449 & $ 4190 \pm 147 $ & no signal &-- &-- &-- \\
HIP 90306 & $ 3767 \pm 77 $ & no signal &-- &-- &-- \\
HIP 93072 & $ 3824 \pm 89 $ & no LC &-- &-- &-- \\
HIP 98007 & $ 4765 \pm 198 $ & no signal &-- &-- &-- \\
HIP 99969 & $ 4467 \pm 161 $ & no LC &-- &-- &-- \\
HIP 100133 & $ 3769 \pm 76 $ & no LC &-- &-- &-- \\
HIP 101852 & $ 6004 \pm 335 $ &$ 6.87 \pm 0.54 $ & 913.87 & 493.41 - 1914.67 & 276.7 - 2693.67 \\
HIP 102119 & $ 4217 \pm 121 $ & no LC &-- &-- &-- \\
HIP 104687 & $ 5832 \pm 319 $ & no LC &-- &-- &-- \\
HIP 105504 & $ 4228 \pm 122 $ & no LC &-- &-- &-- \\
HIP 107317 & $ 3432 \pm 45 $ & $ 6.18 \pm 0.51 $ &-- &-- &-- \\
HIP 108036 & $ 6760 \pm 321 $ & no LC &-- &-- &-- \\
HIP 109807 & $ 4279 \pm 126 $ & no LC &-- &-- &-- \\
HIP 110106 & $ 3841 \pm 93 $ & no signal &-- &-- &-- \\
HIP 110401 & $ 4240 \pm 123 $ & no signal &-- &-- &-- \\
HIP 110663 & $ 5420 \pm 289 $ &$ 7.3 \pm 0.55 $ & 457.37 & 263.89 - 619.51 & 123.59 - 1554.17 \\
HIP 111976 & $ 3904 \pm 134 $ & no signal &-- &-- &-- \\
HIP 113159 & $ 6433 \pm 347 $ & no signal &-- &-- &-- \\
HIP 115280 & $ 6427 \pm 348 $ & no signal &-- &-- &-- \\
HIP 115411 & $ 5515 \pm 331 $ & no signal &-- &-- &-- \\
HIP 116973 & $ 4614 \pm 177 $ & no signal &-- &-- &-- \\
HIP 117159 & $ 5131 \pm 197 $ & no signal &-- &-- &-- \\
HIP 117235 & $ 3947 \pm 151 $ & no signal &-- &-- &-- \\
HIP 117410 & $ 4174 \pm 154 $ &2 periods & 254.07 & 64.24 - 625.41 & 10.11 - 1400.23 \\
HIP 117795 & $ 4342 \pm 137 $ & no signal &-- &-- &-- \\
HIP 118086 & $ 4035 \pm 155 $ & no signal &-- &-- &-- \\
HIP 118310 & $ 4501 \pm 165 $ & no signal &-- &-- &--
\label{tab:gyro}
\end{longtable}

\end{document}